\documentclass[submission, Phys]{SciPost}
\usepackage{lineno}
\usepackage{amsmath}
\usepackage{tikz}
\usepackage{braket}
\usetikzlibrary{3d,arrows.meta,calc,decorations.markings,positioning,fit,backgrounds}
\usepackage{subcaption}
\usepackage{caption}
\usepackage{diagbox}
\usepackage{booktabs,array,longtable,mathrsfs}
\hypersetup{
    colorlinks,
    linkcolor={red!50!black},
    citecolor={blue!50!black},
    urlcolor={blue!80!black}
}

\usepackage[bitstream-charter]{mathdesign}
\DeclareSymbolFont{usualmathcal}{OMS}{cmsy}{m}{n}
\DeclareSymbolFontAlphabet{\mathcal}{usualmathcal}

\newcommand{\sixj}[6]{
    \left[
    \begin{tabular}{@{}c@{}}
        $#1\ $ $#2\ $ $#3$ \\
        $#4\ $ $#5\ $ $#6$
    \end{tabular}
    \right]
}

\newcommand{\Fsymbol}[6]{%
  F_{#5,#6}
  \left[
    \begin{array}{@{}c@{}}
      \scriptstyle #2 \; \scriptstyle #3 \\[-3pt]
      \scriptstyle #1 \; \scriptstyle #4
    \end{array}
  \right]
}

\newcommand{\Ishi}[1]{|#1\rangle\rangle}

\begin{document}

\begin{center}{\Large \textbf{
Discretising 2D Fermionic CFTs
}}\end{center}

\begin{center}
\begin{tabular}{@{}c@{}}
Dazhuo Cheng$^{1,\#}$,
Hexuan Li$^{1,\#}$,
Gong Cheng$^{4,\star}$ and Ling-Yan Hung$^{1,2,3,\dagger}$
\end{tabular}
\end{center}

\begin{center}
\textbf{1} Qiuzhen College, Tsinghua University,
Beijing 100084, China
\\
\textbf{2} Yau Mathematical Sciences Center, Tsinghua University,
Beijing 100084, China
\\
\textbf{3} Beijing Institute of Mathematical Sciences and Applications,
Beijing 101408, China
\\
\textbf{4} Department of Physics, The Chinese University of Hong Kong,
Sha Tin, New Territories, Hong Kong, China
\\[6pt]
\begin{tabular}{@{}c@{}}
{\small \textsuperscript{\#} These authors contributed equally to this work.}
\\[3pt]
${}^{\star}$ {\small\sffamily gongchphy@gmail.com}
\\[2pt]
${}^{\dagger}$ {\small\sffamily lyhung@tsinghua.edu.cn}
\end{tabular}
\end{center}

\begin{center}
November 8, 2026
\end{center}


\definecolor{palegray}{gray}{0.95}

\section*{Abstract}
{\bf
We extend the exact fixed-point tensor construction of Cheng et al.~\cite{Cheng2023} to fermionic minimal models, providing an analytic construction of fixed-point tensors for fermionic rational conformal field theories. The fermionic fixed-point tensor is defined as a correlation function of boundary-changing operators on a triangle, with careful treatment of fermionic parity and spin structure. Boundary tensors are also constructed and combined with triangle tensors to form strips, enabling the extraction of scaling dimensions for boundary-changing operators. The construction incorporates modified weights for shrinkable boundary conditions, systematic depiction of spin structures, recovering both the Neveu-Schwarz and Ramond sectors. As a benchmark, our tensor with a minimal truncation of descendant levels yields convergence to the exact bulk scaling dimensions of many low energy states for the free Majorana fermion ($c=1/2$), 
demonstrating the efficiency of approximate fixed point tensors from boundary changing operators. 
We also make use of a more general description of fermionic CFTs based on fermionisation of rational bosonic CFTs, and make a generic prescription for discretisation making use of their connections to modular tensor categories. We show that the discretised spin structures on a spin surface naturally recover the signs needed to recover the R/NS sectors. We propose that the collection of OPE coefficients of the discretised 2d CFT can be naturally interpreted as a state-sum of spin-TQFTs, analogous to the bosonic case. 
This work provides analytically computed starting points for tensor network simulations of fermionic critical phenomena.

}

\vspace{10pt}
\noindent\rule{\textwidth}{1pt}
\tableofcontents\thispagestyle{fancy}
\noindent\rule{\textwidth}{1pt}
\vspace{10pt}

\section{Introduction}\label{sec:introduction}

Understanding strongly interacting field theory is one of the most daunting and fundamental challenges of theoretical physics. One difficulty is that away from perturbative regimes we do not have a clear definition of the quantum measure of the fields in the path integral. The difficulty of identifying a quantum measure is not merely a formal mathematical nicety; it obstructs the formulation of any concrete computational strategy, even in principle.
There are many powerful numerical methods that can deal with strongly interacting systems, such as the tensor network method, particularly in low dimensions. But they can only be applied if the problem has a natural discrete description---for instance, in lattice models.
To utilize these methodologies in interacting quantum field theory, it is natural to ask how an apparently continuous field theory could acquire a discrete formulation.

A key step was taken in~\cite{Hung2019,Brehm:2021wev,Chen2022,Cheng2023}, where it was realized that, at least for two-dimensional bosonic rational conformal field theories, one can rigorously discretize a path integral by leveraging the idea of cutting and gluing to decompose the path integral over a two-dimensional Riemann surface into open pairs of pants (triangular patches). These are conformally related to three-point correlation functions of boundary-changing operators (BCOs) on the real line bounding the upper half-plane. The real line is colored by conformal boundary conditions except at the points of insertion of the BCOs. The resulting manifold contains holes bounded by conformal boundary conditions.
Then one removes the holes in the manifold by a weighted sum over conformal boundary conditions---these linear combinations of conformal boundary conditions are called ``entanglement brane boundary conditions''~\cite{Hung2019} and were later called either ``cloaking boundaries''~\cite{Brehm:2021wev} or ``shrinkable boundaries''~\cite{Chen2022,Cheng2023}.

Such a decomposition of continuous path integrals is essentially a tensor-network representation of the path integral. It was shown that such a tensor network satisfies the required properties of a ``Fixed-Point Tensor'' (FP tensor) under tensor-network renormalization (TNR) algorithms~\cite{Levin2006,Gu2008,Gu2009,evenbly2015tensor}. The topological holographic principle underlies this success~\cite{gaiotto2015generalized, freed2024topological, ji2020categorical, Kong:2020cie, Kong:2020jne, Chatterjee:2022kxb, apruzzi2023symmetry, Freed:2018cec, Kong:2019byq, bhardwaj2018finite, Albert:2021vts, vanhove2018mapping, aasen2016topological, aasen2020topological}. The discretization makes use of the non-invertible symmetries of the CFT and encodes a three-dimensional Turaev--Viro TQFT that imposes the appropriate symmetries. The FP tensor construction was extended to free boson CFTs beyond rational points~\cite{cheng2026} and inspired a symmetry-preserving TNR framework that stabilizes critical flows under suitable categorical constraints and accurately extracts conformal data~\cite{Bao2025, Meng2026}. In a complementary development, this discretization approach also bridges the two-dimensional Liouville CFT and a path integral over three-dimensional hyperbolic geometries, making a clear and concrete connection between the topological holographic principle and the AdS/CFT correspondence, at least in low dimensions~\cite{Chen:2024unp, Bao:2024ixc}.

It is natural and important to generalize this program to include fermionic CFTs---for unitary relativistic theories, this amounts to studying spin CFTs in two dimensions, which will be the main focus of this paper. In a fermionic CFT, one can assign different spin structures on a non-contractible cycle, which produces different Hilbert spaces in the Neveu-Schwarz (NS) and Ramond (R) sectors, respectively. Each Hilbert space admits a $\mathbb{Z}_2$ grading by fermion parity. Therefore, our discretization program should include prescriptions for encoding these different spin structures---the explicit implementation of these prescriptions in the BCO tensor contraction is a central part of this paper.

To be concrete, in this paper we generalize the exact FP tensor construction to the free Majorana fermion ($c=1/2$), the simplest fermionic CFT, which is part of the fermionic A-type minimal model family. Our work proceeds along five main axes. First, we define the fermionic FP tensor as a correlation function of boundary-changing operators on a triangle, with careful treatment of fermionic parity. Second, we verify that the fermionic FP tensor satisfies the appropriate coarse-graining conditions, thereby establishing its algebraic consistency. Third, we consider the path integral on a cylinder and propose a prescription for encoding the NS or R spin structure along the non-contractible cycle. Fourth, we extend the construction to boundary tensors and open-string spectra, and give the partition functions a topological interpretation via a strange correlator. Last but not least, we give a general prescription to discretising general fermionic CFTs via their connections to bosonic CFTs through fermionisation. This produces as a by-product a state-sum of a spin-TQFTs in 3 dimensions related to a bosonic one by proliferation of defects. 
To test our proposal, we compute the resultant spectrum of the CFT from the cylinder with different spin structures. In a practical computation, we have to truncate the infinite bond dimension by keeping only a finite number of descendants in each Virasoro family, and compute the low-lying bulk conformal dimensions via the transfer matrix on a cylinder. For the free Majorana fermion, our results converge to the exact values ($\Delta = 0, 1/2, 1, 3/2, 2$ in the NS sector) using only a minimal number of descendant levels, demonstrating the efficiency of the BCO basis. The construction generalizes naturally to all fermionic A-series minimal models; the details of this generalization are given in Section~\ref{sec:fermionization} and Appendix~\ref{app:fermionic_models}.

The paper is organised as follows. Section~\ref{sec:discretisation-review} reviews the bosonic discretisation. Section~\ref{sec:fermionic-bcft} develops the free-Majorana boundary data and fixed-point tensors, and tests the truncated cylinder spectrum. Section~\ref{sec:boundary-tensors} constructs boundary tensors and studies open-string propagation on a strip. Section~\ref{sec:fermionization} then gives the general TQFT construction, explains how the Nakayama data encode the spin structure of an open-string closure, and discusses a candidate three-dimensional bulk interpretation. We conclude in Section~\ref{sec:conclusion}. Appendix~\ref{app:combined} collects the algebraic conventions, the detailed A-series boundary and BCO solutions, the discrete spin moves and disk calculations, and the three-dimensional interpretation of boundary tensors.

\section{Discretizing Bosonic CFTs: A Review}\label{sec:discretisation-review}

As outlined in the Introduction (Section~\ref{sec:introduction}), the discretization of a bosonic rational CFT proceeds by cutting the path integral into triangular patches bounded by conformal boundary segments and by removing the resulting holes through weighted sums over boundary conditions. In this section we review this construction in detail, following~\cite{Hung2019,Brehm:2021wev,Chen2022,Cheng2023}; it forms the bosonic template that we generalize to fermionic CFTs in the remainder of the paper.

\subsection{Tensor Structure in Bosonic RCFTs}\label{subsec:bosonic_tensor}

The bosonic FP tensor~\cite{Cheng2023} is a rank-nine object written as $\mathcal{T}_{(i,I)(j,J)(k,K)}^{abc}$.
Its superscripts $a,b,c$ label the conformal boundary conditions of the RCFT,
while the subscripts $i,j,k$ denote the primary boundary-changing operators (BCOs)
living on the segments $ab$, $bc$, and $ca$, respectively.
Within each primary family, $I,J,K$ index the descendant fields.
The indices $a,b,c$ and $i,j,k$ range over a finite set characteristic of the RCFT,
whereas $I,J,K$ inhabit an infinite-dimensional space,
so the exact FP tensor has unbounded bond dimension.

\begin{figure}[htbp]
\centering
\begin{subfigure}{0.55\textwidth}
    \centering
    \includegraphics[width=\linewidth]{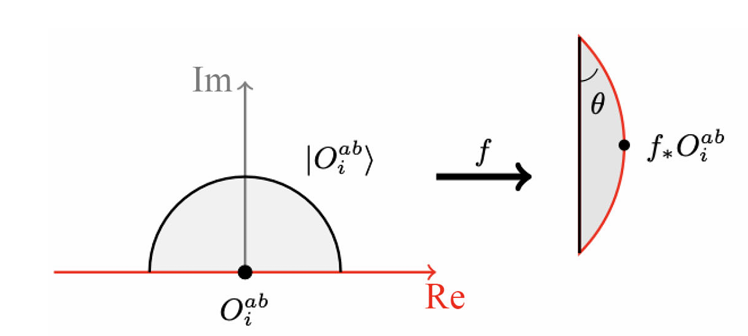}
    \caption{}\label{fig:states}
\end{subfigure}
\begin{subfigure}{0.35\textwidth}
    \centering
    \includegraphics[width=\linewidth]{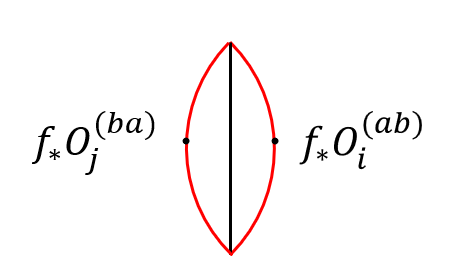}
    \caption{}\label{fig:nut_shape}
\end{subfigure}
    \caption{(a) The semicircle is mapped to a bow-shaped segment through the conformal map $f(z)$. (b) The nut-shaped region obtained by gluing two segments along the vertical edge.} 
\end{figure}
The FP tensor is built in two steps. First, one prepares states on a semicircle by placing a BCO at the origin. Under the conformal map $f(z)$, the semicircle is mapped to a bow-shaped segment (Fig.~\ref{fig:states}), yielding a state $|\phi_i^{ab}\rangle$ along the vertical edge. These states are made orthonormal by demanding that the two-point function on a nut-shaped region (Fig.~\ref{fig:nut_shape}) remains diagonal.
\begin{figure}[htbp]
\centering
\begin{subfigure}{0.44\textwidth}
    \centering
    \includegraphics[width=0.8\linewidth]{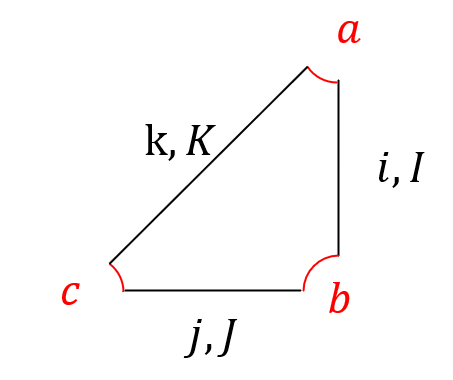}
    \caption{}\label{fig:triangle_tensor}
\end{subfigure}
\begin{subfigure}{0.54\textwidth}
    \centering
    \includegraphics[width=0.8\linewidth]{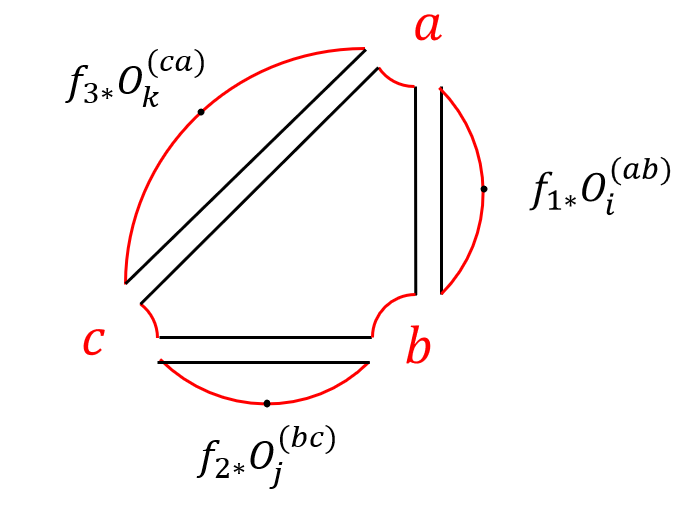}
    \caption{}\label{fig:triangle_fmap}
\end{subfigure}
    \caption{(a) The shape of the tensor. (b) The amplitude of three open states can be calculated by attaching three segment regions along the open boundaries of the triangle.} 
\end{figure}

In the second step, the FP tensor is defined via the CFT path integral on an open triangle with rounded corners that carry boundary conditions $a,b,c$ (Fig.~\ref{fig:triangle_tensor}).
Each edge of the triangle hosts one of the states $|\phi_i^{ab}\rangle$, $|\phi_j^{bc}\rangle$, $|\phi_k^{ca}\rangle$ prepared previously.

\begin{figure}[htbp]
\centering
\includegraphics[width=0.3\textwidth]{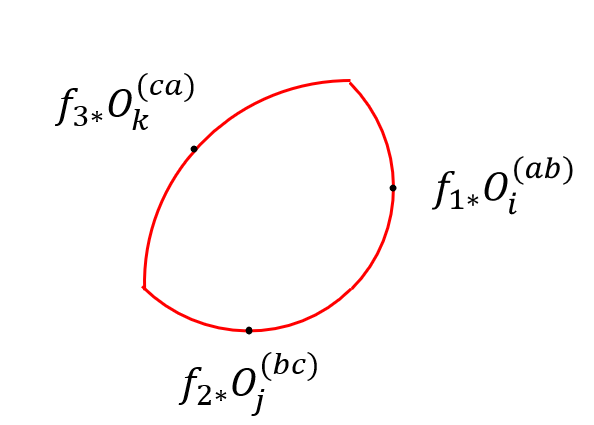}
\caption{The conformal transformation $g$ maps the region to the upper half-plane.}\label{fig:disk_with_two_cusps}
\end{figure}
This path integral can be recast as a three-point correlation function. The maps $f_1,f_2,f_3$ are obtained by translating and rotating $f(z)$ to match the orientation of each edge (Fig.~\ref{fig:triangle_fmap}); each $f_i$ sends a semicircle to a bow-shaped domain. An additional map $g(\xi)$ takes the region of Fig.~\ref{fig:disk_with_two_cusps} to the upper half-plane (UHP):
\begin{equation}
    g(\xi) = {\left( -i \frac{\xi + (1+i)}{\xi - (1+i)} \right)}^{\frac{4}{3}}.
\end{equation}
Composing these maps, $\chi_i = g \circ f_i$ pushes the operators onto the UHP, producing $\chi_{1*} O_{(i,I)}^{ab}(x_1)$, $\chi_{2*} O_{(j,J)}^{bc}(x_2)$, and $\chi_{3*} O_{(k,K)}^{ca}(x_3)$.

Consequently, the FP tensor equals the three-point correlator of these pushed-forward operators on the UHP: 
\begin{equation}
    \mathcal{T}_{(i,I)(j,J)(k,K)}^{abc} =
    \langle \chi_{1*} O_{(i,I)}^{ab}(x_1) \chi_{2*} O_{(j,J)}^{bc}(x_2)
    \chi_{3*} O_{(k,K)}^{ca}(x_3) \rangle_{\mathrm{UHP}}.
\end{equation}

This quantity decomposes into a structure constant $C_{ijk}^{abc}$ multiplied by a three-point conformal block $\alpha_{IJK}^{ijk}$:
\begin{equation}
    \mathcal{T}_{(i,I)(j,J)(k,K)}^{abc} = \alpha_{IJK}^{ijk} C_{ijk}^{abc}.
\end{equation}
The structure constants incorporate the fusion rules and the dependence on boundary conditions, whereas the conformal blocks encode the descendant level information and the triangle geometry.

\subsection{Fixed-Point Conditions and Shrinkable Boundaries}\label{sec:fixed-point-conditions}

\begin{figure}[htbp]
\centering
\begin{subfigure}{0.49\textwidth}
    \centering
    \includegraphics[width=\linewidth]{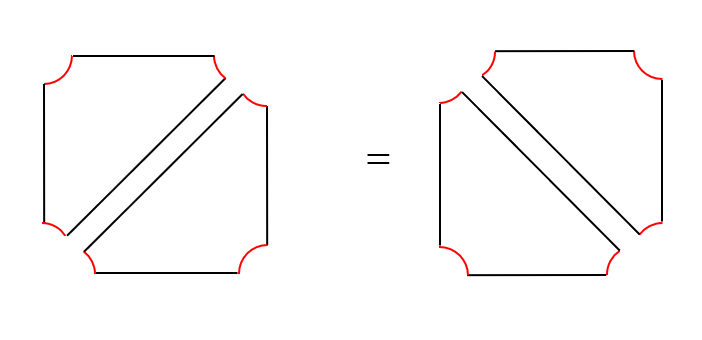}
    \caption{}\label{fig:crossing}
\end{subfigure}
\begin{subfigure}{0.49\textwidth}
    \centering
    \includegraphics[width=\linewidth]{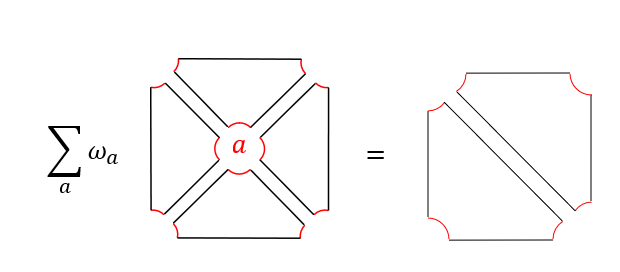}
    \caption{}\label{fig:coarsegrain}
\end{subfigure}
    \caption{(a) Crossing relation for the FP tensor. (b) Coarse-graining condition for the FP tensor.} 
\end{figure}
For a tensor to qualify as a renormalization-group fixed point, it must obey two crucial properties. The first is crossing symmetry: the FP tensor is required to satisfy a crossing relation, diagrammed in Fig.~\ref{fig:crossing} and written as
\begin{equation}
    \sum_{m,M} \mathcal{T}_{(i,I)(j,J)(m,M)}^{abc} \mathcal{T}_{(k,K)(l,L)(m,M)}^{cda}
    = \sum_{n,N} \mathcal{T}_{(l,L)(i,I)(n,N)}^{dab} \mathcal{T}_{(j,J)(k,K)(n,N)}^{bcd}.
\end{equation}
The second is the coarse-graining condition: contracting four FP tensors over their shared pairwise indices must recover a single rank-4 FP tensor up to an overall weight. Illustrated in Fig.~\ref{fig:coarsegrain}, this condition involves summing over intermediate boundary conditions with the weight $w_i = S_{11}^{1/2} S_{i1}$, where $S$ denotes the modular $S$-matrix. The physical interpretation is that four triangles are glued into a square with a central hole; this hole is then eliminated by a weighted boundary-condition sum that makes it shrinkable.

This weighted sum is inspired by the notion of a ``shrinkable boundary condition'' due to Hung and Wan~\cite{Hung2019}.
For diagonal RCFTs, performing the same weighted sum over Cardy states $|i\rangle_c$ produces the vacuum Ishibashi state:
\begin{equation}
    \sum_i w_i |i\rangle = |0\rangle\rangle_c.
\end{equation}
As the size of the central hole goes to zero, the vacuum state dominates the Ishibashi-state expansion. The hole therefore effectively disappears, and contracting the four tensors yields a rank-4 tensor associated with a square without a hole.

\subsection{Tiling the Partition Function and Extracting the Spectrum}

Armed with these properties, one can tile the FP tensors over a triangulated manifold to recover the full CFT partition function:
\begin{equation}\label{eq:partition_function_bosonic}
    Z_M = \sum_{\{(i,I)\}, \{a\}} \prod_v \omega_a \prod_{\triangle} \mathcal{T}_{(i,I)(j,J)(k,K)}^{abc},
\end{equation}
where the sum runs over all possible assignments of $(i,I)$ and $a$ to BCOs and boundary conditions, respectively;
the factor $\prod_v \omega_a$ assigns a vertex weight $\omega_a$ determined by the boundary condition $a$ at that vertex---these are precisely the shrinkable weights $w_a$ introduced in Section~\ref{sec:fixed-point-conditions};
the product $\prod_{\triangle} \mathcal{T}_{(i,I)(j,J)(k,K)}^{abc}$ runs over all triangles, each contributing an FP tensor whose indices match the three BCOs and boundary conditions on its edges and corners.

A key application of this formalism is extracting the bulk conformal data through a cylinder construction~\cite{Cheng2023}.
The FP tensors---interpreted as correlation functions of BCOs on triangular patches---are assembled into a cylinder along the Euclidean time direction.
The resulting cylinder partition function describes Euclidean time evolution; the transfer-matrix eigenvalues take the form $N e^{-(2\pi/n)(\Delta - c/12)}$, where $n$ counts the number of tensor blocks around the circumference, $\Delta$ is the scaling dimension of the corresponding eigenstate and $N$ is an overall factor. 

Diagonalizing this transfer matrix yields both $\Delta$ and the central charge $c$ of the CFT\@.
This method accurately reproduces the spectrum of various minimal models~\cite{Cheng2023}---including the Ising, Yang-Lee, and tricritical Ising models---even with only a few descendant levels kept.

\section{Fermionic BCFT: the Free Majorana Fermion}\label{sec:fermionic-bcft}

The bosonic construction reviewed in Section~\ref{sec:discretisation-review} turns boundary three-point functions into local tensors and removes the holes of the resulting triangulation by a weighted sum over conformal boundary conditions. Extending this construction to a spin CFT requires additional data and consistency conditions. Boundary fields are $\mathbb{Z}_2$-graded, boundary zero modes can require auxiliary boundary degrees of freedom, and gluing fermionic tensors produces signs that must retain the spin-structure dependence of the continuum theory. In this section we make these ingredients explicit for the free Majorana fermion, the simplest fermionic CFT.

We first determine the boundary data in both the open- and closed-string channels. The boundary Majorana mode and Chan--Paton factors lead to a set of fermionic branes and graded boundary-changing operators (BCOs), while the spin Cardy condition fixes the corresponding boundary states. We then obtain the BCO three-point coefficients from the sewing constraints and use them to construct the fixed-point tensor in Section~\ref{sec:FPtensor}. Finally, we formulate the fermionic contraction rules and shrinkable-boundary sum, verify the coarse-graining condition, and assemble the tensors on a cylinder. The signs associated with the non-contractible cycle distinguish the Neveu--Schwarz and Ramond spin structures, providing the prototype for the general A-series construction developed in Section~\ref{sec:fermionization} and Appendix~\ref{app:fermionic_models}.

\subsection{Free Fermion CFT}\label{subsec:ferm_cft}

The free Majorana fermion has central charge $c=1/2$, and its Hilbert space on a cylinder splits into two distinct sectors determined by the periodicity of the fermion around the spatial cycle: the Neveu-Schwarz (NS) sector, where the fermion is antiperiodic, and the Ramond (R) sector, where it is periodic. The NS-sector primary fields are the identity $1$ ($h=\bar h=0$), the energy density $\epsilon$ ($h=\bar h=1/2$), the holomorphic fermion $\psi$ ($h=1/2,\bar h=0$), and its antiholomorphic partner $\bar\psi$ ($h=0,\bar h=1/2$); the R-sector primary fields are the spin field $\sigma$ and the disorder operator $\mu$ (both with $h=\bar h=1/16$).
We define fermion parity $(-1)^F$ in the NS sector by the parity of the number of fermionic modes acting on the NS vacuum. In the R sector, we choose the two Ramond ground states to have opposite fermion parity, and denote the even state as $\sigma$ and the odd state $\mu$. Thus $1$, $\epsilon$, and $\sigma$ are called bosonic ($(-1)^F=+1$), while $\psi$, $\bar\psi$, and $\mu$ are called fermionic ($(-1)^F=-1$).

\subsection{Free Fermion Boundaries and BCOs}\label{sec:free-fermion-bc}

\subsubsection{Boundary conditions and the boundary fermion}\label{sec:free-fermion-boundaries-bcos}

We next turn to the boundary data needed for the tensor construction: for a free Majorana fermion on the upper half-plane, a conformal boundary forces the bulk field to satisfy
\begin{equation}
  \psi(z) = \pm \tilde\psi(z) \qquad (z \in \mathbb{R}),
\end{equation}
which defines the $B_+$ and $B_-$ boundaries, respectively~\cite{Hori2009}.
When both ends carry the same boundary type, the open-string mode expansion has half-integer grading. With a stretch of notation we call that ''Neveu-Schwarz'' type even though here it is not connected to spin structure:
\begin{equation}
  \psi(z) = \sum_{r\in\mathbb{Z}+1/2} \psi_r \, z^{-r-1/2},
  \qquad
  \{\psi_r,\psi_s\} = \delta_{r+s,0},
  \label{eq:mode_expansion_NS_open}
\end{equation}
For a pair of opposite boundaries ($B_-/B_+$ or $B_+/B_-$), the expansion is of integer grading, or ``Ramond type'',
\begin{equation}
  \psi(z) = \sum_{n\in\mathbb{Z}} \psi_n \, z^{-n-1/2},
  \qquad
  \{\psi_n,\psi_m\} = \delta_{n+m,0},
  \label{eq:mode_expansion_R_open}
\end{equation}
which includes a Majorana zero mode $\psi_0$ with $\{\psi_0,\psi_0\} = 1$.

This zero mode obstructs the implementation of fermion parity
within a single ordinary (ungraded) irreducible representation:
no operator $(-1)^F$ satisfying
$\{(-1)^F,\psi_0\}=0$ exists on such a representation~\cite{Hori2009}.
The standard workaround is to introduce an additional free boundary
Majorana fermion $\xi$ localized at the $B_-$ boundary~\cite{Hori2009},
obeying
\begin{equation}
  \{\xi,\xi\} = 1,\qquad \{\xi,\psi_n\} = 0.
\end{equation}

\subsubsection{BCO spectrum}\label{sec:bco-spectrum-boundary-fermion}

With the boundary fermion $\xi$ in place, the boundary field content in each open-string sector follows from a standard mode-counting argument~\cite{Hori2009}; the results are collected in Table~\ref{tab:bco_free_fermion_content}. Note that, up to this point, the labels $1_B$ and $a_B$ denote the two $h = 0$ primary open-string states.

\begin{table}[h]
  \centering
\renewcommand{\arraystretch}{1.15}
  \caption{Boundary fields in each open-string sector of the free-fermion model, after introducing a boundary fermion at $B_-$.}\label{tab:bco_free_fermion_content}
  \begin{tabular}{ll}
\toprule
Sector & Boundary fields (conformal weight)  \\
\midrule
$B_+/B_+$ & $1_B$ ($h=0$),\ $\psi_B$ ($h=1/2$) \\
    $B_-/B_-$ & $1_B$ ($h=0$),\ $\epsilon_B$ ($h=1/2$),\ $a_B$ ($h=0$),\ $\psi_B$ ($h=1/2$) \\
    $B_-/B_+$ & $\sigma_B$ ($h=1/16$),\ $\mu_B$ ($h=1/16$) \\
\bottomrule
\end{tabular}
\end{table}

When one wants to introduce the $(-1)^F$ action in the open channel, one must consider its action on the Chan--Paton vector space~\cite{Hori2009} carried by the boundary. In the following we will adopt string theoretic language and call a conformal boundary a brane. Each brane carries a Chan--Paton (CP) vector space, which corresponds to a representation of the Fermion parity $\mathbb{Z}_2$ group, generated by $\{1,(-1)^F\}$. 
Thus, we have four different branes, namely $B_{\pm,0}$ and $B_{\pm,1}$, where $0$ and $1$ stand for the trivial and standard representations of $\mathbb{Z}_2$.

The $(-1)^F$ action on the open-string sector is then defined as the product of its action on the string itself and its actions on the two CP vector spaces. A BCO is called bosonic if it is even under this action and fermionic if it is odd.

One may notice that if an open string is stretched between two branes and one of them is $B_-$, then the open-sector spectrum does not depend on the choice of representation, meaning that $B_{-,0}$ and $B_{-,1}$ are indistinguishable. In the following, the brane arising from $B_-$ will be denoted simply by $B_-$.

The final BCO spectrum between the branes is summarized in Table~\ref{tab:bco_freeferm_d-branes}. Among these BCOs, $1_B$ (weight $0$), $\epsilon_B$ (weight $1/2$), and $\sigma_B$ (weight $1/16$) are bosonic, while $\psi_B$ (weight $1/2$), $a_B$ (weight $0$), and $\mu_B$ (weight $1/16$) are fermionic.

\begin{table}[h]
  \centering
\renewcommand{\arraystretch}{1.15}
  \caption{Boundary-changing operators between branes of the free-fermion theory.}\label{tab:bco_freeferm_d-branes}
  \begin{tabular}{cccc}
\toprule
& $B_{+,0}$ & $B_{+,1}$ & $B_{-}$\\
\midrule
$B_{+,0}$ & $1_B,\ \psi_B$ & $\epsilon_B,\ a_B$ & $\sigma_B,\ \mu_B$\\
    $B_{+,1}$ & $\epsilon_B,\ a_B$ & $1_B,\ \psi_B$ & $\sigma_B,\ \mu_B$\\
    $B_{-}$ & $\sigma_B,\ \mu_B$ & $\sigma_B,\ \mu_B$ & $1_B,\ \epsilon_B,\ a_B,\ \psi_B$\\
\bottomrule
\end{tabular}
\end{table}
\subsection{Boundary States and the Spin Cardy Condition}\label{sec:boundary-states-spin-Cardy}

So far we have worked in the open-string channel. Conformal boundary forming a closed circle admits a closed-string description. They are expressed in terms of coherent closed-string states annihilated by $L_n - \bar L_{-n}$, where $L_n$ and $\bar L_n$ are the holomorphic and anti-holomorphic Virasoro operators. Ishibashi states form basis of such states, which can be written in terms of the closed-sector mode expansion as
\begin{align}
|B_{\pm}\rangle\rangle_{\text{NS}} &= \exp\!\left( \pm i \sum_{\substack{r>0,\\ r\in \mathbb{Z}+1/2}} \psi_{-r} \tilde{\psi}_{-r} \right) |0\rangle, \\
|B_{\pm}\rangle\rangle_{\text{R}} &= \exp\!\left( \pm i \sum_{\substack{n>0,\\ n\in \mathbb{Z}}} \psi_{-n} \tilde{\psi}_{-n} \right) |\pm\rangle,
\end{align}
where $|0\rangle$ is the Neveu-Schwarz ground state and $|\pm\rangle$ are the two Ramond ground states.  The spin structure around the boundary circle $S^1$ determines the sector: antiperiodic fermions correspond to the NS sector, while periodic fermions correspond to the R sector.

The boundary states associated with the branes must satisfy the spin Cardy consistency condition: the cylinder amplitude computed in the closed-string channel must agree with the partition function computed in the open-string channel~\cite{Hori2009,Ebisu2021}. More specifically, 
\begin{equation}\label{eq:spin_cardy}
  \begin{aligned}
    {}_{\text{NS}}\langle a,\rho | e^{-L_1 H_{\text{closed}}} | b,\sigma\rangle_{\text{NS}} &= \operatorname{Tr}_{\mathcal{H}_{a,\rho|b,\sigma}} e^{-L_2 H_{\text{open}}}, \\
    {}_{\text{R}}\langle a,\rho | e^{-L_1 H_{\text{closed}}} | b,\sigma \rangle_{\text{R}} &= \operatorname{Tr}_{\mathcal{H}_{a,\rho|b,\sigma}} {(-1)}^F e^{-L_2 H_{\text{open}}}.
  \end{aligned}
\end{equation}

Here, $a$ and $b$ label the boundary conditions $B_{\pm}$, while $\rho$ and $\sigma$ label the representations $0$ and $1$. Matching the BCO spectra of Sec.~\ref{sec:bco-spectrum-boundary-fermion} against the closed-channel overlaps fixes the Cardy-state normalizations: 

\begin{equation}
|B_{+,0}\rangle_{\text{NS}} = |B_+\rangle\rangle_{\text{NS}},\qquad
|B_{+,0}\rangle_{\text{R}}   = {\sqrt[4]{2}}|B_+\rangle\rangle_{\text{R}},
\end{equation}
\begin{equation}
|B_{+,1}\rangle_{\text{NS}} = |B_+\rangle\rangle_{\text{NS}},\qquad
|B_{+,1}\rangle_{\text{R}}   = -{\sqrt[4]{2}} |B_+\rangle\rangle_{\text{R}},
\end{equation}
\begin{equation}
|B_-\rangle_{\text{NS}} = \sqrt{2}\,|B_-\rangle\rangle_{\text{NS}},\qquad
|B_-\rangle_{\text{R}} = 0,
\end{equation}
where $|B_-\rangle_{\text{R}} = 0$ because the boundary fermion on $B_-$ has an unsaturated zero mode for periodic boundary conditions; integration over this Grassmann zero mode sets the boundary path integral to zero~\cite{Hori2009}. 

It is instructive to relate these boundary states to the standard Cardy states of the Ising ($c=1/2$) minimal model~\cite{DiFrancesco1997}. The Ising model is related to the free fermion model by the GSO projection of the latter, and the standard Ising Cardy states are
\begin{equation}
\begin{split}
|0\rangle_c =& \frac{1}{\sqrt{2}}\Ishi{0} + \frac{1}{\sqrt{2}}\Ishi{\tfrac12} + \frac{1}{2^{1/4}}\Ishi{\tfrac1{16}},\\
|\tfrac12\rangle_c =& \frac{1}{\sqrt{2}}\Ishi{0} + \frac{1}{\sqrt{2}}\Ishi{\tfrac12} - \frac{1}{2^{1/4}}\Ishi{\tfrac1{16}},\\
|\tfrac1{16}\rangle_c =& \Ishi{0} - \Ishi{\tfrac12}.
\end{split}
\end{equation}
Matching open-string partition functions, the fermion boundary states are related to the Ising Cardy states by~\cite{Hori2009,Runkel2020}
\begin{equation}\label{eq:fermion_ising_matching}
  \begin{aligned}
    |B_{+,0}\rangle_{\text{NS}} &= \frac{1}{\sqrt{2}}\bigl(|0\rangle_c + |\tfrac12\rangle_c\bigr), \\
    |B_{+,0}\rangle_{\text{R}} &= \frac{1}{\sqrt{2}}\bigl(|0\rangle_c - |\tfrac12\rangle_c\bigr), \\
    |B_{+,1}\rangle_{\text{NS}} &= \frac{1}{\sqrt{2}}\bigl(|0\rangle_c + |\tfrac12\rangle_c\bigr), \\
    |B_{+,1}\rangle_{\text{R}} &= -\frac{1}{\sqrt{2}}\bigl(|0\rangle_c - |\tfrac12\rangle_c\bigr), \\
    |B_{-}\rangle_{\text{NS}} &= \sqrt{2}\,|\tfrac1{16}\rangle_c, \\
    |B_{-}\rangle_{\text{R}} &= 0.
  \end{aligned}
\end{equation}

\subsection{BCO Three-point Function and Structure Constants}\label{sec:bco-structure-constants}
In this subsection, we present the  Operator Product Expansion (OPE) and the three-point function of the BCOs in the free-fermion model. The OPE of two BCOs takes the form
\begin{equation}
  O^{ab}_i O^{bc}_j \approx \sum_k C^{(abc)k}_{ij} O^{ac}_k,
\end{equation}
where $O^{ab}_i$ denotes a BCO that changes the boundary condition from $a$ to $b$, and $i$ labels the operator.

In the free-fermion theory, we must additionally specify the Chan--Paton (CP) data at the endpoints, since the CP grading contributes to the fermion parity of the full operator. We therefore use brane labels $A$, $B$, and $C$, which specify both the conformal boundary conditions and their associated $\mathbb{Z}_2$ representations. The OPE then becomes
\begin{figure}[h]
  \centering
    \includegraphics[width=0.5\textwidth]{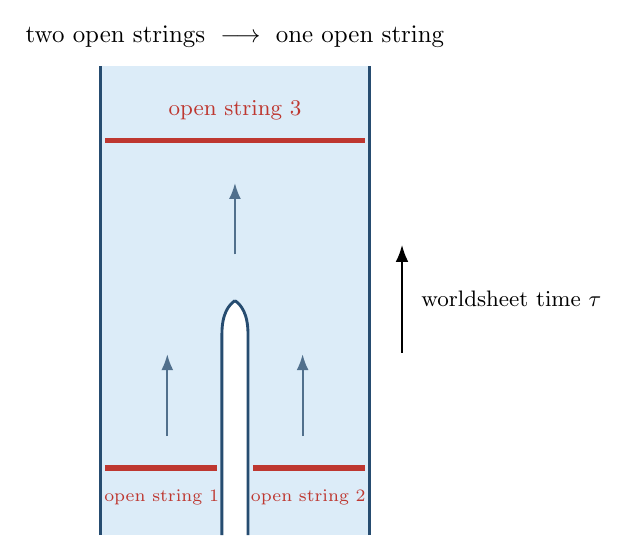}
    \caption{Two open strings fusing into one.}\label{fig:open_string_2_to_1}
\end{figure}
\begin{equation}
  O^{A,B}_i O^{B,C}_j \approx \sum_k C^{(ABC)k}_{ij} O^{A,C}_k,
\end{equation}
where $A$, $B$, and $C$ are branes (boundary conditions equipped with $\mathbb{Z}_2$ representations).

The structure constants that enter the fixed-point tensor are not free: consistency of the boundary OPE imposes the sewing constraints~\cite{Runkel1998}:
\begin{equation}
  C^{(BCD) q}_{j k} C^{(ABD) l}_{i q} C^{(ADA)1}_{ll}=
  \sum_{p} C^{(ABC) p}_{i j} C^{(CDA) p}_{k l} C^{(ACA) 1}_{p p}
  \cdot \Fsymbol{i}{j}{k}{l}{p}{q}
\label{eq:open_4p_crossing}
\end{equation}
\begin{equation}
  C_{ij}^{(ABC)k} \, C_{kk}^{(ACA)1} = C_{jk}^{(BCA)i} \, C_{ii}^{(ABA)1}
  \label{eq:open_3p_crossing}
\end{equation}

We work in a gauge in which all BCO two-point functions are normalized to unity.  Throughout this section, $C^{(ABC)}_{ijk}$ denotes the corresponding three-point coefficient in this normalized basis; these are the coefficients that enter the fixed-point tensor.

In the following, we use $0$, $1$, and $f$ to denote the branes $B_{+,0}$, $B_{+,1}$, and $B_-$, respectively. Table~\ref{tab:free_fermion_cor_structure_constants} lists all nonzero coefficients for the ordered brane triples satisfying $A\leq B\leq C$, under the convention $0<1<f$.
The general method for solving the crossing constraints for arbitrary A-series models is given in Appendix~\ref{app:fermionic_bco_solution}.

\begin{table}[h]
\centering
\renewcommand{\arraystretch}{1.4}
\caption{Free-fermion three-point coefficients $C^{(ABC)}_{ijk}$ in the two-point-normalized BCO basis.}\label{tab:free_fermion_cor_structure_constants}
\begin{tabular}{cccccccc}
\toprule
Coefficient & Value & Coefficient & Value & Coefficient & Value & Coefficient & Value \\
\midrule
$C^{(000)}_{1,1,1}$ & $2^{1/4}$ &
$C^{(000)}_{1,\psi,\psi}$ & $2^{1/4}$ &
$C^{(111)}_{1,1,1}$ & $2^{1/4}$ &
$C^{(111)}_{1,\psi,\psi}$ & $2^{1/4}$ \\
$C^{(001)}_{1,\epsilon,\epsilon}$ & $2^{1/4}$ &
$C^{(001)}_{1,a,a}$ & $2^{1/4}$ &
$C^{(001)}_{\psi,\epsilon,a}$ & $2^{1/4}$ &
$C^{(001)}_{\psi,a,\epsilon}$ & $2^{1/4}$ \\
$C^{(011)}_{\epsilon,1,\epsilon}$ & $2^{1/4}$ &
$C^{(011)}_{\epsilon,\psi,a}$ & $2^{1/4}$ &
$C^{(011)}_{a,1,a}$ & $2^{1/4}$ &
$C^{(011)}_{a,\psi,\epsilon}$ & $2^{1/4}$ \\
$C^{(00f)}_{1,\sigma,\sigma}$ & $2^{1/4}$ &
$C^{(00f)}_{1,\mu,\mu}$ & $2^{1/4}$ &
$C^{(00f)}_{\psi,\sigma,\mu}$ & $2^{-1/4}$&
$C^{(00f)}_{\psi,\mu,\sigma}$ & $2^{-1/4}$ \\
$C^{(11f)}_{1,\sigma,\sigma}$ & $2^{1/4}$ &
$C^{(11f)}_{1,\mu,\mu}$ & $2^{1/4}$ &
$C^{(11f)}_{\psi,\sigma,\mu}$ & $2^{-1/4}$&
$C^{(11f)}_{\psi,\mu,\sigma}$ & $2^{-1/4}$ \\
$C^{(01f)}_{\epsilon,\sigma,\sigma}$ & $2^{-1/4}$ &
$C^{(01f)}_{\epsilon,\mu,\mu}$ & $2^{-1/4}$ &
$C^{(01f)}_{a,\sigma,\mu}$ & $2^{1/4}$ &
$C^{(01f)}_{a,\mu,\sigma}$ & $2^{1/4}$ \\
$C^{(0ff)}_{\sigma,1,\sigma}$ & 1&
$C^{(0ff)}_{\sigma,\epsilon,\sigma}$ & $\frac{1}{\sqrt{2}}$&
$C^{(0ff)}_{\sigma,\psi,\mu}$ & $\frac{1}{\sqrt{2}}i$&
$C^{(0ff)}_{\sigma,a,\mu}$ & 1\\
$C^{(0ff)}_{\mu,1,\mu}$ & 1&
$C^{(0ff)}_{\mu,\epsilon,\mu}$ & $-\frac{1}{\sqrt{2}}$&
$C^{(0ff)}_{\mu,\psi,\sigma}$ & $-\frac{1}{\sqrt{2}}i$&
$C^{(0ff)}_{\mu,a,\sigma}$ & 1\\
$C^{(1ff)}_{\sigma,1,\sigma}$ & 1&
$C^{(1ff)}_{\sigma,\epsilon,\sigma}$ & $-\frac{1}{\sqrt{2}}$&
$C^{(1ff)}_{\sigma,\psi,\mu}$ & $-\frac{1}{\sqrt{2}}i$&
$C^{(1ff)}_{\sigma,a,\mu}$ & 1\\
$C^{(1ff)}_{\mu,1,\mu}$ & 1&
$C^{(1ff)}_{\mu,\epsilon,\mu}$ & $\frac{1}{\sqrt{2}}$&
$C^{(1ff)}_{\mu,\psi,\sigma}$ & $\frac{1}{\sqrt{2}}i$&
$C^{(1ff)}_{\mu,a,\sigma}$ & 1\\
$C^{(fff)}_{1,1,1}$ & 1&
$C^{(fff)}_{1,\epsilon,\epsilon}$ & 1 \\
$C^{(fff)}_{1,\psi,\psi}$ & 1&
$C^{(fff)}_{1,a,a}$ & 1 &
$C^{(fff)}_{\epsilon,\psi,a}$ & $i$ &
$C^{(fff)}_{\epsilon,a,\psi}$ & $-i$\\
\bottomrule
\end{tabular}
\end{table}
\clearpage

\subsection{Exact discrete tensor network representation of  Free Fermion}\label{sec:FPtensor}

In this subsection, we construct the exact fixed-point tensors for the free-fermion theory following the template of Section~\ref{sec:discretisation-review}.
The construction is largely parallel to the bosonic case, with several important modifications to distinguish different spin structures, and accommodate the non-trivial statistics between fermionic states.
For a tensor
\begin{equation}
\mathcal{T}_{(i,I)(j,J)(k,K)}^{ABC},
\end{equation}
the labels $A$, $B$, and $C$ denote the branes discussed in Section~\ref{sec:boundary-states-spin-Cardy} ($0$, $1$, and $f$), while $i$, $j$, and $k$ label boundary-changing operators (BCOs) between the branes $AB$, $BC$, and $CA$, respectively. In contrast to the bosonic case, these BCOs may be either bosonic or fermionic. The indices $I$, $J$, and $K$ label descendants.

As in the bosonic case, the tensor is given by the product of the three-point coefficient $C^{(ABC)}_{ijk}$ of Section~\ref{sec:bco-structure-constants}, evaluated in the two-point-normalized BCO basis, and the corresponding conformal block. The OPE coefficients for fermionic minimal models are determined in Appendix~\ref{app:fermionic_bco_solution}.

\subsubsection{Fermionic tensor contraction rules}
To manipulate these tensors, we have to be careful with how the tensors with shared indices are contracted. They should follow the rules of fermionic tensor networks~\cite{Mortier2024} contraction that takes into account fermi statistics. Consider a generic rank-4 tensor $\mathcal{T}^{ABCD}\in V_1\otimes V_2\otimes V_3^*\otimes V_4^*$ 
\begin{equation}
    \mathcal{T}^{ABCD} = \mathcal{T}^{ABCD}_{ijkl}\ket{i}\ket{j}\bra{k}\bra{l},
\end{equation}
Note that a non-vanishing component of the tensor necessarily has even fermion parity.

To contract a bra and a ket, we adopt this ordering 
\begin{equation}
    \operatorname{tr}_{j,s}\ket{j}\bra{s} ={(-1)}^{|j|}\delta_{js},
\end{equation}
where $|j| \in \{0,1\}$ is an abuse of notation to denote the fermion parity of the state $j$.  It is 0 for bosonic, and 1 for fermionic.

A generic tensor contraction is then carried out by keeping track of the fermionic statistics of the basis states. For instance,
\begin{equation}
\begin{aligned}
    \operatorname{tr}_{j,s}(\mathcal{T}^{ABCD}\otimes \mathcal{T}^{BEFC}) =& \operatorname{tr}_{j,s}\left(\mathcal{T}^{ABCD}_{ijkl}\ket{i}\ket{j}\bra{k}\bra{l}\otimes \mathcal{T}^{BEFC}_{pqrs}\ket{p}\ket{q}\bra{r}\bra{s}\right)\\
    =&\operatorname{tr}_{j,s}\left(\mathcal{T}^{ABCD}_{ijkl}\mathcal{T}^{BEFC}_{pqrs}{(-1)}^{|s|(|p|+|q|+|r|)+|s|(|k|+|l|)}\ket{i}\ket{j}\bra{s}\bra{k}\bra{l}\otimes\ket{p}\ket{q}\bra{r}\right)\\
    =& \sum_j \mathcal{T}^{ABCD}_{ijkl}\mathcal{T}^{BEFC}_{pqrj}{(-1)}^{|j|(|p|+|q|+|r|+|k|+|l|)+|j|}\ket{i}\bra{k}\bra{l}\otimes\ket{p}\ket{q}\bra{r}\\
    =& \left(\sum_j \mathcal{T}^{ABCD}_{ijkl}\mathcal{T}^{BEFC}_{pqrj}\right)\ket{i}\ket{p}\ket{q}\bra{r}\bra{k}\bra{l}.
\end{aligned}
\end{equation}

 To obtain the last line, the remaining factors are reordered from $\ket{i}\bra{k}\bra{l}\otimes\ket{p}\ket{q}\bra{r}$ into the cyclic planar order $\ket{i}\ket{p}\ket{q}\bra{r}\bra{k}\bra{l}$, producing the additional Koszul sign ${(-1)}^{(|p|+|q|+|r|)(|k|+|l|)}$. Together with the even-parity condition
\begin{equation}
    |s|+|p|+|q|+|r|=0 \pmod 2,
\end{equation}
the total phase, including this reordering sign, cancels.

More generally, for a network of parity-even tensors arranged on a planar square lattice, with no edge crossings or self-traces, all fermionic sign factors cancel as long as the dangling boundary legs are ordered according to their cyclic planar order. Therefore, the coefficients of the fully contracted network are obtained using exactly the same index-contraction rule as in the bosonic case.

\begin{figure}
    \centering
    \includegraphics[width=0.5\textwidth]{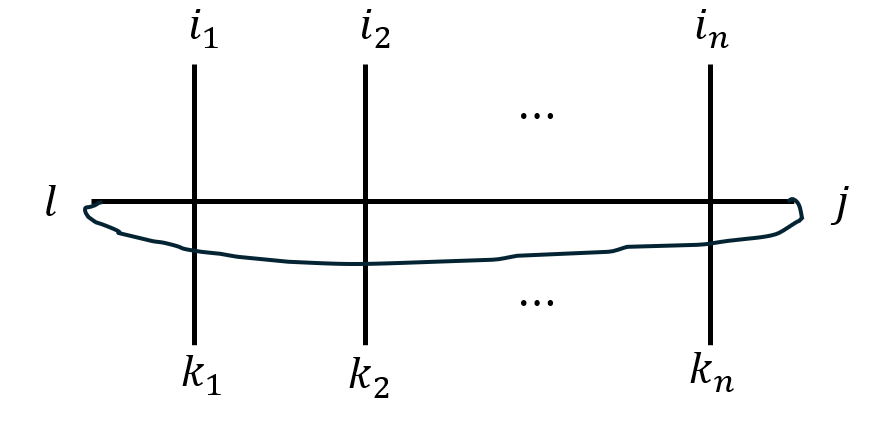}
    \caption{Example of an edge crossing in a tensor network.}\label{fig:edge_crossing}
\end{figure}

Additional signs appear only when the network contains edge crossings or self-traces. An example is shown in Fig.~\ref{fig:edge_crossing}.

For such a configuration, the trace is evaluated using the fermionic contraction rule:
\begin{equation}
    \begin{aligned}
    &\operatorname{tr}_{j,l}\left(\mathcal{T}_{i_1i_2\cdots i_n jk_n k_{n-1}\cdots k_1l}\ket{i_1i_2\cdots i_n}\ket{j}\bra{k_n k_{n-1}\cdots k_1}\bra{l}\right)\\
    =&\operatorname{tr}_{j,l}\left({(-1)}^{|j|(\sum_s |k_s|)}\mathcal{T}_{i_1i_2\cdots i_n j k_n k_{n-1}\cdots k_1l}\ket{i_1i_2\cdots i_n}\bra{k_n k_{n-1}\cdots k_1}\otimes\ket{j}\bra{l}\right)\\
    =&\left(\sum_j {(-1)}^{|j|+|j|(\sum_s |k_s|)}\mathcal{T}_{i_1i_2\cdots i_n jk_n k_{n-1}\cdots k_1j}\right)\ket{i_1i_2\cdots i_n}\bra{k_n k_{n-1}\cdots k_1}.
    \end{aligned}
\end{equation}

The extra sign is essential for producing the bulk fermion spectrum, as we discuss below.

\subsubsection{Fixed point conditions}
Fixed point tensors should satisfy conditions reviewed in section \ref{sec:fixed-point-conditions}. This involves two equations. The coarse-graining equation and the crossing equation. We will check them numerically in turn.

\begin{figure}[h]
    \centering
    \includegraphics[width=0.7\textwidth]{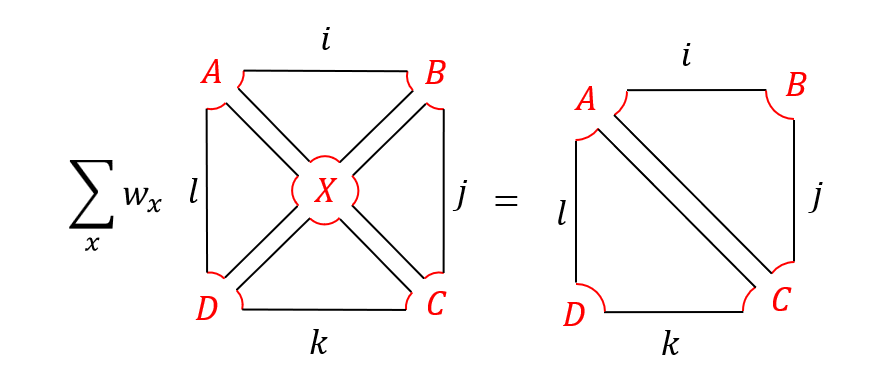}
    \caption{Diagrammatic representation of the coarse-graining condition.
    }\label{fig:fixed}
\end{figure}

{\bf \underline{Coarse-graining condition}} \\
The coarse-graining condition for fermionic blocks is analogous to its bosonic counterpart in Section~\ref{sec:discretisation-review}. Its diagrammatic representation is shown in Fig.~\ref{fig:fixed}, where red arcs $A,B,C,D,X$ denote the branes and black lines $i,j,k,l$ are BCOs connecting them.

The corresponding equation in tensor components is
\begin{equation}\label{eq:fermionic_coarsegraining}
\begin{aligned}
   {[T^{ABCD}_{ijkl}]}_{\text{L.H.S.}} &:= \frac{1}{4\sqrt{2}} \sum_X \omega_X \sum\left(T^{BXA}_{qpi}T^{CXB}_{rqj}T^{DXC}_{srk}T^{AXD}_{psl}\right), \\[4pt]
   {[T^{ABCD}_{ijkl}]}_{\text{R.H.S.}} &:= \frac{1}{2}\left[\sum\left(T^{ABC}_{ijm}T^{CDA}_{klm}\right)+\sum\left(T^{BCD}_{jkn}T^{DAB}_{lin}\right)\right],
\end{aligned}
\end{equation}
where $T^{ABC}_{ijm}$ is shorthand for the tensor $T^{ABC}_{(i,I)(j,J)(m,M)}$, and sums over tensors imply contraction over shared indices. ${[T^{ABCD}_{ijkl}]}_{\text{L.H.S.}}$ and ${[T^{ABCD}_{ijkl}]}_{\text{R.H.S.}}$ are square tensors constructed by gluing triangle tensors.

The weight $\omega_X$ in Eq.~\eqref{eq:fermionic_coarsegraining} is the fermionic shrinkable boundary weight, obtained exactly as in the bosonic case discussed in Section~\ref{sec:discretisation-review} by expressing the vacuum Ishibashi state as a weighted sum over boundary states. For the free fermion, using the boundary states of Section~\ref{sec:boundary-states-spin-Cardy}, this gives Eq.~\eqref{eq:ff_shrinkable}:
\begin{equation}\label{eq:ff_shrinkable}
|0\rangle\rangle =
\frac 1 2\ket{B_{+,0}}_{NS} +
\frac 1 {2\sqrt{2}}\ket{B_-}_{NS}.
\end{equation}

Consequently, the shrinkable boundary weights for the free fermion are
\begin{equation}\label{eq:omega_b_ff}
    \omega_b = \begin{cases}
        \frac 1 2, & b = 0,\\[4pt]
        0, & b = 1,\\[4pt]
        \frac 1 {2\sqrt{2}}, & b = f.
    \end{cases}
\end{equation}
Since the $0$ and $1$ branes have identical NS boundary states, the decomposition in Eq.~\eqref{eq:ff_shrinkable} is not unique: any choice satisfying $\omega_0+\omega_1=1/2$ gives the same shrinkable state. For example, the symmetric choice $\omega_0=\omega_1=1/4$ is equally valid. For convenience, we use the asymmetric choice in Eq.~\eqref{eq:omega_b_ff}.
Here $X$ ranges over the intermediate branes, with the weights of Eq.~\eqref{eq:omega_b_ff}; for the general A-series these are given by the closed formula of Appendix~\ref{sec:shrinkable-weights}.

A numerical check of the coarse-graining condition for the free-fermion model is given in Table~\ref{tab:fixed_check}, where $0$, $1$, and $f$ denote the three branes, and $\mathbb{1}$, $\psi$, $\sigma$, and $\mu$ are BCOs. 

\begin{table}[h]
\centering
\renewcommand{\arraystretch}{1.15}
\caption{Numerical verification of the fixed-point condition. The table lists representative tensor elements
for several choices of branes and BCOs. The first column shows values obtained by contracting two triangles along the long edge,
while the second column shows the same elements computed by contracting four triangles along the short
edges. In both contractions, all internal sums retain Virasoro descendants up to and including level 5 in each conformal family.}
\label{tab:fixed_check}

\begin{subtable}[t]{0.3\textwidth}
\centering
\caption{\rule[-8pt]{0pt}{22pt}$T^{0001}_{\mathbb{1}\mathbb{1}aa}$}
\label{tab:sub1}
\begin{tabular}{cc}
\toprule
$T_{\text{R.H.S.}}$ &$T_{\text{L.H.S.}}$ \\
\midrule
$1.4317$ & $1.4402$ \\
$0.1702$ & $0.1741$ \\
$0.1149$ & $0.1165$ \\
$0.0271$ & $0.0301$ \\
$0.0186$ & $0.0207$ \\
$0.0130$ & $0.0145$ \\
$0.0050$ & $0.0065$ \\
$0.0034$ & $0.0044$ \\
\bottomrule
\end{tabular}
\end{subtable}
\hfill
\begin{subtable}[t]{0.3\textwidth}
\centering
\caption{\rule[-8pt]{0pt}{22pt}$T^{001f}_{\mathbb{1}a\sigma\mu}$}
\label{tab:sub2}
\begin{tabular}{cc}
\toprule
$T_{\text{R.H.S.}}$ &$T_{\text{L.H.S.}}$ \\
\midrule
$1.2178$ & $1.2307$ \\
$0.1498$ & $0.1549$ \\
$0.1010$ & $0.1036$ \\
$0.0257$ & $0.0292$ \\
$0.0176$ & $0.0200$ \\
$0.0122$ & $0.0140$ \\
$0.0031$ & $0.0046$ \\
$0.0029\,\mathrm{i}$ & $0.0032\,\mathrm{i}$ \\
\bottomrule
\end{tabular}
\end{subtable}
\hfill
\begin{subtable}[t]{0.3\textwidth}
\centering
\caption{\rule[-8pt]{0pt}{22pt}$T^{00ff}_{\psi\sigma \mathbb{1} \mu}$}
\label{tab:sub3}
\begin{tabular}{cc}
\toprule
$T_{\text{R.H.S.}}$ &$T_{\text{L.H.S.}}$ \\
\midrule
$0.4306$ & $0.4440$ \\
$0.1609$ & $0.1698$ \\
$0.1026$ & $0.1086$ \\
$-0.0510\,\mathrm{i}$ & $-0.0582\,\mathrm{i}$ \\
$0.0409$ & $0.0465$ \\
$-0.0345\,\mathrm{i}$ & $-0.0398\,\mathrm{i}$ \\
$0.0279$ & $0.0320$ \\
$-0.0159$ & $-0.0197$ \\
\bottomrule
\end{tabular}
\end{subtable}

\end{table}

{\bf \underline{Crossing condition}} \\

\begin{figure}[htbp]
    \centering
    \includegraphics[width=0.6\textwidth]{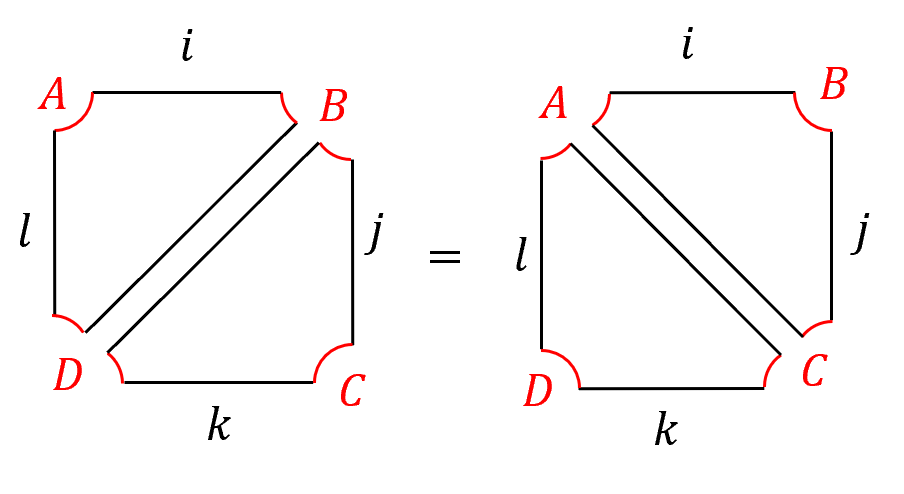}
    \caption{Crossing relation for the free-fermion fixed-point tensor.
    The two decompositions join triangle tensors along the $B$--$D$
    diagonal (left) and the $A$--$C$ diagonal (right).
    Red labels denote branes, while black labels denote BCOs.}
    \label{fig:fermionic-crossing}
\end{figure}

The two diagonal decompositions entering the crossing equation are
illustrated in Fig.~\ref{fig:fermionic-crossing}.
A numerical check of this relation for the free-fermion fixed-point
tensor is given in Table~\ref{tab:crossing_check}.  We use the same
normalization, representative choices of branes and BCOs, and descendant
cutoffs as in Table~\ref{tab:fixed_check}.  The two columns are obtained by
contracting two triangle tensors along the two possible diagonal directions.

\begin{table}[h]
\centering
\renewcommand{\arraystretch}{1.15}
\caption{Numerical verification of the crossing equation.  The first and
second columns show representative tensor components obtained by contracting
the two triangle tensors along the $B$--$D$ and $A$--$C$ diagonals,
respectively.  The normalization is the same as in
Table~\ref{tab:fixed_check}.}
\label{tab:crossing_check}

\begin{subtable}[t]{0.3\textwidth}
\centering
\caption{\rule[-8pt]{0pt}{22pt}$T^{0001}_{\mathbb{1}\mathbb{1}aa}$}
\label{tab:crossing_sub1}
\begin{tabular}{cc}
\toprule
$T_{B\text{--}D}$ & $T_{A\text{--}C}$ \\
\midrule
$1.4317$ & $1.4317$ \\
$0.1702$ & $0.1702$ \\
$0.1149$ & $0.1149$ \\
$0.0248$ & $0.0294$ \\
$0.0165$ & $0.0207$ \\
$0.0109$ & $0.0151$ \\
$0.0050$ & $0.0050$ \\
$0.0033$ & $0.0035$ \\
\bottomrule
\end{tabular}
\end{subtable}
\hfill
\begin{subtable}[t]{0.3\textwidth}
\centering
\caption{\rule[-8pt]{0pt}{22pt}$T^{001f}_{\mathbb{1}a\sigma\mu}$}
\label{tab:crossing_sub2}
\begin{tabular}{cc}
\toprule
$T_{B\text{--}D}$ & $T_{A\text{--}C}$ \\
\midrule
$1.2126$ & $1.2230$ \\
$0.1503$ & $0.1493$ \\
$0.1014$ & $0.1005$ \\
$0.0243$ & $0.0272$ \\
$0.0161$ & $0.0191$ \\
$0.0107$ & $0.0138$ \\
$0.0034$ & $0.0028$ \\
$0.0006\,\mathrm{i}$ & $0.0053\,\mathrm{i}$ \\
\bottomrule
\end{tabular}
\end{subtable}
\hfill
\begin{subtable}[t]{0.3\textwidth}
\centering
\caption{\rule[-8pt]{0pt}{22pt}$T^{00ff}_{\psi\sigma \mathbb{1}\mu}$}
\label{tab:crossing_sub3}
\begin{tabular}{cc}
\toprule
$T_{B\text{--}D}$ & $T_{A\text{--}C}$ \\
\midrule
$0.4306$ & $0.4306$ \\
$0.1609$ & $0.1609$ \\
$0.1026$ & $0.1026$ \\
$-0.0623\,\mathrm{i}$ & $-0.0396\,\mathrm{i}$ \\
$0.0332$ & $0.0486$ \\
$-0.0441\,\mathrm{i}$ & $-0.0250\,\mathrm{i}$ \\
$0.0213$ & $0.0345$ \\
$-0.0085$ & $-0.0233$ \\
\bottomrule
\end{tabular}
\end{subtable}

\end{table}

\subsubsection{Spectrum from tensors: encoding the NS and R sectors}

The exact, untruncated correlators obey sewing, while the calculations above test their descendant-truncated tensor approximations. The spin-marked sewing construction of Section~\ref{sec:fermionization} supplies the geometric prescription for changing a decomposition.  
We can therefore choose a convenient tiling of any surface using the FP tensors in complete analogy with the bosonic case~\eqref{eq:partition_function_bosonic}, to reconstruct the full CFT partition function:
\begin{equation}\label{eq:partition_function_fermionic}
    Z_M = \sum_{\{(i,I)\}, \{A\}} \prod_v \omega_A \prod_{\triangle} \mathcal{T}_{(i,I)(j,J)(k,K)}^{ABC},
\end{equation}
where the sum runs over all labels $(i,I)$ for boundary-changing operators and branes $A$;
$\prod_v \omega_A$ assigns a weight $\omega_A$ to each vertex $v$ according to its brane label;
and $\prod_{\triangle} \mathcal{T}_{(i,I)(j,J)(k,K)}^{ABC}$ runs over all triangles of the triangulation.
When the triangulated manifold has nontrivial topology (e.g.\ a torus), additional sign factors from fermionic contractions must be included following the rules discussed above.

\begin{figure}
    \centering
    \includegraphics[width = 0.5 \textwidth]{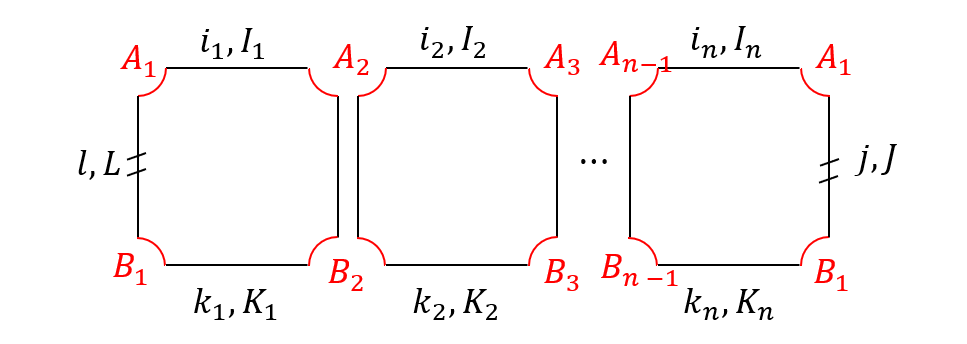}
    \caption{Cylinder geometry constructed by gluing fixed-point tensors. The vertical direction corresponds to Euclidean time propagation, while the horizontal direction represents the spatial circumference.}\label{fig:cylinder}
\end{figure}

We now extract the conformal spectrum from these tensors. As in the bosonic case, we glue the tensors together to form a cylinder, as illustrated in Fig.~\ref{fig:cylinder}. As it is well-known, the path-integral over a cylinder can be expressed as $\exp(-\tau H)$, where $H$ is the Hamiltonian of the CFT.  Covering a cylinder by these fixed point tensors produce a transfer matrix whose spectrum should recover the CFT spectrum. This is the consistency check we will perform. 

In the cylinder geometry, each vertex hosts a small hole that must be shrunk by summing over intermediate boundary conditions with the fermionic shrinkable boundary weight $\omega_b$ defined in Eq.~\eqref{eq:omega_b_ff} (for the general A-series, see Appendix~\ref{sec:shrinkable-weights}). In our triangulation however, we do not have any internal holes. Conformal boundary labels of ''half holes'' located at the top and bottom boundaries of the cylinder are treated as part of the dangling indices alongside the labels of boundary changing operators at the edges.

Eigenvalues of the cylinder transfer matrix are given by $N e^{-(2\pi/n)(\Delta - c/12)}$, where $N$ is a normalization constant, $n$ is the number of tensor blocks along the circumference, and $c$ is the central charge. To obtain the correct spectrum of the CFT, careful attention must be paid to fermionic sign factors arising from tensor contractions on the cylinder.

As discussed above, fermionic tensor contractions introduce additional sign factors. Specifically, for configurations involving edge crossings or self-traces, the contraction rule essentially yields an extra sign ${(-1)}^{|j|(\sum_s |k_s|)+|j|}$. For the transfer cylinder of Fig.~\ref{fig:cylinder}, the spin structure around the spatial circumference selects the Hilbert space: NS means antiperiodic and R means periodic fermions around that circle. If Euclidean time is subsequently closed, its spin structure separately selects a trace or a supertrace. Section~\ref{subsec:spin-open-closed} derives the closure rule from the boundary Nakayama insertion.

In the NS sector, the anti-periodicity produces an additional sign factor ${(-1)}^{|j|}$. Remarkably, this extra sign exactly cancels the ${(-1)}^{|j|}$ factor in the fermionic contraction rule. Consequently, the net sign factor, denoted as $S_{\text{NS}}$,  for the NS sector spectrum becomes
\begin{equation}
    S_{\text{NS}} = {(-1)}^{|j|(\sum_s |k_s|)},
\end{equation}

In contrast, for the R sector with periodic boundary conditions, no such cancellation occurs. The full sign factor $S_R$  including both contributions from the fermionic contraction rule and the periodic boundary conditions is
\begin{equation}
    S_{\text{R}} = {(-1)}^{|j|+|j|(\sum_s |k_s|)}.
\end{equation}

The numerical implementation of this procedure for the free-fermion model yields the NS- and R-sector scaling-dimension estimates shown in Tables~\ref{tab:data_fermion_cylinder} and~\ref{tab:data_fermion_cylinder_R}, respectively. The columns of both tables are labeled
according to the cutoff in descendant levels for the fields on the legs of the triangle tensor. For example, the label 117 means that the short edges of the
triangle have descendant level cutoff 1 and the long edge has cutoff 7.

\begin{table}[h]
    \centering
\renewcommand{\arraystretch}{1.15}
    \caption{Scaling-dimension for the free fermion on a cylinder with
    $n=3$ in the NS sector. The four CFT states at $\Delta=2$ are grouped together because their discrete translation eigenvalues do not distinguish such states at this circumference.}\label{tab:data_fermion_cylinder}

    \begin{tabular}{cccccc}
\toprule
CFT states & exact $\Delta$ & 007 & 117 & 227 & 337\\
\midrule
1                   & 0.0000 & 0.0000 & 0.0000 & 0.0000 & 0.0000\\[4pt]
        $\psi,\tilde \psi$  & 0.5000 & 0.5064 & 0.5005 & 0.5010 & 0.4999\\[4pt]
        $\epsilon$          & 1.0000 & 1.0592 & 0.9952 & 1.0130 & 1.0021\\[4pt]
        $\partial \psi$,
        $\bar \partial \tilde \psi$ & 1.5000 & 1.7864 & 1.5059 & 1.5404& 1.4986\\[4pt]
        \begin{tabular}{@{}c@{}}$L_{-2}1,\ \bar L_{-2}1,$\\[-2pt]$\partial\epsilon,\ \bar\partial\epsilon$\end{tabular}
        & 2.0000
        & $\left\{\begin{gathered}2.1515\\[-2pt]2.3343\end{gathered}\right\}$
        & $\left\{\begin{gathered}1.9874\\[-2pt]2.0129\end{gathered}\right\}$
        & $\left\{\begin{gathered}2.0469\\[-2pt]2.0715\end{gathered}\right\}$
        & $\left\{\begin{gathered}2.0038\\[-2pt]2.0052\end{gathered}\right\}$ \\
\bottomrule
\end{tabular}
\end{table}

\begin{table}[h]
    \centering
\renewcommand{\arraystretch}{1.15}
    \caption{Scaling-dimension estimates for the free fermion on a cylinder with $n=3$ in the R sector, using the same cutoffs as Table~\ref{tab:data_fermion_cylinder}. The R ground-state dimension is extracted relative to the NS vacuum.}
    \label{tab:data_fermion_cylinder_R}

    \begin{tabular}{cccccc}
\toprule
CFT states & exact $\Delta$ & 007 & 117 & 227 & 337\\
\midrule
$\sigma,\mu$ & 0.1250 & 0.1243 & 0.1249 & 0.1249 & 0.1250\\[3pt]
        \begin{tabular}{@{}c@{}}$\partial\sigma,\ \bar\partial\sigma,$\\[-3pt]$\partial\mu,\ \bar\partial\mu$\end{tabular}
        & 1.1250 & 1.2044 & 1.1260 & 1.1391 & 1.1254\\[5pt]
        \begin{tabular}{@{}c@{}}$\partial^2\sigma,\ \bar\partial^2\sigma,$\\[-3pt]$\partial^2\mu,\ \bar\partial^2\mu$\end{tabular}
        & 2.1250 & 2.3931 & 2.1343 & 2.2325 & 2.1234\\[7pt]
        $\partial\bar\partial\sigma,\ \partial\bar\partial\mu$
        & 2.1250 & 2.2687 & 2.1042 & 2.1915 & 2.1357 \\
\bottomrule
\end{tabular}
\end{table}

\medskip
\noindent This completes the discretization of the free fermion: the FP tensor satisfies the coarse-graining condition, and the cylinder, with the spin structure encoded in the contraction signs $S_{\text{NS}}$ and $S_{\text{R}}$, yields low-lying scaling dimensions close to the expected NS and R values. Individual descendant assignments within the NS $\Delta=2$ group remain unresolved at $n=3$.
The next section constructs the strip and its boundary tensors. Section~\ref{sec:fermionization} then places these constructions in the general TQFT framework; the A-series formulas are collected in Appendix~\ref{app:fermionic_models}.

\section{Boundary Tensors and Open-String Spectra}\label{sec:boundary-tensors}
\subsection{From Manifolds without Boundary to Manifolds with Boundary}

In Section~\ref{sec:FPtensor} we constructed fixed-point tensors for fermionic minimal models by triangulating closed manifolds (without boundary). A key application was the cylinder geometry, where the tensor network transfer matrix reproduces the bulk conformal spectrum. However, many physical situations involve manifolds \emph{with} boundaries---for instance, a strip with two parallel edges, or a disk with a circular boundary. To extend the tensor network approach to such geometries, we need additional tensors that encode the physics at conformal boundaries. Such tensor networks have been considered before, and boundary fixed-point tensors have been obtained numerically starting from 2d statistical lattice models with different boundary conditions \cite{Iino2019BTRG,Iino2020BTNR,Iino2021TricriticalPotts}. In this section we will obtain a field-theoretic expression of these boundary fixed point tensors. 
To make the geometric role of a physical boundary explicit, we
demonstrate the boundary-tensor construction in the bosonic Ising
model. This example isolates how the boundary patch and its conformal
map modify the tensor, and the strip spectra presented below provide
bosonic Ising benchmarks for the construction. For fermionic theories,
the same patch geometry is used, together with the graded boundary
and BCO data of Sections~3 and~5 and the spin-dependent sewing signs described there.


The triangle tensors introduced in Sections~\ref{subsec:bosonic_tensor} and~\ref{sec:FPtensor} describe regions with three ``open'' edges that are later glued to neighboring triangles. A boundary, on the other hand, is a physical edge where specific conformal boundary conditions are imposed.

In this section we introduce a new type of tensor, the \emph{boundary tensor}, associated with a patch containing three open gluing edges and an extended physical boundary segment, as shown in Fig.~\ref{fig:boundary_tensor}. By combining boundary tensors with the triangle tensors of Sections~\ref{subsec:bosonic_tensor} and~\ref{sec:FPtensor}, we can assemble manifolds with boundaries, the simplest example being a strip. The tensor network on a strip approximates the transfer matrix of an open string propagating between two parallel boundaries. Diagonalizing this transfer matrix yields the scaling dimensions of boundary-changing operators, i.e., the \emph{open-string spectrum}.

The construction proceeds in three logical steps. First, we define the boundary tensor through a three-point correlation function of BCOs on a boundary patch with three open gluing edges and an extended physical boundary segment (Sec.~\ref{sec:boundary_tensor_def}). Second, we glue boundary tensors together with triangle tensors to form a strip and interpret the resulting tensor network as open-string propagation (Sec.~\ref{sec:strip_open_string}). Third, we extract the open-string spectrum from the eigenvalues of the strip transfer matrix and compare it with exact BCFT results (Sec.~\ref{sec:open_string_spectrum}).

\subsection{Definition of the Boundary Tensor}\label{sec:boundary_tensor_def}

\begin{figure}[htbp]
\centering
\begin{subfigure}{0.49\textwidth}
    \centering
    \includegraphics[width=0.7\linewidth]{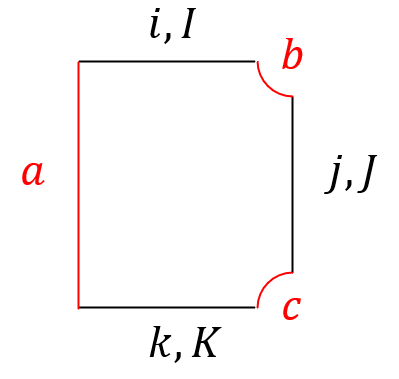}
    \caption{}\label{fig:boundary_tensor}
\end{subfigure}
\begin{subfigure}{0.49\textwidth}
    \centering
    \includegraphics[width=0.7\linewidth]{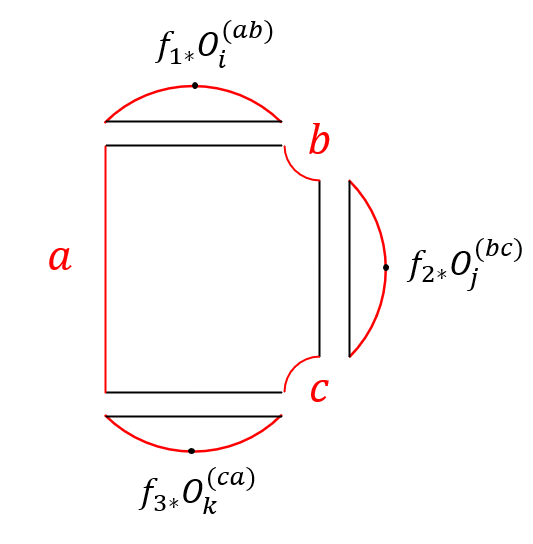}
    \caption{}\label{fig:f_map}
\end{subfigure}
\caption{(a) The shape of the boundary tensor. Three boundary segments carry boundary conditions $a,b,c$; the open legs correspond to BCOs $i,j,k$ with descendants $I,J,K$. (b) The three-point amplitude that defines the boundary tensor.} 
\end{figure}

The boundary tensor $\mathcal{B}^{abc}_{(i,I)(j,J)(k,K)}$, depicted in Fig.~\ref{fig:boundary_tensor}, is the basic building block for manifolds with physical boundaries.
In this figure, $a,b,c$ denote conformal boundary conditions, $i,j,k$ are boundary-changing operators (BCOs) associated with the pairs $ab$, $bc$, and $ca$, respectively, and $I,J,K$ are descendant labels.
As in the case of the triangle tensor defined in Section~\ref{subsec:bosonic_tensor}, the black lines in Figure~\ref{fig:boundary_tensor} represent states prepared by the map $f$ (shown in Figure~\ref{fig:f_map}), while the red lines indicate the physical conformal boundaries.

The value of the tensor is defined as the three-point function shown in Figure~\ref{fig:f_map}.
It factorizes into the product of the three-point structure constant $C^{abc}_{ijk}$ and the conformal block $\tilde{\alpha}^{ijk}_{IJK}$:
\[
\mathcal{B}^{abc}_{(i,I)(j,J)(k,K)} = C^{abc}_{ijk}\,\tilde{\alpha}^{ijk}_{IJK}.
\]
The structure constant $C^{abc}_{ijk}$ is the same as the three-point correlation structure constant that appears in the triangle tensor of Section~\ref{subsec:bosonic_tensor}. The conformal block $\tilde\alpha^{ijk}_{IJK}$ differs from $\alpha^{ijk}_{IJK}$ because the boundary patch has a different
geometry.

\begin{figure}[htbp]
    \centering
    \includegraphics[width=0.6\textwidth]{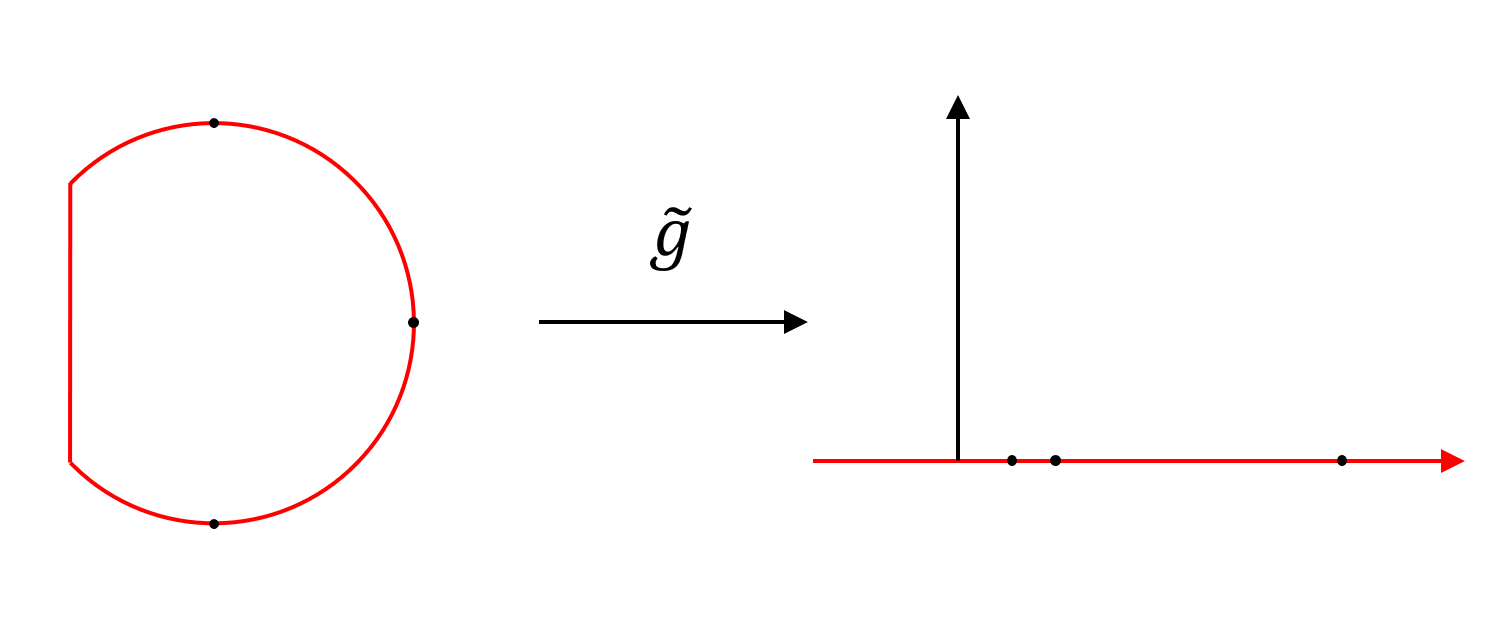}
    \caption{The conformal map $\tilde g$ from the region $\{z \mid \operatorname{Re}(z) > 0,\ |z-1|<\sqrt{2}\}$ to the upper half-plane.}\label{fig:g_map}
\end{figure}

To compute the boundary conformal blocks, we note that the map $f(z)$ appearing below is the same conformal map introduced in Section~\ref{subsec:bosonic_tensor} (see Fig.~\ref{fig:states}), which sends a semicircle to a bow-shaped segment. The conformal transformation from the region $\{z \mid \operatorname{Re}(z) > 0,\ |z-1|<\sqrt{2}\}$ to the upper half-plane is given by
$\tilde g(\xi) = {\bigl(e^{-i\pi/4}\frac{\xi+i}{\xi-i}\bigr)}^{4/3}$ (see Figure~\ref{fig:g_map}).
This map $\tilde g$ is different from the map $g$ used in Section~\ref{subsec:bosonic_tensor} due to the different geometry of the three-point function; the tilde notation emphasizes that this is a boundary-related conformal transformation.

Consider the maps
\begin{equation}
\begin{aligned}
    f_1(z) &= i f(z) + 1 + i, \\
    f_2(z) &= f(z) + 2, \\
    f_3(z) &= -i f(z) + 1 - i,
\end{aligned}
\end{equation}
and define $\chi_i = \tilde g \circ f_i$ for $i = 1,2,3$. The points $x_1, x_2, x_3$ are the images of the BCO insertion points on the real axis of the upper half-plane under the composition $\chi_i = \tilde g \circ f_i$.
The value of the tensor is then given by
\begin{equation}
    C^{abc}_{ijk}\,\tilde{\alpha}^{ijk}_{IJK} = 
    \bigl\langle \chi_{1*}\, O^{ab}_{(i,I)}(x_1) \;
            \chi_{2*}\, O^{bc}_{(j,J)}(x_2) \;
            \chi_{3*}\, O^{ca}_{(k,K)}(x_3) \bigr\rangle_{\text{UHP}}.
\end{equation}

\subsection{Strip Geometry and Open-String Propagation}\label{sec:strip_open_string}

\begin{figure}[htbp]
    \centering
    \includegraphics[width=0.7\textwidth]{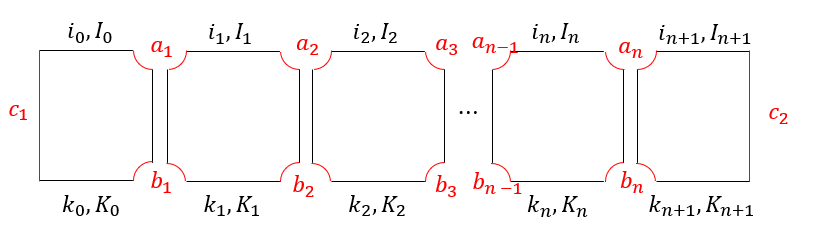}
    \caption{Tensor network for a strip. Boundary tensors are placed at the two physical edges, while the interior is filled with pairs of triangle tensors glued together along their long edges. The resulting network approximates the transfer matrix of an open string propagating between the two boundaries.}\label{fig:strip}
\end{figure}

Once the boundary tensor $\mathcal{B}^{abc}_{(i,I)(j,J)(k,K)}$ is defined, we can combine it with the triangle tensor $\mathcal{T}^{abc}_{(i,I)(j,J)(k,K)}$ of Section~\ref{subsec:bosonic_tensor} to build manifolds with boundaries. The simplest such geometry is a strip---a rectangle with two parallel boundaries, as shown in Figure~\ref{fig:strip}.

The tensor network on the strip has a natural interpretation as propagation in Euclidean time. If we regard the direction along the strip (vertical in Fig.~\ref{fig:strip}) as Euclidean time, the network represents the partition function of an \emph{open string} whose endpoints are constrained to lie on the two boundaries. In BCFT language, this corresponds to the open-string channel where the strip partition function is written as a trace over the open-string Hilbert space $\mathcal{H}_{ab}$ with boundary conditions $a$ and $b$:
\[
Z_{\text{strip}} = \operatorname{Tr}_{\mathcal{H}_{ab}} e^{-L_1 H_{\text{open}}}.
\]
Here $l_1$ is the length of the strip (time direction) and $H_{\text{open}}$ is the Hamiltonian for the open string.

The open-string Hamiltonian on a strip of width $l_2$ (the spatial extent) is given by the Virasoro generator $L_0$ through the conformal mapping from the strip to the upper half-plane:

\begin{equation}
    H_{\text{open}} = \frac{\pi}{l_2}(\hat{L}_0 - \frac{c}{24}),
\end{equation}
where $\hat{L}_0$ is the zeroth generator of the Virasoro algebra and $c$ is the central charge. 

The tensor network transfer matrix $\mathcal{M}$ along the time direction is proportional to the exponential of $H_{\text{open}}$:
\[
\mathcal{M} \propto e^{-l_1 H_{\text{open}}}.
\]
Consequently, the eigenvalues $\{\lambda_i\}$ of $\mathcal{M}$ are related to the scaling dimensions $\Delta_i$ of boundary-changing operators via
\begin{equation}
    \lambda_i = N\exp\!\left( \frac{\pi l_1}{l_2}\left(\frac{c}{24} - \Delta_i\right) \right),
\end{equation}
Thus, diagonalizing the tensor network transfer matrix yields the scaling dimensions $\Delta_i$ of boundary-changing operators (BCOs). These $\Delta_i$ constitute the \emph{open-string spectrum} of the theory for the chosen pair of boundary conditions.

\subsection{Open-String Spectrum from the Strip}\label{sec:open_string_spectrum}

The eigenvalues of the strip transfer matrix give the scaling dimensions of boundary-changing operators. Table~\ref{tab:data_Ising_strip_free} shows numerical results for the Ising model with free--free boundary conditions. The columns indicate different descendant-level cutoffs: for example, ``117'' means the short edges of triangle tensors have cutoff 1, the long edge has cutoff 7, and the boundary-tensor edges have cutoff 1. 

\begin{table}[htbp]
    \centering
\renewcommand{\arraystretch}{1.15}
    \caption{Open-string spectrum for an Ising strip with free--free boundary conditions.}\label{tab:data_Ising_strip_free}
    \begin{tabular}{lcccc}
\toprule
Operator            & 007      & 117      & 227      & 337      \\
\midrule
$1$                 & 0.0000   & 0.0000   & 0.0000   & 0.0000   \\
        $\epsilon$          & 0.5060   & 0.5150   & 0.5047   & 0.5076   \\
        $L_{-1}\epsilon$    & 1.5599   & 1.5449   & 1.5248   & 1.5239   \\
        $L_{-2} 1$          & 2.0856   & 2.0574   & 2.0344   & 2.0328   \\
        $L_{-1}^2 \epsilon$ & 2.7386   & 2.5778   & 2.5709   & 2.5412   \\
        $L_{-3} 1$          & 3.2841   & 3.0909   & 3.0854   & 3.0524   \\
\bottomrule
\end{tabular}
\end{table}

\begin{table}[htbp]
    \centering
\renewcommand{\arraystretch}{1.15}
    \caption{Open-string spectrum for an Ising strip with fixed(+)--fixed(+) boundary conditions.}\label{tab:data_Ising_strip_fixed}
    \begin{tabular}{lcccc}
\toprule
Operator            & 007      & 117      & 227      & 337      \\
\midrule
$1$                 & 0.0000   & 0.0000   & 0.0000   & 0.0000   \\
        $L_{-2} 1$          & 2.0856   & 2.0574   & 2.0344   & 2.0328   \\
        $L_{-3} 1$          & 3.2841   & 3.0909   & 3.0854   & 3.0524   \\
        $L_{-4} 1$          & 4.4122   & 4.1159   & 4.1213   & 4.0653   \\
        $L_{-2}^2 1$        & 4.8783   & 4.1079   & 4.1773   & 4.0713   \\
\bottomrule
\end{tabular}
\end{table}
The data show convergence to the exact scaling dimensions ($\Delta(1)=0$, $\Delta(\epsilon)=0.5$, $\Delta(L_{-1}\epsilon)=1.5$, etc.) as the cutoff increases. Table~\ref{tab:data_Ising_strip_fixed} shows the analogous data for fixed(+)--fixed(+) boundary conditions, which isolate the identity tower. This verifies that the boundary tensor construction correctly reproduces the open-string spectrum.
\section{Fermionic CFT from topological defects}
\label{sec:fermionization}

The constructions above use boundary conditions and boundary-changing operators (BCOs) to build the fermionic CFT from open-string amplitudes. Their generalisation is most transparent in the three-dimensional topological field theory (TQFT) description of rational CFT. The essential data are a bosonic chiral category, a parity degree of freedom, and an algebra of topological defects:
\begin{equation}
 \mathcal C=\operatorname{Rep}(\mathcal V),\qquad
 \widehat{\mathcal C}=\mathcal C\boxtimes\mathcal{SV}ect,
 \qquad (F,N_F),\qquad N_F^2=\mathrm{id}_F .
 \label{eq:main-tft-input}
\end{equation}
Here $\mathcal V$ is a rational bosonic chiral algebra whose representation category $\mathcal C$ is modular; $F$ is a special Frobenius algebra in $\widehat{\mathcal C}$, and $N_F$ is its Nakayama automorphism. This is the extension of the Fuchs--Runkel--Schweigert construction~\cite{FRS1,FRS4} to cover fermionic and spin CFTs \cite{Runkel2020,Runkel2022,Novak2015}. For fixed chiral data, the full fermionic CFT is specified by the generally non-symmetric Frobenius algebra $F$, which determines its bulk and boundary spectra and OPE coefficients. In the ribbon-graph construction of its correlators, the world-sheet spin structure determines the Nakayama insertions, and the geometrical dependence of the path-integral is recorded in the conformal blocks. This section explains that organisation, while Appendix~\ref{app:combined} gives the algebraic and model-specific calculations.

\subsection{The algebra of fermionisation}
\label{subsec:1+PiG-family}

The category $\mathcal C$ contains the Moore--Seiberg data of the bosonic chiral theory: fusion, associators and braiding. Passing to its Deligne product with $\mathcal{SV}ect$ assigns each chiral representation an even and an odd version,
\begin{equation}
 \hat{\imath}=i\boxtimes\Pi^p,\qquad p\in\mathbb Z_2,
\end{equation}
where $\Pi$ is the one-dimensional odd super-vector space. Exchanging two odd labels now gives the additional sign $-1$. This operation keeps the chiral conformal blocks and adds the parity bookkeeping needed to assemble fermionic correlators. The product category is not itself modular: its transparent $\mathcal{SV}ect$ factor must be retained. Accordingly, the TQFT used below is the super-vector-space-valued extension $\widehat Z_{\mathcal C}$ of the Reshetikhin--Turaev theory of $\mathcal C$, rather than an ordinary modular TQFT constructed from $\widehat{\mathcal C}$~\cite[Sections~1.2 and~4]{Runkel2022}.

An algebra $F$ is charactered by a product that joins two $F$-lines, a coproduct splits one line, and a unit and a counit that create or terminate the corresponding network. Associativity and the Frobenius relation, together with specialness allows one to convert between different graphs that represent the same spin-structure. Particularly, like in the bosonic case, conformal boundary conditions are encoded as modules of the algebra $F$.

The new geometric ingredient relative to the symmetric Frobenius algebras that characterise modular invariance bosonic CFT is the possibility that the invariant pairing of $F$ is non-symmetric. Rotating a junction then leaves an automorphism $N_F$ on its incident line. This is the Nakayama automorphism. For an oriented theory it is trivial; for the spin construction it may be nontrivial but obeys $N_F^2=1$ \cite{Runkel2020,Runkel2022,Novak2015}. It records the change under a full turn of the local frame, while two such turns are trivial. Bosonic algebras may also possess automorphisms: the distinction here is the nontrivial \emph{canonical} automorphism of the Frobenius pairing and its coupling to spin structure. Non-symmetric in this sense should also not be confused with noncommutative multiplication.

The examples in this paper use
\begin{equation}
 F=\mathbf1\oplus\Pi G,
 \qquad G\otimes G\cong\mathbf1,
 \qquad N_F=\mathrm{id}_{\mathbf1}\oplus(-\mathrm{id}_{\Pi G}),
 \label{eq:main-fermionisation-algebra}
\end{equation}
with $d_G=1$. The multiplication identifies the product of two odd $G$-lines with the vacuum channel. Existence of this multiplication requires that the associator restricted to the order-two symmetry be trivialisable; for a self-dual invertible object this is equivalent to $\theta_G^2=1$~\cite[Lemma~A.22]{Runkel2022}.  We note that this defect-network description is closely related to the notion of fermion condensation \cite{WanWang2017,Lou2021DummyGuide,AasenLakeWalker2019} when $\theta_G = -1$. 
 More general algebras satisfying~\eqref{eq:main-tft-input} give the route beyond simple-current examples. In this section ``general'' means within this rational, semisimple chiral setting; extension to a non-rational CFT requires additional input.

The presentation above starts from bosonic chiral data and includes the diagonal bosonic parent used in our examples.  The data of the bosonic parent is encoded in the symmetric Frobenius algebra $\mathcal{A} = 1 \subset F$. If one instead fixes a non-diagonal bosonic full CFT described by an algebra $A$, and $G$ would have to be generally a bi-module of $A$.

\subsection{Boundaries, BCOs and TQFT expectation values}
\label{subsec:boundaries-tft-main}

An $F$-module $M$ is a boundary on which an $F$-line can end. Its action $\rho_M:F\otimes M\to M$ specifies that endpoint. To use this boundary on a spin world sheet one also chooses a lift $N_M$ of the Nakayama action:
\begin{equation}
 N_M^2=\mathrm{id}_M,\qquad
 N_M\rho_M=\rho_M(N_F\otimes N_M).
 \label{eq:main-module-equivariance}
\end{equation}
The resulting objects are the Nakayama-equivariant modules of Ref.~\cite[Section~6.1]{Runkel2022}. One can enumerate the underlying simple modules by decomposing the induced modules $F\otimes\hat{\imath}$, and then supply the compatible spin lifts. This distinguishes the algebraic boundary condition from the extra data that describe its transport under changes of spin frame.

A BCO from $M$ to $N$, of chiral type $i$ and parity $p$, is labelled by
$\hat{\imath}=i\boxtimes\Pi^p\in\widehat{\mathcal C}$ and represented by an $F$-module morphism 
\begin{equation}
 \mathcal H_{MN}^{\hat{\imath}}
 =\operatorname{Hom}_F\!\left(M\otimes\hat{\imath},N\right),
 \qquad
 \mathcal H_{MN}^{\mathrm{open}}
 =\bigoplus_{\hat{\imath}=i\boxtimes\Pi^p}\mathcal H_{MN}^{\hat{\imath}}\otimes\mathcal V_i.
 \label{eq:main-bco-hom}
\end{equation}
In this formula $\mathcal V_i$ includes the descendants, and the finite-dimensional first factor gives the multiplicity of each boundary primary.
We choose basis morphisms $\psi_{M\hat{\imath}}^{N}:M\otimes\hat{\imath}\to N$ in the BCO spaces of~\eqref{eq:main-bco-hom}. In particular, $\psi_{N\hat{\jmath}}^{L}:N\otimes\hat{\jmath}\to L$ represents a BCO from boundary $N$ to boundary $L$, and $\psi_{M\hat{k}}^{L}:M\otimes\hat{k}\to L$ represents a possible OPE output. Basis indices are suppressed when the relevant BCO and fusion spaces are one-dimensional.

For multiplicity-free labels different compositions of these morphisms give
\begin{equation}
 \psi_{N\hat{\jmath}}^{L}\circ(\psi_{M\hat{\imath}}^{N}\otimes\mathrm{id}_{\hat{\jmath}})
 =\sum_{\hat{k}} C^{(MNL)\hat{k}}_{\hat{\imath}\hat{\jmath}}\,
 \psi_{M\hat{k}}^{L}\circ(\mathrm{id}_M\otimes\lambda_{\hat{\imath}\hat{\jmath}}^{\hat{k}})
 \circ a_{M,\hat{\imath},\hat{\jmath}},
 \label{eq:main-module-ope}
\end{equation}
where $\hat{\imath},\hat{\jmath},\hat{k}\in\widehat{\mathcal C}$ are the parity-labelled objects, $\lambda_{\hat{\imath}\hat{\jmath}}^{\hat{k}}:\hat{\imath}\otimes\hat{\jmath}\to\hat{k}$ is a chosen fusion vertex, $a$ is the associator, and $C^{(MNL)\hat{k}}_{\hat{\imath}\hat{\jmath}}$ is the OPE coefficient between the BCOs. 
Multiplicity indices and the relation to the conformal-block normalisation are given in Appendix~\ref{app:subsec:ope_6j}. The OPE is a module fusing-matrix element, and not in general a bare $6j$ symbol of $\mathcal C$.

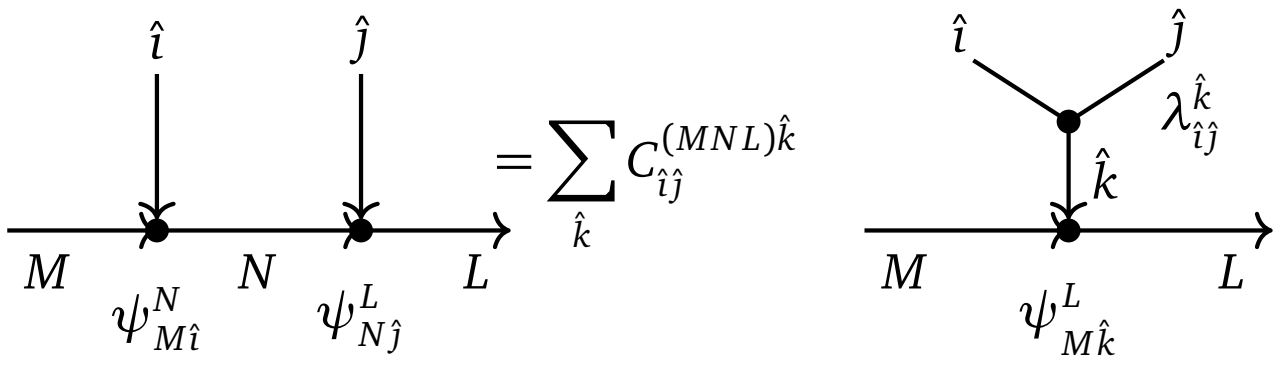
\begin{figure}[htbp]
\centering
\begin{tikzpicture}[scale=0.9, every node/.style={font=\small}, line width=0.8pt]
 \draw[->] (-3.9,0)--(-2.8,0);
 \draw[->] (-2.8,0)--(-1.3,0);
 \draw[->] (-1.3,0)--(-0.2,0);
 \draw[->] (-2.8,1.15)--(-2.8,0.08);
 \draw[->] (-1.3,1.15)--(-1.3,0.08);
 \fill (-2.8,0) circle (2.5pt); \fill (-1.3,0) circle (2.5pt);
 \node at (-3.6,-0.3) {$M$}; \node at (-2.05,-0.3) {$N$};
 \node at (-0.45,-0.3) {$L$};
 \node at (-2.8,1.4) {$\hat{\imath}$}; \node at (-1.3,1.4) {$\hat{\jmath}$};
 \node at (-2.8,-0.65) {$\psi_{M\hat{\imath}}^{N}$};
 \node at (-1.3,-0.65) {$\psi_{N\hat{\jmath}}^{L}$};
 \node at (0.8,0.35) {$=\displaystyle\sum_{\hat{k}} C^{(MNL)\hat{k}}_{\hat{\imath}\hat{\jmath}}$};
 \draw[->] (2.4,0)--(3.9,0); \draw[->] (3.9,0)--(5.4,0);
 \draw (3.2,1.25)--(3.9,0.8)--(4.6,1.25);
 \draw[->] (3.9,0.8)--(3.9,0.08);
 \fill (3.9,0.8) circle (2.5pt); \fill (3.9,0) circle (2.5pt);
 \node at (2.7,-0.3) {$M$}; \node at (5.1,-0.3) {$L$};
 \node at (3.1,1.45) {$\hat{\imath}$}; \node at (4.7,1.45) {$\hat{\jmath}$};
 \node at (4.17,0.42) {$\hat{k}$};
 \node at (4.8,0.85) {$\lambda_{\hat{\imath}\hat{\jmath}}^{\hat{k}}$};
 \node at (3.9,-0.65) {$\psi_{M\hat{k}}^{L}$};
\end{tikzpicture}
\caption{Boundary OPE as a change of basis of module junctions. Thickened lines are understood as ribbons with the chosen framing. This schematic depicts~\eqref{eq:main-module-ope}; it does not exchange the two insertions. Parity signs arise when ribbons or ordered insertions are exchanged.}
\label{fig:module-ope-main}
\end{figure}

To obtain the full correlator, thicken the world sheet $\Sigma$ to its connecting three-manifold $M_\Sigma$ and insert the ribbon graph $\Gamma_{F,\mathfrak s}$ determined by the algebra, boundary actions, field coupons and spin structure $\mathfrak s$. (A coupon is a common terminology in TQFT literature denoting a small labelled box in a ribbon diagram representing a morphism. The ribbons entering and leaving the box specify its domain and codomain.)
Then
\begin{equation}
 \operatorname{Corr}_{F}(\Sigma,\mathfrak s)
 =\widehat Z_{\mathcal C}\!\left(M_\Sigma,\Gamma_{F,\mathfrak s}\right)
 \in\widehat Z_{\mathcal C}(\widehat\Sigma).
 \label{eq:main-tft-correlator}
\end{equation}
Here $\widehat\Sigma=\partial M_\Sigma$ is the double of the world sheet. The TQFT state space $\widehat Z_{\mathcal C}(\widehat\Sigma)$ is identified with the space of conformal blocks on $\widehat\Sigma$. The TQFT vector specifies a linear combination of conformal blocks, which supply the dependence on insertion positions. For a disk, $M_\Sigma$ is a three-ball, so the three-BCO amplitude used in each triangle tensor follows from ribbon insertions in this manifold.

There is a module pentagon equation that expresses associativity of the boundary OPE. The full TQFT prescription additionally includes bulk projectors, bulk--boundary couplings and gluing maps. Together with its factorisation and monodromy properties imply the spin versions of the Cardy--Lewellen sewing constraints~\cite{FRS4,Runkel2020,Runkel2022}.  Therefore the TQFT with the extra input of $F$ provides a solution to the Cardy-Lewellen problem, allowing us to generalise the free fermion discretisation to more general spin CFTs. 

\subsection{Ising and the A-series}
\label{subsec:ising-main}
The A-series fermionic minimal models can be obtained from the A-series minimal models by introducing $F$ described above. 
For example, the  free fermion theory discussed in section 3 is well known to be obtainable from the Ising model. 
For Ising, $\mathcal C$ has objects $\mathbf1,\epsilon,\sigma$, with
$\epsilon^2=\mathbf1$, $\epsilon\sigma=\sigma$ and $\sigma^2=\mathbf1\oplus\epsilon$. The algebra is $F=\mathbf1\oplus\Pi\epsilon$. Decomposing induced modules gives three underlying simple graded modules, up to parity-preserving isomorphism:
\begin{equation}
 M_1=\mathbf1\oplus\Pi\epsilon,\qquad
 M_2=\Pi\mathbf1\oplus\epsilon,\qquad
 M_3=\sigma\oplus\Pi\sigma.
 \label{eq:main-ising-modules}
\end{equation}
Comparing with our previous field theoretic review of the free fermion, $M_1,M_2$ are the two parity-related $B_+$ boundaries and $M_3$ is the $B_-$ boundary carrying a boundary fermion. The first two are of $m$-type, with no odd dimension-zero boundary automorphism; the third is of $q$-type, with an odd dimension-zero generator whose square can be normalised to one \cite{AasenLakeWalker2019}. The two summands of $M_3$ therefore form one elementary graded boundary, not two independent boundaries. Conversely, $M_1$ and $M_2$ should not be identified when parity-preserving boundary equivalence is required. 

For example,
\[
 \operatorname{Hom}_F(M_1\otimes\Pi\epsilon,M_1)\neq0,
 \qquad
 \operatorname{Hom}_F(M_3\otimes\Pi\mathbf1,M_3)\neq0.
\]
By Eq.~\eqref{eq:main-bco-hom}, these nonzero multiplicity spaces imply that the open-string spectra on $M_1$ and $M_3$ contain odd boundary primaries of conformal weights $h_\epsilon=1/2$ and $h_{\mathbf1}=0$, respectively. The factor $\Pi$ records the field parity, while the underlying chiral label determines the conformal weight. These fields are $\psi_B$ and the boundary Majorana fermion $a_B$ of Section~\ref{sec:bco-spectrum-boundary-fermion}, respectively. The bulk NS/R projections and their relation to contractions of these open-string fields require the spin information discussed in the next subsection.

The same steps apply to the unitary A-series $M(p,p+1)$. There, one picks $G=\phi_{p-1,1}\sim\phi_{1,p}$. Appendix~\ref{app:fermionic_models} reviews the minimal-model notation, the Kac labels and their identification, and the simple current $G$. One then constructs $F = 1+\Pi G$, obtains its modules, its module morphisms and the fusion between them. The detailed boundary states, BCO spectra and OPE coefficients of these models are collected in Appendix~\ref{app:A_2}, including the parity convention in the Ramond sector and the possible Arf-theory change. The explicit shrinkable weights needed in the tensors are given in Appendix~\ref{sec:shrinkable-weights}. Thus the extension uses the same local categorical construction as Ising; the model dependence resides in the chiral data and the resulting module solutions.

\begin{table}[tbp]
 \centering
\renewcommand{\arraystretch}{1.15}
 \caption{Dictionary between the free-fermion notation of Section~3
 and the module description of Section~5.3. We choose the ordering
 of the parity-related modules so that $M_1$ corresponds to $B_{+,0}$
 and $M_2$ to $B_{+,1}$, with canonical spin lifts $N_M=P_M$.
 In the lower panel, $\Pi i$ denotes $i\boxtimes\Pi$.
 A field of type $\hat\imath=i\boxtimes\Pi^p$ between boundaries
 $A$ and $B$ is represented by a morphism in
 $\operatorname{Hom}_F(M_A\otimes\hat\imath,M_B)$.
 Thus the lower panel identifies field types; their occurrence
 depends on the chosen boundary pair.}
 \label{tab:ising-boundary-module-dictionary}

 \begin{tabular}{cclc}
\toprule
Brane $A$ & Boundary & $F$-module & Type \\
\midrule
$0$ & $B_{+,0}$ & $M_1=\mathbf1\oplus\Pi\epsilon$ & $m$ \\
  $1$ & $B_{+,1}$ & $M_2=\Pi\mathbf1\oplus\epsilon$ & $m$ \\
  $f$ & $B_-$     & $M_3=\sigma\oplus\Pi\sigma$     & $q$ \\
\bottomrule
\end{tabular}

 \medskip

 \begin{tabular}{cccc}
\toprule
Boundary field & Chiral and parity label $\hat\imath$
                 & Parity $p$ & Weight $h_i$ \\
\midrule
$1_B$        & $\mathbf1$     & $0$ & $0$    \\
  $\epsilon_B$ & $\epsilon$     & $0$ & $1/2$  \\
  $\sigma_B$   & $\sigma$       & $0$ & $1/16$ \\
  $\psi_B$     & $\Pi\epsilon$  & $1$ & $1/2$  \\
  $a_B$        & $\Pi\mathbf1$  & $1$ & $0$    \\
  $\mu_B$      & $\Pi\sigma$    & $1$ & $1/16$ \\
\bottomrule
\end{tabular}

\end{table}

\subsection{Spin structures and the reconstruction of closed states}\label{subsec:spin-open-closed}
Spin structures admit a discretised graphical representation. The framework is constructed by a series of works \cite{Novak2014,Novak2015,RunkelSzegedy2018}. In particular, non-contractible closed cycles characterised by NS or R spin structures admit a natural description in this framework. The spin structures of open manifolds are also considered in \cite{SternSzegedy2020}. 
In our discrete representation however, we are essentially reconstructing closed manifolds with manifolds with conformal boundaries. Therefore, we are asking the question: what kind of spin-structures (or linear combinations thereof) assigned at the boundaries (in the presence of BCOs) that topologically form a closed cycle would uniquely couple {\it exclusively} to either NS or R closed states. This is beyond the scope of existing literature.  

We first describe the discrete spin structure in terms of edge
labels, then translate these labels into Nakayama insertions and
follow their action through the boundary sewing to derive the
NS/R cylinder contraction signs.

\begin{figure}[htbp]
\centering
\begin{tikzpicture}[
 font=\small,
 spinprimal/.style={draw=black!75,line width=.65pt},
 spindual/.style={draw=blue!65!black,line width=.9pt,densely dashed},
 spinmark/.style={draw=orange!85!black,line width=2.3pt,line cap=round},
 spinout/.style={postaction={decorate},decoration={markings,mark=at position .7 with {\arrow{Latex}}}},
 spinlabel/.style={fill=white,inner sep=1.5pt}
]
\begin{scope}
 \coordinate (sv) at (0,0);
 \coordinate (sp1) at (2.4,0);
 \coordinate (sp2) at (-1.2,2.07846);
 \coordinate (sp3) at (-1.2,-2.07846);
 \coordinate (sd1) at (.4,.69282);
 \coordinate (sd2) at (-.8,0);
 \coordinate (sd3) at (.4,-.69282);
 \node[font=\small\bfseries] at (.3,2.85) {(a) Triangulation and dual graph};
 \draw[spinprimal] (sp1)--(sp2)--(sp3)--cycle;
 \draw[spinprimal,spinout] (sv)--(sp1);
 \draw[spinprimal,spinout] (sv)--(sp2);
 \draw[spinprimal,spinout] (sp3)--(sv);
 \draw[spindual] (sd1)--(sd2)--(sd3)--cycle;
 \draw[spindual] (sd1)--(.9,1.55885);
 \draw[spindual] (sd2)--(-1.75,0);
 \draw[spindual] (sd3)--(.9,-1.55885);
 \foreach \spinpoint in {sd1,sd2,sd3}
  {\fill[blue!65!black] (\spinpoint) circle (2pt);}
 \foreach \spinpoint in {sv,sp1,sp2,sp3}
  {\fill[black!75] (\spinpoint) circle (1.5pt);}
 \node[below right,inner sep=2pt] at (sv) {$v$};
 \node[spinlabel] at (1.83,.24) {$s_{e_1}$};
 \node[spinlabel] at (-1.15,1.58) {$s_{e_2}$};
 \node[spinlabel] at (-1.15,-1.58) {$s_{e_3}$};
 \draw[spinmark] (.67,.14)--(.99,.14);
 \draw[spinmark] (-1.08,.32)--(-1.08,.64);
 \draw[spinmark] (1.08,-.65)--(1.36,-.48834);
 \draw[spinprimal] (-1.65,-2.48)--(-.98,-2.48);
 \node[anchor=west] at (-.85,-2.48) {triangulation};
 \draw[spindual] (-1.65,-2.81)--(-.98,-2.81);
 \fill[blue!65!black] (-1.315,-2.81) circle (2pt);
 \node[anchor=west] at (-.85,-2.81) {dual graph};
 \draw[spinmark] (-1.48,-3.14)--(-1.16,-3.14);
 \node[anchor=west] at (-.85,-3.14) {marked side of a face};
\end{scope}
\begin{scope}[xshift=6.7cm]
 \coordinate (hv) at (0,0);
 \coordinate (h1) at (1.7,0);
 \coordinate (h2) at (-.85,1.47224);
 \coordinate (h3) at (-.85,-1.47224);
 \node[font=\small\bfseries] at (.25,2.85) {(b) Half-edges at $v$};
 \node at (.25,2.30) {$s_{h_i}=s_{e_i}\in\mathbb Z_2$};
 \draw[spinprimal,spinout] (hv)--(h1);
 \draw[spinprimal,spinout] (hv)--(h2);
 \draw[spinprimal,spinout] (h3)--(hv);
 \fill[black!75] (hv) circle (1.6pt);
 \foreach \spinpoint in {h1,h2,h3}
  {\draw[spinprimal,fill=white] (\spinpoint) circle (1.6pt);}
 \node[below right,inner sep=2pt] at (hv) {$v$};
 \node[right] at (h1) {$h_1$};
 \node[above left] at (h2) {$h_2$};
 \node[below left] at (h3) {$h_3$};
 \node[spinlabel] at (1.03,.36) {$s_{h_1}$};
 \node[spinlabel] at (-.96,.80) {$s_{h_2}$};
 \node[spinlabel] at (-.96,-.80) {$s_{h_3}$};
 \draw[spinmark] (.68,.14)--(1.00,.14);
 \draw[black!50,-{Latex[length=1.5mm]}] (.43,.1565)
  arc[start angle=20,end angle=100,radius=.4576];
 \node at (.25,-2.14) {$H_v=\{h_1,h_2,h_3\},\quad D_v=\{h_1\}$};
 \node at (.25,-2.65) {$\widehat s_{h_1}=s_{h_1},\qquad\widehat s_{h_2}=s_{h_2}$};
 \node at (.25,-3.14) {$\widehat s_{h_3}=-1-s_{h_3}\pmod2$};
\end{scope}
\end{tikzpicture}
\caption{Marked triangulation and spin indices. (a) Three triangles meet at an interior vertex $v$. The black arrows orient their common edges; the dashed blue dual graph has one vertex in each triangle and crosses each common edge once. Short orange strokes mark one side of each face, on that face's side of the edge. The outer triangle bounds a patch of a larger triangulation, not a physical boundary; the dual graph continues beyond it, and other edge indices are omitted. (b) The incident half-edges inherit the same indices, $s_{h_i}=s_{e_i}$. The curved arrow indicates the anticlockwise sense used to define $D_v$. Only the face immediately anticlockwise from $h_1$ marks its edge, so $D_v=\{h_1\}$. The arrows make $h_1,h_2$ outgoing and $h_3$ incoming, giving the displayed contributions $\widehat s_{h_i}$. The indices must satisfy the admissibility conditions in Appendix~\ref{app:subsec:spin_structures}.}
\label{fig:spin-triangulation-dual}
\end{figure}
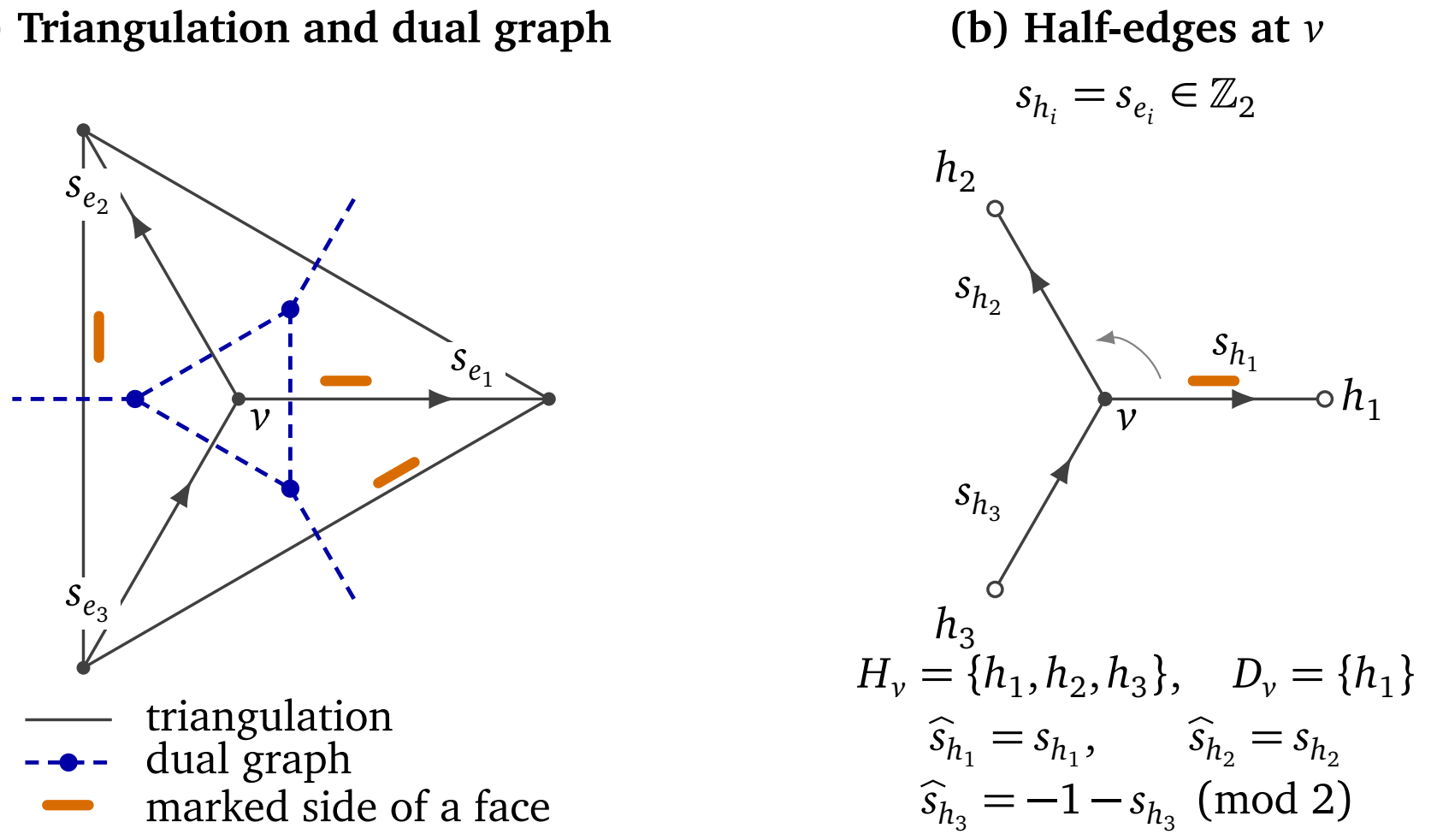

To correctly assemble BCOs into a given spin structure of a cycle, we first describe the geometric construction underlying the marked polygonal framework of Ref.~\cite[Section~2.2]{Runkel2022}. Start with an oriented world sheet $\Sigma$ and a triangulation compatible with its boundaries; neighbouring triangles may also be combined into polygonal faces. Each face is contractible and carries a unique spin structure up to isomorphism. The global choice lies in how these local spin structures are glued across edges: a gluing of oriented frames has two possible lifts to spin frames, differing by a sign. Figure~\ref{fig:spin-triangulation-dual} shows the triangulation, its dual graph, and the edge indices used below.

We distinguish a \emph{free boundary}, which is a physical conformal boundary, from a \emph{gluing boundary}, a parametrised circle or interval along which world sheets are sewn. Removing a small disk around a bulk insertion produces a closed gluing boundary. The edges and vertices on a gluing boundary must agree with its parametrisation so that the decompositions can be matched when sewing.

Choose one marked edge on each face, before identifying its edges with those of neighbouring faces. Every edge $e$ outside the free boundary also receives an orientation and an index $s_e\in\mathbb Z_2$. With the orientations and marked edges fixed, this index records which of the two spin lifts is used in the gluing. Edges on the free boundary carry no such index. 

To formulate the consistency of the gluing near a vertex $v$, split each indexed edge into two half-edges and let $H_v$ be those incident at $v$. Let $D_v\subset H_v$ consist of the half-edges whose edge is marked for the face immediately anticlockwise from that half-edge. The marked side is distinguished before edge identifications; a loop at $v$ contributes two half-edges to $H_v$. Writing $s_h=s_{e(h)}$ for the index of the edge containing $h$, the orientation-dependent contribution at $v$ is
\[
 \widehat s_h=
 \begin{cases}
  s_h,&h\text{ leaves }v,\\
  -1-s_h,&h\text{ enters }v,
 \end{cases}
 \qquad\text{with all indices understood modulo }2.
\]
The incoming-edge shift includes the change of frame convention associated with reversing the edge. Note that these indices are not the fermion parities of the BCOs.

The glued spin structure must extend over each interior vertex. This requirement gives the admissibility condition~\eqref{eq:app-spin-admissibility}. This is reviewed and illustrated in more detail in Appendix~\ref{app:subsec:spin_structures}. On the other hand, at a closed gluing boundary, the prescribed NS or R type spin structure is encoded by the holonomy condition reviewed in ~\eqref{eq:app-spin-holonomy}. In our convention a bounding (NS) circle has holonomy $1$, whereas a nonbounding (R) circle has holonomy $0$. Admissible edge indices, modulo the local changes of presentation, therefore specify the spin surface. The graph moves reviewed in the appendix transform these indices when the triangulation is changed.

These markings of the spin surface get translated into appropriate $F$ line insertions in the path-integral. The dual graph is decorated by $F$-lines carrying powers of $N_F$, and on a physical boundary these become powers of the Nakayama action $N_M$. The power is determined by the edge index $s_e$. The compatibility $N_M\rho_M=\rho_M(N_F\otimes N_M)$ permits these insertions to be moved through the algebra action. At a boundary field, however, this transport also acts on the coupon. 
 In a basis diagonalising that action,
\[
 N_N\psi_{\hat{\imath}}
 =\eta_\psi\,\psi_{\hat{\imath}}(N_M\otimes\mathrm{id}_{\hat{\imath}}),
 \qquad \eta_\psi=\pm1.
\]
\nopagebreak[4]
\begin{center}
\begin{minipage}{.95\linewidth}
\centering
\begin{tikzpicture}[
 font=\small,
 >={Latex[length=1.8mm]},
 liftmodule/.style={draw=blue!65!black,line width=1pt,->},
 liftchiral/.style={draw=black!70,line width=.7pt,->},
 liftcoupon/.style={draw=blue!65!black,fill=blue!4,minimum width=.85cm,minimum height=.50cm,inner sep=2pt},
 liftbox/.style={draw=orange!80!black,fill=orange!7,minimum width=.70cm,minimum height=.44cm,inner sep=2pt}
]
\begin{scope}[xshift=-3.1cm]
 \node[liftcoupon] (slidepsiL) at (0,0) {$\psi_{\hat{\imath}}$};
 \node[liftbox] (slideNL) at (1.35,0) {$N_N$};
 \draw[liftmodule] (-2.3,0)--(slidepsiL.west);
 \draw[liftmodule] (slidepsiL.east)--(slideNL.west);
 \draw[liftmodule] (slideNL.east)--(2.3,0);
 \draw[liftchiral] (0,-1.15)--(slidepsiL.south);
 \node[below] at (0,-1.15) {$\hat{\imath}$};
 \node[above] at (-1.8,.08) {$M$};
 \node[above] at (2.05,.08) {$N$};
\end{scope}
\node at (0,0) {$=\;\eta_\psi$};
\begin{scope}[xshift=3.1cm]
 \node[liftcoupon] (slidepsiR) at (0,0) {$\psi_{\hat{\imath}}$};
 \node[liftbox] (slideMR) at (-1.35,0) {$N_M$};
 \draw[liftmodule] (-2.3,0)--(slideMR.west);
 \draw[liftmodule] (slideMR.east)--(slidepsiR.west);
 \draw[liftmodule] (slidepsiR.east)--(2.3,0);
 \draw[liftchiral] (0,-1.15)--(slidepsiR.south);
 \node[below] at (0,-1.15) {$\hat{\imath}$};
 \node[above] at (-2.05,.08) {$M$};
 \node[above] at (1.8,.08) {$N$};
\end{scope}
\end{tikzpicture}
\par\smallskip
{\small Compositions are read from left to right, with inputs ordered as $M\otimes\hat{\imath}$. For a spin-lift eigenfield $\psi_{\hat{\imath}}$, transporting the lift from the outgoing $N$-ribbon to the incoming $M$-ribbon contributes $\eta_\psi$. The incoming chiral leg carries $\mathrm{id}_{\hat{\imath}}$.\par}
\end{minipage}
\end{center}
For the canonical parity lifts in our Ising example, $\eta_\psi=(-1)^{|\hat{\imath}|}$. 

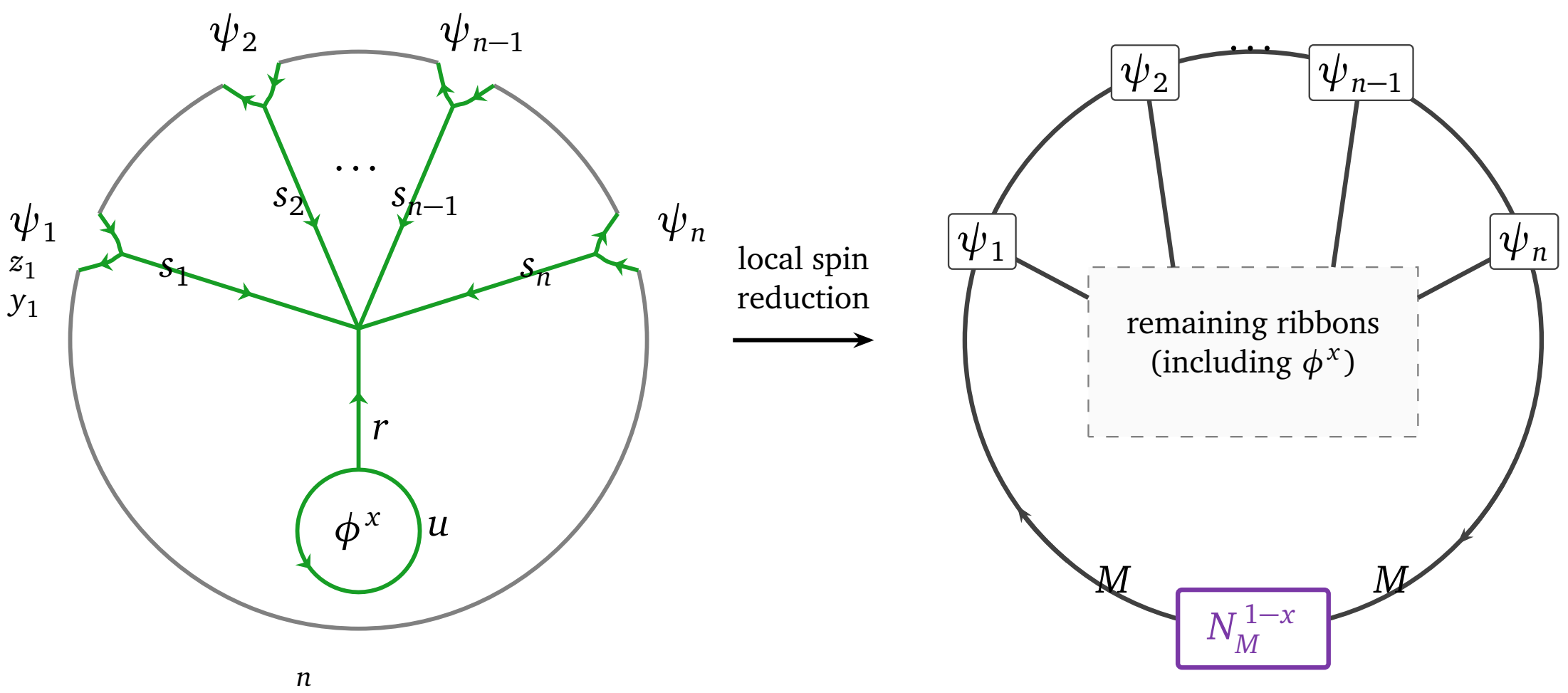
\begin{figure}[!htbp]
\centering
\begingroup
\definecolor{SpinDiskGreen}{rgb}{0.09021,0.61569,0.14511}
\definecolor{SpinDiskPurple}{rgb}{0.48,0.22,0.65}
\resizebox{0.97\textwidth}{!}{%
\begin{tikzpicture}[
  every node/.style={font=\small},
  sd-boundary/.style={draw=gray,line width=.996pt},
  sd-edge/.style={draw=SpinDiskGreen,line width=.996pt,
    postaction={decorate},decoration={markings,
      mark=at position .55 with {\arrow{Stealth[length=2.9pt,width=4.6pt]}}}},
  sd-ribbon/.style={draw=black!75,line width=1.1pt},
  sd-coupon/.style={draw=black!75,fill=white,rounded corners=1pt,
    minimum width=.54cm,minimum height=.36cm,inner sep=2pt}
]
\def\R{2.45}
\def\Rin{2.14}
\def\Rs{0.52}

\begin{scope}
  \draw[sd-boundary] (0,0) circle (\R);
  \coordinate (O) at (0,.10);
  \coordinate (A1) at (166:\R);
  \coordinate (B1) at (154:\R);
  \coordinate (V1) at (160:\Rin);
  \coordinate (A2) at (118:\R);
  \coordinate (B2) at (106:\R);
  \coordinate (V2) at (112:\Rin);
  \coordinate (A3) at (74:\R);
  \coordinate (B3) at (62:\R);
  \coordinate (V3) at (68:\Rin);
  \coordinate (A4) at (26:\R);
  \coordinate (B4) at (14:\R);
  \coordinate (V4) at (20:\Rin);
  \draw[white,line width=3.4pt] (A1) arc[start angle=166,end angle=154,radius=\R];
  \draw[white,line width=3.4pt] (A2) arc[start angle=118,end angle=106,radius=\R];
  \draw[white,line width=3.4pt] (A3) arc[start angle=74,end angle=62,radius=\R];
  \draw[white,line width=3.4pt] (A4) arc[start angle=26,end angle=14,radius=\R];
  \draw[sd-edge] (V1) .. controls (163:2.25) and (161:2.12) .. (A1);
  \draw[sd-edge] (B1) .. controls (159:2.12) and (157:2.25) .. (V1);
  \draw[sd-edge] (V2) .. controls (115:2.25) and (113:2.12) .. (A2);
  \draw[sd-edge] (B2) .. controls (111:2.12) and (109:2.25) .. (V2);
  \draw[sd-edge] (V3) .. controls (71:2.25) and (69:2.12) .. (A3);
  \draw[sd-edge] (B3) .. controls (67:2.12) and (65:2.25) .. (V3);
  \draw[sd-edge] (V4) .. controls (23:2.25) and (21:2.12) .. (A4);
  \draw[sd-edge] (B4) .. controls (19:2.12) and (17:2.25) .. (V4);
  \draw[sd-edge] (V1)--(O);
  \draw[sd-edge] (V2)--(O);
  \draw[sd-edge] (V3)--(O);
  \draw[sd-edge] (V4)--(O);
  \coordinate (C) at (0,-1.62);
  \coordinate (Ct) at ($(C)+(90:\Rs)$);
  \draw[sd-edge] ($(C)+(20:\Rs)$)
    arc[start angle=20,end angle=380,radius=\Rs];
  \draw[sd-edge] (Ct)--(O);
  \node at (-1.55,.58) {$s_1$};
  \node at (-.58,1.18) {$s_2$};
  \node at (.58,1.18) {$s_{n-1}$};
  \node at (1.52,.58) {$s_n$};
  \node at (0,1.43) {$\cdots$};
  \node at (.19,-.76) {$r$};
  \node at (.68,-1.58) {$u$};
  \node at (C) {$\phi^x$};
  \node at (160:2.93) {$\psi_1$};
  \node at (112:2.82) {$\psi_2$};
  \node at (68:2.82) {$\psi_{n-1}$};
  \node at (20:2.93) {$\psi_n$};
  \node[font=\scriptsize] at (-2.84,.62) {$z_1$};
  \node[font=\scriptsize] at (-2.84,.29) {$y_1$};
  \node at (0,-3.30) {(a) $\displaystyle\sum_{i=1}^{n}s_i=1-x$};
\end{scope}

\draw[-{Stealth[length=5pt]},line width=1pt] (3.18,0)--(4.38,0);
\node[align=center,font=\scriptsize] at (3.78,.53)
  {local spin\\reduction};

\begin{scope}[xshift=7.6cm]
  \draw[sd-ribbon] (0,0) circle (\R);
  \draw[sd-ribbon,-{Stealth[length=4pt]}]
    (230:\R) arc[start angle=230,end angle=215,radius=\R];
  \draw[sd-ribbon,-{Stealth[length=4pt]}]
    (330:\R) arc[start angle=330,end angle=315,radius=\R];
  \draw[gray,dashed,fill=gray!4] (-1.40,-.82) rectangle (1.40,.62);
  \node[align=center,font=\scriptsize] at (0,-.04)
    {remaining ribbons\\(including $\phi^x$)};
  \draw[sd-ribbon] (160:\R)--(-1.4,.36);
  \draw[sd-ribbon] (112:\R)--(-.68,.62);
  \draw[sd-ribbon] (68:\R)--(.68,.62);
  \draw[sd-ribbon] (20:\R)--(1.4,.36);
  \node[sd-coupon] at (160:\R) {$\psi_1$};
  \node[sd-coupon] at (112:\R) {$\psi_2$};
  \node[sd-coupon] at (68:\R) {$\psi_{n-1}$};
  \node[sd-coupon] at (20:\R) {$\psi_n$};
  \node at (0,2.45) {$\cdots$};
  \node[draw=SpinDiskPurple,fill=white,line width=1.1pt,
    rounded corners=1pt,minimum width=1.28cm,minimum height=.53cm,
    text=SpinDiskPurple] at (0,-\R) {$N_M^{\,1-x}$};
  \node at (-1.18,-2.03) {$M$};
  \node at (1.18,-2.03) {$M$};
  \node at (0,-3.30) {(b) One residual boundary lift};
\end{scope}
\end{tikzpicture}%
}
\endgroup
\caption{Spin reduction for one bulk insertion and cyclically ordered BCOs.
(a) The many-insertion disk decomposition of
Figure~\ref{fig:app-spin-multiple-decomposition}, with the fields indicated.
The spin-label constraint is $\sum_i s_i=1-x$ modulo~$2$.
(b) Schematic reduced boundary ribbon: the residual spin decoration is
concentrated into $N_M^{1-x}$ on a segment labelled by $M$.
The cyclic order of the BCOs and the remaining $F$-ribbon and bulk-coupling
data are retained; the latter are suppressed in the dashed region.
The reduction leaves $N_M$ for NS ($x=0$) and $\mathrm{id}_M$ for R ($x=1$).}
\label{fig:spin-open-closed-closure}
\end{figure}

We find the correct spin-decoration at the collection of boundaries that form an $S^1$ such that it recovers the NS or R spin structure. 
For a disk with a bulk insertion of type $x$, where $x=0$ denotes NS and $x=1$ denotes R, it would exclusively couple to a boundary with multiple conformal boundary condition and BCOs separating them such that the spin decoration (i.e. spin indices on the boundary that controls powers of $N_F$ or $N_M$ insertions) can be reduce to a single residual insertion on a chosen boundary segment $M$:
\begin{equation}
 \boxed{
 N_M^{\,1-x}=
 \begin{cases}
  N_M,&x=0\quad(\mathrm{NS}),\\
  \mathrm{id}_M,&x=1\quad(\mathrm R).
 \end{cases}}
 \label{eq:new-spin-residual}
\end{equation}
This remains true for any number of cyclically ordered BCOs. Figure~\ref{fig:spin-open-closed-closure} uses the disk with one bulk field and $n$ boundary fields from Appendix~\ref{app:subsec:disk_correlators} to illustrate the reduction. In this marked decomposition the admissibility equations give
\begin{equation}
 t:=\sum_{i=1}^{n}s_i=1-x\pmod2.
 \label{eq:new-spin-sector-bit}
\end{equation}
One can choose a representative with $s_1=t$ and all other $s_i=0$; the remaining boundary lift is then $N_M^t$. 

\begin{figure}[!htbp]
\centering
\begin{tikzpicture}[
 font=\small,
 >={Latex[length=1.8mm]},
 seammod/.style={draw=blue!65!black,line width=.9pt,->},
 seamfield/.style={draw=black!75,line width=.8pt,->},
 seamcoupon/.style={draw=blue!65!black,fill=blue!4,minimum width=.70cm,minimum height=.48cm,inner sep=2pt},
 seamlift/.style={draw=orange!85!black,fill=orange!7,minimum width=.68cm,minimum height=.46cm,inner sep=2pt}
]
\begin{scope}[xshift=-3.0cm]
 \node[seamcoupon] (pL) at (0,0) {$\psi_{\hat{\jmath}}$};
 \node[seamlift] (nL) at (1.22,0) {$N_N^t$};
 \draw[seammod] (-2.2,0)--(pL.west);
 \draw[seammod] (pL.east)--(nL.west);
 \draw[seammod] (nL.east)--(2.2,0);
 \draw[seamfield] (0,-.95)--(pL.south);
 \node[below] at (0,-.95) {$\hat{\jmath}$};
 \node[above] at (-1.9,.08) {$M$};
 \node[above] at (1.98,.08) {$N$};
\end{scope}
\node at (0,0) {$=\;(-1)^{pt}$};
\begin{scope}[xshift=3.0cm]
 \node[seamcoupon] (pR) at (0,0) {$\psi_{\hat{\jmath}}$};
 \node[seamlift] (mR) at (-1.22,0) {$N_M^t$};
 \draw[seammod] (-2.2,0)--(mR.west);
 \draw[seammod] (mR.east)--(pR.west);
 \draw[seammod] (pR.east)--(2.2,0);
 \draw[seamfield] (0,-.95)--(pR.south);
 \node[below] at (0,-.95) {$\hat{\jmath}$};
 \node[above] at (-1.98,.08) {$M$};
 \node[above] at (1.9,.08) {$N$};
\end{scope}
\begin{scope}[xshift=-3.0cm,yshift=-2.15cm]
 \node[seamcoupon] (qL) at (0,0) {$\psi_{\hat{\jmath}}$};
 \node[seamlift] (qmL) at (-1.22,0) {$N_M^t$};
 \node[seamlift] (qnL) at (1.22,0) {$N_N^t$};
 \draw[seammod] (-2.2,0)--(qmL.west);
 \draw[seammod] (qmL.east)--(qL.west);
 \draw[seammod] (qL.east)--(qnL.west);
 \draw[seammod] (qnL.east)--(2.2,0);
 \draw[seamfield] (0,-.95)--(qL.south);
 \node[below] at (0,-.95) {$\hat{\jmath}$};
 \node[above] at (-1.98,.08) {$M$};
 \node[above] at (1.98,.08) {$N$};
\end{scope}
\node at (0,-2.15) {$=\;(-1)^{pt}$};
\begin{scope}[xshift=3.0cm,yshift=-2.15cm]
 \node[seamcoupon] (qR) at (0,0) {$\psi_{\hat{\jmath}}$};
 \draw[seammod] (-2.2,0)--(qR.west);
 \draw[seammod] (qR.east)--(2.2,0);
 \draw[seamfield] (0,-.95)--(qR.south);
 \node[below] at (0,-.95) {$\hat{\jmath}$};
 \node[above] at (-1.9,.08) {$M$};
 \node[above] at (1.9,.08) {$N$};
\end{scope}
\end{tikzpicture}
\caption{The boundary spin transition acting on the open seam state. Inputs are ordered as $M\otimes\hat{\jmath}$, and compositions are read from left to right. Top: for the canonical parity lifts, moving $N_N^t$ through $\psi_{\hat{\jmath}}$ leaves $N_M^t$ on the incoming module segment and contributes $(-1)^{pt}$, with $p=|\hat{\jmath}|$. Bottom: acting at both endpoints gives $U_t$; after the top-row move, the two lifts on $M$ cancel. The coupon is even as a categorical morphism, while the field parity is carried by its $\hat{\jmath}$-leg.}
\label{fig:new-coupon-seam-transport}
\end{figure}
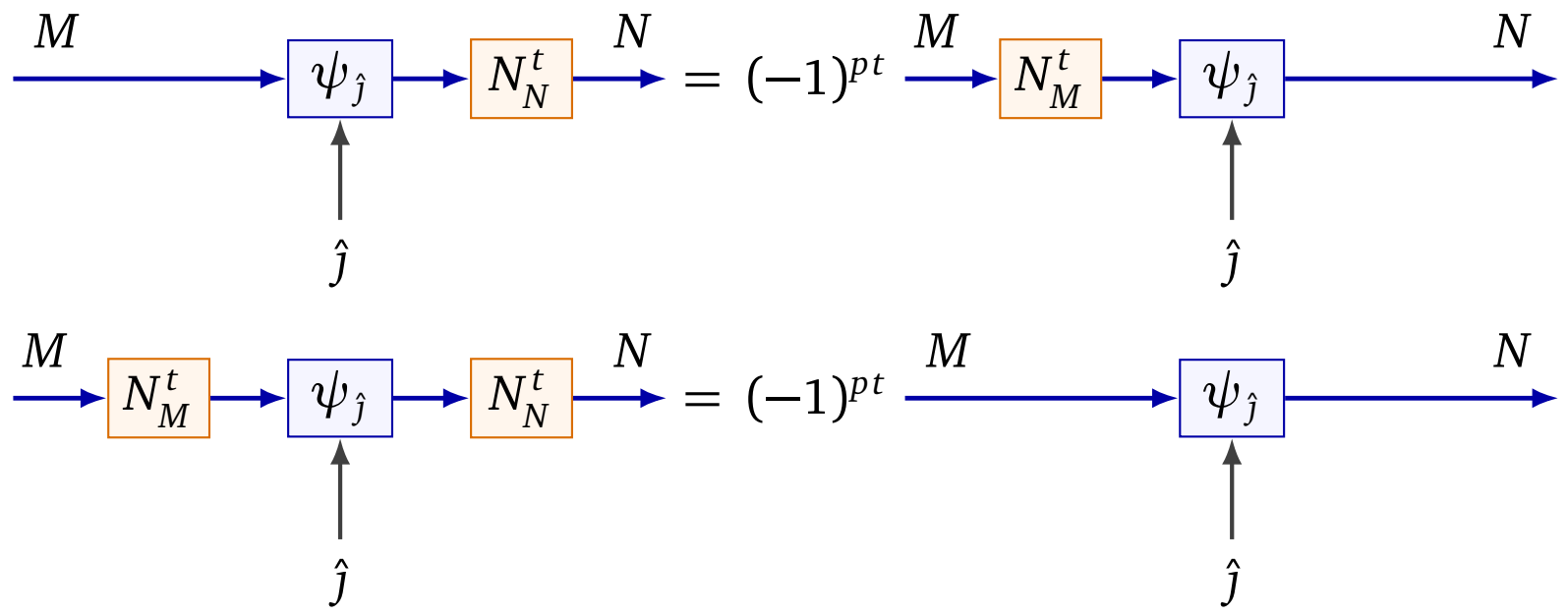

We now connect this boundary prescription to the cylinder signs used in the spectrum computation. Cut the cylinder along the computational seam. The state $\hat{\jmath}$ on that cut is an open state between two boundary modules $M,N$, represented by an even categorical coupon
\begin{equation}
 \psi_{\hat{\jmath}}:M\otimes\hat{\jmath}\longrightarrow N,
 \qquad p=|\hat{\jmath}|.
 \label{eq:new-cylinder-seam-coupon}
\end{equation}
The field parity is carried by the $\hat{\jmath}$-leg. Choose local spin frames so that the transition with exponent $t$ is represented at this seam. It acts through the lifts at the two endpoints of the open interval. For the canonical parity lifts $N_M=P_M$ and $N_N=P_N$, evenness gives $P_N\psi_{\hat{\jmath}}=\psi_{\hat{\jmath}}(P_M\otimes P_{\hat{\jmath}})$, and hence
\begin{equation}
 \boxed{
 N_N^t\psi_{\hat{\jmath}}
 =(-1)^{pt}\psi_{\hat{\jmath}}(N_M^t\otimes\mathrm{id}_{\hat{\jmath}}),
 \qquad p=|\hat{\jmath}|.}
 \label{eq:new-coupon-seam-transport}
\end{equation}
Figure~\ref{fig:new-coupon-seam-transport} shows this local move and the resulting action at both endpoints. A transported lift remains on the adjacent module segment. In the simultaneous endpoint action the two lifts on $M$ cancel because $N_M^2=\mathrm{id}_M$:
\begin{equation}
 U_t(\psi_{\hat{\jmath}}):=N_N^t\psi_{\hat{\jmath}}(N_M^t\otimes\mathrm{id}_{\hat{\jmath}})
 =(-1)^{pt}\psi_{\hat{\jmath}}.
 \label{eq:new-seam-endpoint-action}
\end{equation}
Thus the residual boundary lift and $U_t$ represent the same spin transition on, respectively, a boundary module and the open-state space of the cut.

Finally, the ket--bra ordering at the seam uses the right evaluation on the super-vector-space factor,
\begin{equation}
 \widetilde{\mathrm{ev}}\bigl(|\hat{\jmath}\rangle\otimes\langle\hat{l}|\bigr)
 =(-1)^p\delta_{\hat{\jmath}\hat{l}},
 \qquad
 \widetilde{\mathrm{ev}}\bigl(U_t|\hat{\jmath}\rangle\otimes\langle\hat{l}|\bigr)
 =(-1)^{p(1+t)}\delta_{\hat{\jmath}\hat{l}},
 \label{eq:new-seam-graded-evaluation}
\end{equation}
in the duality convention of Ref.~\cite[Section~4.1]{Runkel2022}. Separately, moving the seam leg past the bottom legs $\hat{k}_s$ in the cylinder tensor ordering contributes $(-1)^{pK}$, where $K=\sum_s|\hat{k}_s|\pmod2$. Combining these factors with $t=1-x$ gives
\begin{equation}
 S_x=
 \underbrace{(-1)^{pK}}_{\text{leg reordering}}\quad
 \underbrace{(-1)^p}_{\text{right evaluation}}\quad
 \underbrace{(-1)^{pt}}_{\text{spin transition}}
 =(-1)^{p(K+x)},
 \label{eq:new-cylinder-spin-sign}
\end{equation}
and therefore
\begin{equation}
 \boxed{
 S_{\mathrm{NS}}=(-1)^{|\hat{\jmath}|\sum_s|\hat{k}_s|},\qquad
 S_{\mathrm R}=(-1)^{|\hat{\jmath}|(1+\sum_s|\hat{k}_s|)}.}
 \label{eq:new-cylinder-ns-r-signs}
\end{equation}
Conversely, dividing either cylinder sign by the fixed reordering and right-evaluation factors leaves $(-1)^{pt}$, the action of the spin transition on the open seam state. This recovers the same circle type $x=1-t$ when the full graded seam space is retained. The two prescriptions therefore implement the same spatial NS/R choice. 

\subsection{Quantum dimensions and shrinkable NS boundaries}
\label{subsec:shrinkable-main}

A conformal boundary is labelled by an $F$-module $M$ with a compatible spin lift $N_M$. Closing its identity morphism with the NS spin structure defines its boundary quantum dimension,
\begin{equation}
 d_M^{\mathrm{NS}}:=\operatorname{Tr}^{\mathrm{NS}}_M(\mathrm{id}_M).
 \label{eq:new-ns-module-dimension}
\end{equation}
We use the unnormalised loop dimension $D(M)$ defined by the equivariant evaluation and coevaluation maps of Ref.~\cite[Section~6.1, Eqs.~(6.10)--(6.13)]{Runkel2022}, regarding $M$ as an $(F,\mathbf1)$-bimodule. This NS trace includes the spin-lift insertion described above; it is not the signed categorical dimension of the underlying object in $\widehat{\mathcal C}$.

The spin structure induced on the boundary of an empty disk is NS. Accordingly, the dimension-weighted prescription for filling a hole is the following sum of NS conformal boundary states:
\begin{equation}
 |B_{\mathrm{sh}}\rangle_{\mathrm{NS}}
 =\kappa \sum_{M\in\mathcal I_F}d_M^{\mathrm{NS}}|M\rangle_{\mathrm{NS}},
 \label{eq:new-dimension-weighted-shrinking}
\end{equation}
up to some normalisation $\kappa$. 
Here $\mathcal I_F$ contains all elementary graded boundary modules, up to parity-preserving isomorphism, with the chosen compatible lifts. Shrinkability requires this sum to resolve the vacuum Ishibashi state, $|B_{\mathrm{sh}}\rangle_{\mathrm{NS}}=|0\rangle\rangle$, with $\kappa$ fixed by the vacuum and boundary-state normalisations. The trace and boundary two-point pairing must therefore use consistent conventions.

The Ising shrinkable boundary used in Section~\ref{sec:FPtensor} is also of the form~\eqref{eq:new-dimension-weighted-shrinking}. For the three modules in Eq.~\eqref{eq:main-ising-modules}, $d_M^{\mathrm{NS}}=(2,2,2\sqrt2)$ and $\kappa=1/8$ give the weights $(1/4,1/4,1/(2\sqrt2))$. The first two modules have identical NS boundary states, so combining their contributions reproduces the choice $(1/2,0,1/(2\sqrt2))$ for the branes $(0,1,f)$ used there.

Appendix~\ref{sec:shrinkable-weights} verifies this identity explicitly for the unitary A-series. For the induced module $M_b=F\otimes b$, one finds $d_{M_b}^{\mathrm{NS}}=2d_b$, where $d_b=S_{b1}/S_{11}$ is the bosonic chiral quantum dimension, and $\kappa=S_{11}^{3/2}/(2\sqrt2)$. Combining the identical NS states of two parity-related $m$-type modules adds their weights, whereas a $q$-type module is counted once. This accounts for the relative factor of two in the reduced boundary-label convention used in our tensors.

\subsection{A candidate bulk with topological and dynamical boundaries}
\label{subsec:bulk-candidate}

Following the fermionic brane notation of Section~\ref{sec:bco-structure-constants}, let $A,B,C$ label branes represented by graded $F$-modules $M_A,M_B,M_C$ with the chosen compatible spin lifts of Section~\ref{subsec:boundaries-tft-main}. Write
$\hat{\imath}=i\boxtimes\Pi^{p_i}$, $\hat{\jmath}=j\boxtimes\Pi^{p_j}$ and $\hat{k}=k\boxtimes\Pi^{p_k}$. The triangle tensor separates into a conformal block and a boundary three-point coefficient:
\begin{equation}
 \mathcal{T}_{(\hat{\imath},I)(\hat{\jmath},J)(\hat{k},K)}^{ABC}
 =\alpha_{IJK}^{ijk} C_{\hat{\imath}\hat{\jmath}\hat{k}}^{(ABC)}.
 \label{eq:new-topological-conformal-split}
\end{equation}
Here $\hat{\imath},\hat{\jmath},\hat{k}$ label the parity-decorated primary BCOs between the brane pairs $AB,BC,CA$, respectively, and $I,J,K$ label descendants of their underlying chiral representations $i,j,k\in\mathcal C$. Fusion multiplicity indices are suppressed. The coefficient $C_{\hat{\imath}\hat{\jmath}\hat{k}}^{(ABC)}$ is obtained from the boundary OPE using the boundary two-point pairing; in mutually dual boundary-field bases it is the corresponding module fusing-matrix element of Section~\ref{subsec:boundaries-tft-main}. The conformal block $\alpha_{IJK}^{ijk}$ depends only on the underlying chiral labels and contains the descendant and triangle-geometry dependence.

Using the same sums over labels and products over vertices and triangles as in Section~\ref{sec:discretisation-review}, the partition function on a triangulated surface $\Sigma$ with spin structure represented by admissible edge labels $\{s\}=\{s_e\}$ is
\begin{equation}
 \begin{aligned}
 Z_{F,\{s\}}(\Sigma)
 &=\sum_{\{(\hat{\imath},I)\},\{A\}}(-1)^{f(\{s\};\{A\},\{\hat{\imath}\})}
   \prod_v\omega_{A_v}
   \prod_{\triangle}\mathcal{T}_{(\hat{\imath},I)(\hat{\jmath},J)(\hat{k},K)}^{ABC}\\
 &=\sum_{\{(\hat{\imath},I)\},\{A\}}(-1)^{f(\{s\};\{A\},\{\hat{\imath}\})}
   \prod_v\omega_{A_v}
   \prod_{\triangle}\alpha_{IJK}^{ijk}C_{\hat{\imath}\hat{\jmath}\hat{k}}^{(ABC)}.
 \end{aligned}
 \label{eq:new-exact-contraction}
\end{equation}
Here $\{A\}=\{A_v\}$ assigns a brane to each vertex $v$, and $A,B,C$ in each triangle tensor are the branes at that triangle's three vertices in the prescribed order. In homogeneous BCO bases diagonalising the spin-lift action, $f(\{s\};\{A\},\{\hat{\imath}\})\in\mathbb Z_2$ collects the additional signs from fermionic reordering, graded pairings and spin-lift insertions during sewing, as prescribed in Section~\ref{subsec:spin-open-closed}. It is evaluated for the fixed marked graph and contraction ordering, and depends on the brane labels and BCO parities as well as the spin marking; the relevant eigenvalue and multiplicity labels are included in $\{\hat{\imath}\}$. This sign is now explicit, so the remaining products use the dual-normalised pairings of the bosonic review without including it a second time. With all elementary graded boundary modules retained, Section~\ref{subsec:shrinkable-main} gives $\omega_{A_v}=\kappa d_{M_{A_v}}^{\mathrm{NS}}$.

Collecting the OPE coefficients and shrinking weights into a ket, and the conformal blocks into a bra, gives
\begin{align}
 |\Psi_{F,\{s\}}\rangle
 &=\sum_{\{\hat{\imath}\}}\sum_{\{A\}}(-1)^{f(\{s\};\{A\},\{\hat{\imath}\})}
   \prod_v\omega_{A_v}
   \prod_{\triangle}C_{\hat{\imath}\hat{\jmath}\hat{k}}^{(ABC)}\,|\{\hat{\imath}\}\rangle,
 \label{eq:new-topological-state}\\
 \langle\Omega_\Sigma|
 &=\sum_{\{\hat{\imath}\}}\langle\{\hat{\imath}\}|\sum_{\{I\}}
   \prod_{\triangle}\alpha_{IJK}^{ijk}.
 \label{eq:new-defined-strange-states}
\end{align}
Here $\{\hat{\imath}\}$ denotes the assignment of parity-decorated primary labels to all edges, while $\{I\}$ denotes their descendant labels; the conformal blocks use the underlying chiral labels $\{i\}$. The product of conformal blocks includes the descendant pairings of the review. All brane-label sums, shrinking weights and spin-dependent sewing factors are in $|\Psi_{F,\{s\}}\rangle$. Pairing the two states in the dual label basis therefore gives
\begin{equation}
 \boxed{Z_{F,\{s\}}(\Sigma)=\langle\Omega_\Sigma\mid\Psi_{F,\{s\}}\rangle.}
 \label{eq:new-strange-correlator}
\end{equation}
The OPE coefficients and module dimensions are independent of the metric; all conformal-block and descendant dependence has been collected into $\langle\Omega_\Sigma|$.

With a suitable change of basis and normalisation, the wavefunction $|\Psi_{F,\{s\}}\rangle$ can be written in a form that more closely resembles the usual quantum $6j$-symbol state sum. The chiral and module basis conventions are reviewed in Appendices~\ref{app:subsec:Fsymbol_conventions} and~\ref{app:subsec:ope_6j}, respectively.

\begin{figure}[htbp]
\centering
\begin{tikzpicture}[scale=.88,>=Stealth,font=\small]
 \fill[blue!5] (0,0)--(7,0)--(8.2,1)--(1.2,1)--cycle;
 \fill[green!10] (0,-2.4)--(7,-2.4)--(8.2,-1.4)--(1.2,-1.4)--cycle;
 \draw (0,0)--(7,0)--(8.2,1)--(1.2,1)--cycle;
 \draw (0,-2.4)--(7,-2.4)--(8.2,-1.4)--(1.2,-1.4)--cycle;
 \foreach \x/\y in {0/0,7/0,8.2/1,1.2/1}{\draw[gray] (\x,\y)--(\x,{\y-2.4});}
 \node at (4.1,.50) {gluing surface: conformal functional $\langle\Omega|$};
 \node at (4.0,-.70) {bulk recoupling data};
 \draw[green!45!black,thick] (1.0,-2.13)--(2.5,-1.75)--(4,-2.13)--(5.5,-1.75)--(7,-2.13);
 \draw[green!45!black,thick] (2.5,-1.75)--(2.5,-2.35) (5.5,-1.75)--(5.5,-2.35);
 \fill[green!45!black] (2.5,-1.75) circle (2pt) (5.5,-1.75) circle (2pt);
 \node[draw,fill=white,inner sep=2pt] at (4,-2.13) {$N_F^{s}$};
 \node at (4,-2.92) {physical boundary: $F$-network and modules $M,N,\ldots$};
\end{tikzpicture}
\caption{Proposed bulk interpretation of the 2d CFT path-integral. The lower boundary carries the spin-structure dependent $F$-network which is expressed in terms of the $F$-modules; the upper boundary is paired with the conformal blocks. }
\label{fig:new-bulk-slab}
\end{figure}
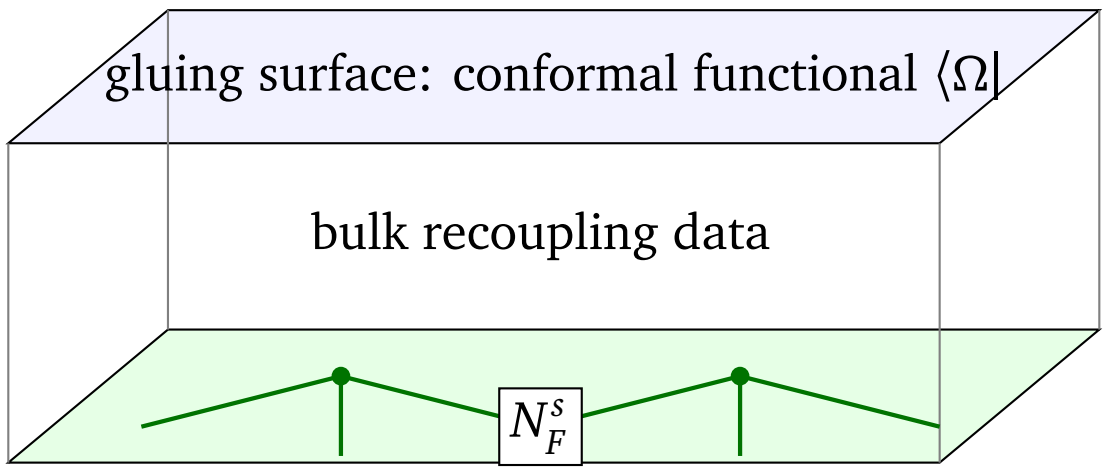

Motivated by the results ~\cite{Lootens2020} which describes the topological boundary of bosonic Turaev-Viro/Levin-Wen model with modules of a Frobenius algebra $A \in \mathcal{C}$ (the counterpart of  $F \in \widehat{\mathcal C}$ here), we propose that $|\Psi_{F,\{s\}}\rangle$ in (\ref{eq:new-topological-state}) is the discrete state-sum representation of $\widehat{\mathcal C}=\mathcal C\boxtimes\mathrm{SVect}$ over a ``sandwich'', with one topological boundary condition characterised by the proliferation of $F$, and the other dynamical boundary condition controlled by the conformal blocks. The fermionic partition function over a spin surface admits a 3d interpretion much like the bosonic one, as depicted in figure \ref{fig:new-bulk-slab}

\section{Conclusion and Discussion}\label{sec:conclusion}

In this paper, we have presented an analytic construction of exact fixed-point (FP) tensors for fermionic rational conformal field theories. Our work generalizes the bosonic construction of Cheng \emph{et al.}~\cite{Cheng2023} to theories with fermionic degrees of freedom, where spin structure dependence, namely the appearance of the Neveu-Schwarz and Ramond sectors, and Grassmann statistics must be carefully incorporated. We focused on the free Majorana fermion ($c=1/2$) as the primary example; the construction extends to the whole fermionic A-series (Section~\ref{sec:fermionization} and Appendix~\ref{app:fermionic_models}). Given that fermionic CFTs can be obtained from fermionisation of bosonic CFTs, the construction presented in this paper should be applicable quite generally. 

The main results of this paper are fourfold.

First, we constructed the fermionic FP tensor in Section~\ref{sec:FPtensor} as a three-point correlation function of boundary-changing operators (BCOs) on a triangular region, with boundary conditions and BCOs labeled by the fermionic boundary states discussed in Section~\ref{sec:fermionic-bcft}. The tensor factorizes into a product of a structure constant $C_{ijk}^{ABC}$  and a conformal block $\alpha_{IJK}^{ijk}$, mirroring the bosonic case. The fermionic nature manifests itself through (i) modified weights for shrinkable boundary conditions distinguishing $\mathbb{Z}_2$-invariant and $\mathbb{Z}_2$-breaking boundaries; (ii) additional sign factors ${(-1)}^{|j|(\sum_s |k_s|)}$ (NS) and ${(-1)}^{|j|(1+\sum_s |k_s|)}$ (R) arising from fermionic tensor contraction rules; and (iii) the vanishing of Ramond components for $\mathbb{Z}_2$-invariant boundaries due to the Ramond zero-mode obstruction (Section~\ref{sec:boundary-states-spin-Cardy} and Appendix~\ref{app:subsec:boundary_states_fermionic_minimal}).

Second, we verified the algebraic consistency of our construction. The fermionic FP tensor satisfies crossing symmetry and the coarse-graining condition, confirming that it describes a fixed point under real-space renormalization. The coarse-graining condition involves a weighted sum over boundary conditions with modified weights $\omega_b$.

Third, we extended the construction to boundary tensors and open-string spectra. By introducing boundary tensors and combining them with triangle tensors, we assembled a strip geometry whose transfer matrix reproduces the open-string spectrum. Numerical results for the Ising model showed that the low-lying boundary scaling dimensions ($\Delta = 0, 0.5, 1.5, 2.0, \dots$) can be extracted with reasonable accuracy under a moderate truncation of descendant levels.

Fourth, we formulate our discretisation in the three-dimensional TQFT framework for rational spin CFTs. This framework organises the boundary OPE data and shrinkable-boundary weights in terms of Frobenius algebras and their modules, and explains the NS/R contraction signs through Nakayama insertions associated with the discrete spin structure. We further express the partition function as a strange correlator, separating the conformal blocks from the OPE data, boundary weights and spin-dependent sewing factors. This factorisation motivates a candidate three-dimensional bulk interpretation with a topological boundary determined by \(F\) and a dynamical boundary carrying the conformal-block data.

Our work suggests several promising directions for future research. The analytic FP tensors constructed here can serve as optimal initial tensors for Grassmann TNR algorithms~\cite{Mortier2024}, potentially improving convergence in simulations of fermionic critical phenomena. 

In conclusion, this paper has taken a significant step toward a comprehensive tensor network formulation of fermionic conformal field theory, bridging fermionic CFT, fixed-point tensor network, and spin topological quantum field theories. We hope that the analytic framework developed here will provide a useful foundation for both theoretical studies and numerical simulations of fermionic critical systems.

\section*{Acknowledgements}

We thank Xuzhihan Ding, Kaixin Ji and Yu Liu for useful discussions. Particularly we thank Yu for collaboration in the early stage of this project.
LYH acknowledges the support of NSFC (Grant No. 11922502, 11875111). 
We have used AI in improving the text of this manuscript. Computations in this manuscript are done by ourselves.

\bibliography{main}
\appendix
\section{Algebraic data, A-series solutions and spin sewing}
\label{app:combined}
This appendix supplies the detailed data for Section~\ref{sec:fermionization}. The algebraic construction comes first, followed by the complete A-series boundary and BCO formulas, the spin-marked disk reductions, and the three-dimensional interpretation of boundary tensors. The explicit coefficient formulas use the specified bases; spin-sector statements use the invariant marked-graph description.

\subsection{Algebra, modules and ribbon evaluations}
\subsubsection{Chiral, parity and ribbon conventions}
\label{app:subsec:Fsymbol_conventions}

We use the ribbon structure on $\mathcal{SV}ect$ of Ref.~\cite[Section~4.1]{Runkel2022}: the braiding is the Koszul braiding, the twist is trivial, and the categorical dimension of the odd line is $\dim(\Pi)=-1$. Consequently, for $\hat{\imath}=i\boxtimes\Pi^{p_i}$,
\begin{align}
 d_{\hat{\imath}}&=(-1)^{p_i}d_i,&
 \theta_{\hat{\imath}}&=\theta_i,&
 R^{\hat{\imath}\hat{\jmath}}_{\hat{k}}
 &=(-1)^{p_i p_j}R^{ij}_k .
 \label{eq:app-super-ribbon}
\end{align}
Only channels with $p_k=p_i+p_j$ are allowed. With the standard strict choice for the parity factor, associator matrices in $\widehat{\mathcal C}$ are numerically those of the underlying labels in $\mathcal C$. This does not allow one to discard the parity ribbons when taking traces, reversing orientation or exchanging insertions.

Choose fusion vertices $\lambda_{ij}^{k}\in\operatorname{Hom}(i\otimes j,k)$. In the multiplicity-free conventions of the manuscript, the fusing matrix acting on conformal blocks is denoted by
\begin{equation}
 [F^{ijk}_l]^{\mathrm{blocks}}_{mn}
 =\Fsymbol{i}{j}{k}{l}{m}{n}.
\end{equation}
The tetrahedral gauge and the blocks gauge are related by
\begin{equation}
 [F^{ijk}_l]_{mn}
 =[F^{ijk}_l]^{\mathrm{blocks}}_{mn}
 \frac{\mathcal N_{jkn}\mathcal N_{inl}}
 {\mathcal N_{ijm}\mathcal N_{mkl}},
 \qquad
 [F^{ijk}_l]_{mn}
 =\sqrt{d_m d_n}\sixj{i}{j}{m}{k}{l}{n}.
 \label{eq:F_gauge_app}
\end{equation}
For the self-dual, multiplicity-free minimal-model labels and the vertex normalisation used here,
\begin{equation}
 \mathcal N_{ijk}
 =\sqrt{\frac{\Theta(i,j,k)}{\sqrt{d_i d_j d_k}}},
 \qquad
 \Theta(i,j,k)=\frac{d_i}{[F^{jjk}_{k}]^{\mathrm{blocks}}_{1i}}.
 \label{eq:N_ijk_app}
\end{equation}
The square-root choices are part of the gauge. The symbol $\Theta$ denotes the theta-graph evaluation and is distinct from the twist eigenvalue $\theta_i$. The $d_i$ in these gauge formulas are dimensions in $\mathcal C$; replacing them mechanically by signed superdimensions would change the conventions. In the parity extension we keep $\mathcal N_{\hat{\imath}\hat{\jmath}\hat{k}}=\mathcal N_{ijk}$ and handle parity through~\eqref{eq:app-super-ribbon}.

Our braiding convention is
\begin{equation}
 \lambda_{ji}^{k}\circ c_{i,j}=R^{ij}_{k}\lambda_{ij}^{k},
 \qquad
 R^{ij}_{k}R^{ji}_{k}=\frac{\theta_k}{\theta_i\theta_j}.
 \label{eq:app-R-definition}
\end{equation}
For fusion multiplicities larger than one all these vertices and matrices carry additional basis indices. The basis-independent constructions below remain valid, but a scalar $6j$ notation is then insufficient.

\subsubsection{Frobenius algebra and Nakayama action}
\label{app:subsec:frobenius-details}

A Frobenius algebra consists of $F$ and even morphisms
\begin{equation}
 \mu:F\otimes F\to F,\quad \eta:\mathbf1\to F,
 \quad \Delta:F\to F\otimes F,\quad \varepsilon:F\to\mathbf1.
\end{equation}
Suppressing associators, the algebra and coalgebra axioms are supplemented by
\begin{equation}
 (\mu\otimes\mathrm{id})(\mathrm{id}\otimes\Delta)
 =\Delta\mu
 =(\mathrm{id}\otimes\mu)(\Delta\otimes\mathrm{id}).
 \label{eq:app-frobenius-identity}
\end{equation}
We use normalised specialness,
\begin{equation}
 \mu\Delta=\mathrm{id}_{F},\qquad
 \varepsilon\eta\ne0,
 \label{eq:app-special-normalisation}
\end{equation}
as in Ref.~\cite[Section~5.1]{Runkel2022}. The pairing $b=\varepsilon\mu$ is nondegenerate. Comparing the two identifications $F\to F^\vee$ obtained from this pairing and the pivotal dualities defines the Nakayama automorphism $N_F$; we use the orientation convention of that reference. Its defining property expresses the effect of moving an input of the pairing around the other side of a coupon. A symmetric Frobenius algebra has $N_F=\mathrm{id}$, while the spin construction requires only $N_F^2=\mathrm{id}$.

For the family in this paper, write $g=\Pi G$ and fix a nonzero product map $\mu_{gg}:g\otimes g\to\mathbf1$. The unit channels and this map define $\mu$ on $F=\mathbf1\oplus g$. Associativity must be imposed with the associator of $\widehat{\mathcal C}$; in particular the two products of three copies of $g$ must coincide. If $G$ is self-dual and invertible, the condition is $\theta_G^2=1$. In an associative gauge, the remaining Frobenius data are
\begin{equation}
 \Delta=\frac12\left(
 \mu_{11}^{-1}+\mu_{1g}^{-1}
 +\mu_{g1}^{-1}+\mu_{gg}^{-1}\right),
 \qquad \varepsilon=2\pi_1,
 \label{eq:app-explicit-frobenius}
\end{equation}
where each inverse is embedded into its indicated summand of $F\otimes F$, and the unit inclusion is normalised by $\pi_1\eta=1$. These formulas imply~\eqref{eq:app-special-normalisation}. For general pivotal dimension $d_G=\pm1$ one obtains
\begin{equation}
 N_F=\mathrm{id}_{\mathbf1}\oplus(-d_G)\mathrm{id}_g,
 \qquad \dim(F)=1-d_G,
 \qquad\varepsilon\eta=2.
 \label{eq:app-nakayama-simple-current}
\end{equation}
The dimension and automorphism formulas are those of Ref.~\cite[Lemma~A.22]{Runkel2022}. In the unitary examples $d_G=1$, and $N_F$ acts as the fermion parity operator. It returns $+1$ and $-1$ for even and respectively odd fermion parity state. This gives $\dim(F)=0$. The nonzero specialness normalisation is nevertheless perfectly well-defined. In particular, $\varepsilon\eta$ must not be replaced by $\dim(F)$, and no averaging formula may divide by $\dim(F)$.

For comparison, the product obeys
\begin{equation}
 \mu_{gg}c_{g,g}=-d_G\theta_G\,\mu_{gg}.
\end{equation}
Thus $F$ is commutative precisely when $d_G\theta_G=-1$, whereas it is symmetric precisely when $d_G=-1$. For Ising, $d_G=1$ and $\theta_G=-1$: its algebra is commutative but non-symmetric. Both $\theta_G=1$ and $\theta_G=-1$ can occur in the unitary A-series, and the construction does not require commutativity.

\subsubsection{Modules, spin lifts and boundary fields}
\label{app:subsec:computing_associator}

Write an underlying left module as $(\dot M,\rho_M)$, with
\begin{align}
 \rho_M(\mu\otimes\mathrm{id}_M)
 &=\rho_M(\mathrm{id}_F\otimes\rho_M),&
 \rho_M(\eta\otimes\mathrm{id}_M)&=\mathrm{id}_M.
 \label{eq:app-spin-module-axioms}
\end{align}
Associators in this and subsequent formulas are understood. A spin boundary additionally specifies
\begin{equation}
 M=(\dot M,\rho_M,N_M),\qquad N_M^2=\mathrm{id}_M,
 \qquad N_M\rho_M=\rho_M(N_F\otimes N_M).
 \label{eq:app-spin-module}
\end{equation}
The lift $N_M$ is extra structure, rather than a Nakayama automorphism intrinsically determined by a module. The canonical choice for the modules used here is the parity operator when $N_F=P_F$. Other compatible lifts should not be silently identified. Morphisms \emph{of equivariant boundary objects} commute with the lifts; boundary-field coupons are more general and occupy the spaces
\begin{equation}
 \psi_{\hat{\imath}}\in\operatorname{Hom}_F(M\otimes\hat{\imath},N),
 \qquad \hat{\imath}=i\boxtimes\Pi^p.
 \label{eq:app-spin-boundary-space}
\end{equation}
In the duality conventions of Ref.~\cite[Section~6.3]{Runkel2022}, this is equivalently the ingoing multiplicity space
$\operatorname{Hom}_{\widehat{\mathcal C}}(\hat{\imath},M^\dagger\otimes_F N)$.
The spin adjoint $M^\dagger$ and its evaluation maps involve $N_M$; replacing them by an unqualified ordinary dual can remove a required Nakayama insertion. Equation~\eqref{eq:app-spin-boundary-space} uses the corresponding fixed orientation of the boundary coupon.

There is a finite algorithm for determining these spaces. Normalised specialness makes $F$ separable, so every simple module occurs as a direct summand of an induced module $F\otimes\hat{\imath}$. Decompose those induced modules using their even endomorphism algebras and retain representatives up to parity-preserving isomorphism. The equivariance equations then determine the allowable lifts.

For completeness, the averaging step can be written without a trace or a quantum-dimension denominator. Set $e=\Delta\eta:\mathbf1\to F\otimes F$, let $X=M\otimes\hat{\imath}$ with its induced left action, and take an arbitrary even $f:\dot X\to\dot N$. Define
\begin{equation}
 P_{X,N}(f)
 =\rho_N\circ(\mathrm{id}_F\otimes f)
 \circ(\mathrm{id}_F\otimes\rho_X)
 \circ(e\otimes\mathrm{id}_X).
 \label{eq:app-module-projector}
\end{equation}
Every composition is typed as
$X\to F\otimes F\otimes X\to F\otimes X\to F\otimes N\to N$.
The Frobenius identities make $P_{X,N}(f)$ $F$-linear. If $f$ is already $F$-linear, the module axiom and $\mu e=\eta$ give $P_{X,N}(f)=f$. Hence $P^2=P$ and its image is exactly $\operatorname{Hom}_F(X,N)$. This is the separability-projector version of the averaging construction used in Ref.~\cite[Section~2.1]{FRS4}; it also applies when $\dim(F)=0$.


For Ising, induction on $\mathbf1,\epsilon,\sigma$ gives the modules
in~\eqref{eq:main-ising-modules}. Each has a one-dimensional space
of even $F$-module endomorphisms. The super-endomorphism algebra
$\operatorname{End}^{\mathrm{super}}_F(M)$ also includes the odd
endomorphisms in the vacuum chiral channel, represented by
$\operatorname{Hom}_F(M\otimes\Pi\mathbf1,M)$. One finds
\begin{equation}
 \operatorname{End}^{\mathrm{super}}_F(M_1)
 \cong\operatorname{End}^{\mathrm{super}}_F(M_2)\cong\mathbb C,
 \qquad
 \operatorname{End}^{\mathrm{super}}_F(M_3)\cong\mathrm{Cl}_1.
 \label{eq:app-ising-mq}
\end{equation}
Thus $M_1$ and $M_2$ are of $m$-type, while $M_3$ is of $q$-type:
its odd generator $\gamma$ satisfies $\gamma^2=1$ after a choice
of normalisation. The parity doubling of $M_3$ is intrinsic to
this odd automorphism and does not represent an independent
Chan--Paton multiplicity.

The odd boundary field of conformal weight $1/2$ on $M_1$
belongs to $\operatorname{Hom}_F(M_1\otimes\Pi\epsilon,M_1)$,
rather than the vacuum chiral channel, and therefore does not
contribute to its super-endomorphism algebra. For the canonical
spin lifts, moving a parity-$p$ boundary-field coupon past a lift
introduces a factor $(-1)^p$. This covariance determines the
sign associated with such a move at a spin-sensitive gluing seam.

\subsubsection{Module associators and OPE normalisation}
\label{app:subsec:ope_6j}

Choose bases
\begin{equation}
 \psi_{M\hat{\imath}}^{N,\alpha}
 \in\operatorname{Hom}_F(M\otimes\hat{\imath},N)
\end{equation}
and splitting maps $\bar\psi_{N,\alpha}^{M\hat{\imath}}$ such that
\begin{equation}
 \psi_{M\hat{\imath}}^{N,\alpha}\bar\psi_{N,\beta}^{M\hat{\imath}}
 =\delta^\alpha_\beta\mathrm{id}_{N},\qquad
 \sum_{N,\alpha}\bar\psi_{N,\alpha}^{M\hat{\imath}}
 \psi_{M\hat{\imath}}^{N,\alpha}=\mathrm{id}_{M\otimes\hat{\imath}}.
 \label{eq:app-module-bases}
\end{equation}
These are algebraic splitting normalisations. They do not by themselves fix every physical boundary two-point function.

The matrix $\mathsf G_F$ is defined by expanding one composite coupon in the other basis:
\begin{align}
 &\psi_{N\hat{\jmath}}^{L,\beta}
 (\psi_{M\hat{\imath}}^{N,\alpha}\otimes\mathrm{id}_{\hat{\jmath}})
 \nonumber\\
 &\quad=\sum_{\hat{k},\gamma,\delta}
 [\mathsf G_F]_{N,\alpha\beta}^{\hat{k},\gamma\delta}
 \psi_{M\hat{k}}^{L,\gamma}
 (\mathrm{id}_M\otimes\lambda_{\hat{\imath}\hat{\jmath}}^{\hat{k},\delta})
 a_{M,\hat{\imath},\hat{\jmath}}.
 \label{eq:app-module-associator-definition}
\end{align}
This explicit equation fixes which fusing matrix or inverse fusing matrix is meant; Ref.~\cite{FRS4} uses $\mathsf G[A]$ for this direction. Resolving the compositions with the splitting maps evaluates the matrix through ribbon graphs. It avoids a division by the pivotal dimension of a simple module, which may vanish in a supercategory.

In a compatible conformal-block and field basis,
\begin{equation}
 C^{(MNL)\hat{k},\gamma\delta}_{\hat{\imath},\alpha;
 \hat{\jmath},\beta}
 =[\mathsf G_F^{\mathrm{blocks}}]_{N,\alpha\beta}^{\hat{k},\gamma\delta}.
 \label{eq:OPE_coeff_from_F_symbol_of_modules}
\end{equation}
In the multiplicity-free notation this is the quantity denoted $\mathbf F^{\mathrm{blocks}}[F]$ in the main discussion. Evaluating the same three-point disk with cyclic rather than OPE normalisation additionally contracts the outgoing index with the boundary two-point pairing. Thus an OPE coefficient and a cyclic three-point coefficient coincide only after that pairing has been included or normalised consistently.

For example, under a multiplicity-free change of vertices
$\lambda_{\hat{\imath}\hat{\jmath}}^{\hat{k}}\mapsto g_{\hat{\imath}\hat{\jmath}}^{\hat{k}}\lambda_{\hat{\imath}\hat{\jmath}}^{\hat{k}}$ and boundary bases
$\psi_{M\hat{\imath}}^{N}\mapsto u_{M\hat{\imath}}^{N}\psi_{M\hat{\imath}}^{N}$, the coefficients defined by~\eqref{eq:app-module-associator-definition} transform as
\begin{equation}
 C^{(MNL)\hat{k}}_{\hat{\imath}\hat{\jmath}}\mapsto
 \frac{u_{M\hat{\imath}}^{N}u_{N\hat{\jmath}}^{L}}{u_{M\hat{k}}^{L}g_{\hat{\imath}\hat{\jmath}}^{\hat{k}}}
 C^{(MNL)\hat{k}}_{\hat{\imath}\hat{\jmath}}.
 \label{eq:app-module-gauge-change}
\end{equation}
Consequently, there is no universal conversion factor between a module fusing matrix and a tetrahedral $6j$ symbol independent of the chosen module bases. In the Cardy case $F=\mathbf1$, the module category is $\mathcal C$ itself and the familiar chiral formula is recovered. For nontrivial $F$, the action coefficients and boundary bases must be included. This qualification matters when converting an OPE tensor into a state-sum tensor.

Associativity of three successive module actions implies the mixed pentagon for $\mathsf G_F$ and the chiral associator. In the multiplicity-free blocks convention it reads
\begin{align}
 &\mathbf F^{\mathrm{blocks}}[F]^{(N\hat{\jmath}\hat{k})P}_{L,\hat{q}}
 \mathbf F^{\mathrm{blocks}}[F]^{(M\hat{\imath}\hat{q})P}_{N,\hat{r}}
 \nonumber\\
 &\quad=\sum_{\hat{p}}
 \mathbf F^{\mathrm{blocks}}[F]^{(M\hat{\imath}\hat{\jmath})L}_{N,\hat{p}}
 \mathbf F^{\mathrm{blocks}}[F]^{(M\hat{p}\hat{k})P}_{L,\hat{r}}
 [F^{\hat{\imath}\hat{\jmath}\hat{k}}_{\hat{r}}]^{\mathrm{blocks}}_{\hat{p}\hat{q}}.
 \label{eq:F_pentagon_standard_gauge_app}
\end{align}
The label is retained for cross-references, but the displayed equation explicitly uses the blocks gauge. Changing gauge everywhere using the chosen bulk and module bases gives the equivalent identity. Ordered boundary sewing does not introduce an additional arbitrary parity sign; signs arise from the specified braiding, cyclic transport and spin-framing moves.

The bulk multiplicity spaces require a further projector. With a chosen incoming orientation and fusion convention they can be represented schematically as
\begin{equation}
 \phi^{x}_{\hat{\imath}\hat{\jmath}}
 \in\operatorname{im}Q_x^{(\mathrm{in})}
 \subset\operatorname{Hom}_{\widehat{\mathcal C}}
 (\hat{\imath}\otimes\hat{\jmath},F),
 \qquad x=0\ (\mathrm{NS}),\quad x=1\ (\mathrm R).
 \label{eq:app-spin-bulk-space}
\end{equation}
The detailed braiding and dual conventions in $Q_x^{(\mathrm{in})}$ are those of Ref.~\cite[Eq.~(5.10) and Section~6.2]{Runkel2022}. This image condition is essential: fusion into a summand of $F$ alone does not decide whether a local channel represents an NS or an R bulk field. The following spin and disk calculations implement these projectors together with the boundary lifts.

\subsection{Detailed solutions for the fermionic A-series}
\label{app:A_2}
\subsubsection{A-series data: the simple current and the four sectors}
\label{app:fermionic_models}

We collect the explicit data needed to apply the TQFT construction of Section~\ref{sec:fermionization} to the unitary A-series. Let $q=p+1$, $p\geq3$. The bosonic diagonal model $M(p,q)$ has
\begin{equation}
 c=1-\frac{6(p-q)^2}{pq},\qquad
 h_{r,s}=\frac{(qr-ps)^2-(p-q)^2}{4pq},\qquad
 (r,s)\sim(p-r,q-s),
\end{equation}
where $1\leq r<p$ and $1\leq s<q$. Its simple objects, denoted by the same labels, are self-dual and have multiplicity-free fusion. We denote the irreducible Virasoro representation of label $a$ by $\mathcal V_a$, to distinguish it from the induced boundary module $M_b=F\otimes b$ below. We write $1=(1,1)$ for the vacuum label; an index $0$ on an Ishibashi state also denotes this vacuum.
The nontrivial order-two simple current is
\begin{equation}
 \begin{gathered}
 G=(1,q-1)\sim(p-1,1),\qquad G^2=1,\\
 G(r,s)=(r,q-s)\sim(p-r,s),\qquad h_G=\frac{(p-2)(q-2)}4.
 \end{gathered}
 \label{eq:aseries_G}
\end{equation}
Set $\theta_G=e^{2\pi i h_G}=\pm1$. Then $F=1\oplus\Pi G$ is a particularly tractable instance of the algebra construction in the main text: fusion by $G$ permutes primary labels, and the multiplication and module data reduce to one-dimensional fusion spaces.

The symmetry eigenvalue of an untwisted diagonal primary is
\begin{equation}
 Q(a):=\frac{S_{G,a}}{S_{1,a}}=(-1)^{qr+ps+1}\in\{+1,-1\},
 \qquad a=(r,s).
 \label{eq:aseries_charge}
\end{equation}
Here $S_{G,a}$ denotes the matrix entry with indices $G$ and $a$, whereas $Ga$ denotes the object obtained by fusing $G$ with $a$. The simple-current identity is
\begin{equation}
 S_{Gi,a}=Q(a)S_{i,a},\qquad
 \text{equivalently }\quad S_{i,a}=Q(a)S_{Gi,a}.
 \label{eq:S_matrix_properties_main}
\end{equation}
It follows either by inserting the eigenvalues of the fusion matrix $N_G$ in the Verlinde formula or directly from the minimal-model $S$ matrix.

The $G$-twisted space is $\bigoplus_a\mathcal V_a\otimes\overline{\mathcal V}_{Ga}$. The charge of the summand labelled by $a$ is
\begin{equation}
 Q_{\mathrm{tw}}(a)=e^{2\pi i(h_a-h_{Ga})}
 =\theta_G^{-1}Q(a).
 \label{eq:aseries_twisted_charge}
\end{equation}
The first equality is the non-anomalous defect winding rule~\cite{Fukusumi2021}; the second follows from simple-current monodromy. It can also be checked directly: the Kac weights give
\[
 h_a-h_{Ga}=\frac{(q-2s)(2r-p)}4,\qquad
 2(h_G+h_a-h_{Ga})=qr+ps-2rs-p-q+2.
\]
Since $p+q$ is odd, the last integer has the same parity as $qr+ps+1$, proving Eq.~\eqref{eq:aseries_twisted_charge}.

We fix the Ramond parity convention by retaining the diagonal Ramond-even fields used in this paper. In the notation of Fukusumi--Tachikawa--Zheng this is $\mathsf F'$; their $\mathsf F$ differs by stacking with the Arf theory~\cite{Fukusumi2021}. The two charges above split the bosonic parent into four spaces: $S,T$ are its even and odd untwisted spaces, while $U,V$ are its even and odd $G$-twisted spaces. Table~\ref{tab:states} assigns them to the fermionic sectors in this convention. The $G$ action belongs to the bosonic parent; its relation to fermion parity is specified by this dictionary.
\begin{table}[htbp]
\centering
\renewcommand{\arraystretch}{1.15}
\caption{Sector dictionary in the Ramond parity convention used by the boundary states below. Interchanging $U$ and $T$ in the Ramond column gives the other Arf-related convention.}
\label{tab:states}
\begin{tabular}{ccc}
\toprule
Bosonic parent A & untwisted & $G$-twisted\\
\midrule
$\mathbb Z_2$ even & $S$ & $U$\\
$\mathbb Z_2$ odd & $T$ & $V$ \\
\bottomrule
\end{tabular}
\qquad
\begin{tabular}{ccc}
\toprule
Fermionic theory of this paper & NS & R\\
\midrule
$(-1)^F=+1$ & $S$ & $T$\\
$(-1)^F=-1$ & $V$ & $U$ \\
\bottomrule
\end{tabular}
\end{table}

Combining these charge decompositions with Table~\ref{tab:states} gives the spectrum for the whole unitary family:
\begin{equation}
\begin{aligned}
 \mathcal H_{\mathrm{NS}}^{\mathrm{ev}}
 &=\bigoplus_{Q(a)=+1}\mathcal V_a\otimes\overline{\mathcal V}_a=S,&
 \mathcal H_{\mathrm R}^{\mathrm{ev}}
 &=\bigoplus_{Q(a)=-1}\mathcal V_a\otimes\overline{\mathcal V}_a=T,\\
 \mathcal H_{\mathrm{NS}}^{\mathrm{odd}}
 &=\bigoplus_{Q(a)=-\theta_G}\mathcal V_a\otimes\overline{\mathcal V}_{Ga}=V,&
 \mathcal H_{\mathrm R}^{\mathrm{odd}}
 &=\bigoplus_{Q(a)=\theta_G}\mathcal V_a\otimes\overline{\mathcal V}_{Ga}=U.
\end{aligned}
\label{eq:aseries_general_spectrum}
\end{equation}
The fermionic chiral extensions considered in Section~4 of Ref.~\cite{Runkel2020} have $h_G\in\mathbb Z+\tfrac12$, hence $\theta_G=-1$. In this case, define
\[
 I_{\mathrm{NS}}=\{(r,s):qr+ps\text{ odd}\},\qquad
 I_{\mathrm R}=\{(r,s):qr+ps\text{ even}\}.
\]
Then the even and odd subspaces of the NS sector (respectively, the R sector) are both labelled by $a\in I_{\mathrm{NS}}$ (respectively, $a\in I_{\mathrm R}$), with diagonal labels for even fields and $(a,Ga)$ for odd fields, recovering the spectrum of Ref.~\cite{Runkel2020}. The full unitary A-series also includes integer-spin simple currents: for example, $M(5,6)$ has $h_G=3$ and $\theta_G=+1$. Their odd sectors are given by Eq.~\eqref{eq:aseries_general_spectrum}.

\subsubsection{Branes, boundary states, and boundary-changing fields}
\label{app:subsec:boundary_states_fermionic_minimal}

The Cardy states of the parent theory are
\begin{equation}
 |b\rangle^{\mathrm A}=\sum_a\frac{S_{ba}}{\sqrt{S_{1a}}}|a\rangle\rangle.
 \label{eq:cardy_ishibashi_app}
\end{equation}
Multiplying each Ishibashi component by $Q(a)$ and using Eq.~\eqref{eq:S_matrix_properties_main} gives $|b\rangle^{\mathrm A}\mapsto|Gb\rangle^{\mathrm A}$. Write $i$ for a fixed point $Gi=i$, and write $a\pm$ for the pair $a,Ga$ when $Ga\ne a$, with $a+=a$ and $a-=Ga$. Since the primaries are self-dual, $Gi=i$ is equivalent to $N_{ii}^{G}=1$. Choose one representative from each non-fixed pair and include all fixed points in a set $B$, with $1\in B$. This set labels the underlying boundary conditions; the BCOs and OPE coefficients below instead have \emph{branes} as their endpoints.

For a non-fixed boundary condition $a$, retain the two branes $(a,\rho)$, $\rho\in\{0,1\}$, carrying the two one-dimensional Chan--Paton representations of fermion parity. At a fixed-point boundary $i$, an odd weight-zero boundary field relates the two CP choices, and we retain one brane $i$. Thus the elementary brane set is
\begin{equation}
 \mathcal B=\{(a,0),(a,1):a\in B,\ Ga\ne a\}
 \cup\{i\in B:Gi=i\}.
 \label{eq:aseries_brane_set}
\end{equation}
It is convenient to identify $\mathcal B$ with the full set $\mathcal I$ of bosonic primary labels: $(a,0)$ is labelled by $a$, $(a,1)$ by $Ga$, and a fixed-point brane by $i$. The primary label $b\in\mathcal I$ specifies a brane through this identification. In module language it denotes $M_b=F\otimes b$ with the canonical spin lift. The two non-fixed labels give distinct parity-related $m$-type branes, whereas a fixed label gives one $q$-type brane. Fusion expressions involving $b$ always use its associated bosonic primary label.

In the convention of Table~\ref{tab:states}, the uninserted boundary states of these branes are~\cite{Ebisu2021,Fukusumi2021}
\begin{equation}
\begin{aligned}
 |i\rangle^{\mathrm F}_{\mathrm{NS}}&=\sqrt2|i\rangle^{\mathrm A},&
 |a,\rho\rangle^{\mathrm F}_{\mathrm{NS}}&=\frac{|a+\rangle^{\mathrm A}+|a-\rangle^{\mathrm A}}{\sqrt2},\\
 |i\rangle^{\mathrm F}_{\mathrm R}&=0,&
 |a,\rho\rangle^{\mathrm F}_{\mathrm R}&=(-1)^\rho\frac{|a+\rangle^{\mathrm A}-|a-\rangle^{\mathrm A}}{\sqrt2}.
\end{aligned}
\label{eq:F_boundary_states}
\end{equation}
The two CP choices on a non-fixed boundary have identical NS states and opposite R states. A fixed-point brane has an odd weight-zero boundary field. Its uninserted Ramond amplitude vanishes because of this zero mode; a Ramond boundary state with a zero-mode insertion can be nonzero. The vanishing in Eq.~\eqref{eq:F_boundary_states} refers only to the uninserted state. Equivalently, in the full brane labelling, for every $b\in\mathcal I$,
\begin{equation}
 |b\rangle^{\mathrm F}_{\mathrm{NS}}
 =\sqrt2\sum_{Q(a)=+1}\frac{S_{ba}}{\sqrt{S_{1a}}}|a\rangle\rangle,
 \qquad
 |b\rangle^{\mathrm F}_{\mathrm R}
 =\sqrt2\sum_{Q(a)=-1}\frac{S_{ba}}{\sqrt{S_{1a}}}|a\rangle\rangle.
 \label{eq:aseries_boundary_ishibashi}
\end{equation}
Since $S_{Gb,a}=Q(a)S_{b,a}$, replacing $b$ by $Gb$ leaves the NS expression unchanged and reverses the sign of the R expression, implementing the CP label. For a fixed point $Gi=i$, Eq.~\eqref{eq:S_matrix_properties_main} gives $S_{i,a}=Q(a)S_{i,a}$, so $S_{i,a}=0$ for $Q(a)=-1$ and the R expression vanishes. Thus the brane labels retain the parity choices needed for BCOs and gluing, together with the spin-lift data of Section~\ref{sec:fermionization}.

\paragraph{Spin Cardy condition}
\label{app:subsec:spin_cardy_minimal}
Let $Z_{x|y}(q)=\operatorname{Tr}_{\mathcal H^{\mathrm A}_{x|y}}q^{L_0-c/24}=\sum_jN_{xy}^{j}\chi_j(q)$ be the bosonic annulus partition function in the open channel. With the closed-channel evolution understood, the overlaps following from Eq.~\eqref{eq:F_boundary_states}, for non-fixed representatives $a,b$ and fixed points $i,j$, are
\begin{equation}
\begin{aligned}
 {}^{\mathrm F}_{\mathrm{NS}}\langle i|j\rangle^{\mathrm F}_{\mathrm{NS}}
 &=2Z_{i|j},&
 {}^{\mathrm F}_{\mathrm R}\langle i|j\rangle^{\mathrm F}_{\mathrm R}&=0,\\
 {}^{\mathrm F}_{\mathrm{NS}}\langle a,\rho|b,\tau\rangle^{\mathrm F}_{\mathrm{NS}}
 &=Z_{a+|b+}+Z_{a+|b-},&
 {}^{\mathrm F}_{\mathrm R}\langle a,\rho|b,\tau\rangle^{\mathrm F}_{\mathrm R}
 &=(-1)^{\rho+\tau}(Z_{a+|b+}-Z_{a+|b-}),\\
 {}^{\mathrm F}_{\mathrm{NS}}\langle i|a,\rho\rangle^{\mathrm F}_{\mathrm{NS}}
 &=Z_{i|a+}+Z_{i|a-},&
 {}^{\mathrm F}_{\mathrm R}\langle i|a,\rho\rangle^{\mathrm F}_{\mathrm R}
 &=0.
\end{aligned}
\label{eq:aseries_spin_cardy}
\end{equation}
The NS overlap is an ordinary open-channel trace and the R overlap is a supertrace. The parity of a BCO between two CP-decorated branes is the product of its string parity and the two CP actions, or, additively,
\begin{equation}
 p_{\mathrm{BCO}}=p_{\mathrm{string}}+\rho+\tau\pmod2.
 \label{eq:aseries_bco_cp_parity}
\end{equation}
Changing the CP representation at one endpoint therefore exchanges the even and odd BCO spaces. In the full labelling $b_1,b_2\in\mathcal I$, this is already encoded by replacing that endpoint label by $Gb_1$ or $Gb_2$. The spectrum takes the uniform form
\begin{equation}
 \mathcal H^{\mathrm{ev}}_{b_1|b_2}
 =\bigoplus_{j\in b_1\times b_2}\mathcal V_j,
 \qquad
 \mathcal H^{\mathrm{odd}}_{b_1|b_2}
 =\bigoplus_{j\in b_1\times b_2}\Pi\mathcal V_{Gj}.
 \label{eq:aseries_bco_spectrum}
\end{equation}
Indeed, $N_{Gb_1,b_2}^{j}=N_{b_1,b_2}^{Gj}$, so the endpoint relabelling exchanges the two gradings without changing the underlying ungraded spectrum. For a fixed-point brane, the even and odd character sums agree, explaining the vanishing supertrace and the absence of a distinct CP partner.

For $(p,q)=(3,4)$, the labels $(1,1),(1,2),(1,3)$ are $1,\sigma,\epsilon$ and $G=\epsilon$. The underlying boundary set is $B=\{1,\sigma\}$, while the full brane set has labels $\mathcal I=\{1,\epsilon,\sigma\}$. Their correspondence with Section~\ref{sec:boundary-states-spin-Cardy} is $1\leftrightarrow B_{+,0}$, $\epsilon\leftrightarrow B_{+,1}$, $\sigma\leftrightarrow B_-$, or $0,1,f$ in the main-text shorthand. The vacuum primary label $1$ must therefore be distinguished from the main-text brane shorthand $1$ for $B_{+,1}$.

\subsubsection{Boundary OPE conventions and normalization}\label{app:fermionic_bco_solution}

For each $j\in b_1\times b_2$, with $b_1,b_2$ now labelling branes in $\mathcal I$, write $j$ for the even BCO of Virasoro weight $h_j$ and $gj$ for the odd BCO of weight $h_{Gj}$ in Eq.~\eqref{eq:aseries_bco_spectrum}. Thus $g1$, abbreviated $g$, has chiral label $G$. The prefix $g$ records the \emph{total BCO parity}, including CP factors, and the simple-current action together; it is not itself an additional Kac label. The allowed $j$ depends on the two brane labels. For example, the weight-zero field is even between $B_{+,0}$ and itself but odd between $B_{+,0}$ and $B_{+,1}$. Table~\ref{tab:bco_free_fermion_notation} gives the complete Ising brane spectrum.

\begin{table}[h]
  \centering
\renewcommand{\arraystretch}{1.15}
  \caption{Notation for BCOs between branes in the free-fermion theory. The upper table retains both CP choices of $B_+$, and the lower table lists total BCO parity and conformal weight. The correspondence with Section~\ref{sec:free-fermion-bc} is: $1\equiv 1_B$, $g\equiv\psi_B$, $\epsilon\equiv\epsilon_B$, $\sigma\equiv\sigma_B$, $g\epsilon\equiv a_B$, $g\sigma\equiv\mu_B$.}\label{tab:bco_free_fermion_notation}
  \begin{tabular}{cccc}
\toprule
& $B_{+,0}$ & $B_{+,1}$ & $B_-$ \\
\midrule
$B_{+,0}$ & $1,g$ & $\epsilon,g\epsilon$ & $\sigma,g\sigma$ \\
    $B_{+,1}$ & $\epsilon,g\epsilon$ & $1,g$ & $\sigma,g\sigma$ \\
    $B_-$ & $\sigma,g\sigma$ & $\sigma,g\sigma$ & $1,\epsilon,g,g\epsilon$ \\
\bottomrule
\end{tabular}
  \par\medskip
  \begin{tabular}{cccc}
\toprule
Parity & $h=0$ & $h=1/2$ & $h=1/16$ \\
\midrule
${(-1)}^F = 1$ & $1$ & $\epsilon$ & $\sigma$ \\
    ${(-1)}^F = -1$ & $g\epsilon$ & $g$ & $g\sigma$ \\
\bottomrule
\end{tabular}
\end{table}

In this appendix we label branes directly by $a,b,c,d\in\mathcal I$, using the identification of Section~\ref{app:subsec:boundary_states_fermionic_minimal}. These are the bosonic primary labels associated with the branes denoted $A,B,C,D$ in Section~\ref{sec:bco-structure-constants}. Each endpoint ranges over the full set $\mathcal I$: for a non-fixed representative $b_0\in B$, either $b_0$ or $Gb_0$ may occur at any endpoint, retaining the two CP choices. A fixed-point brane occurs once.

Write $O_x^{ab}$ for a BCO from brane $a$ to brane $b$, with $x$ specifying the field and its total parity. Its OPE is
\begin{equation}
 O_x^{ab}O_y^{bc}\approx\sum_z C^{(abc)z}_{xy}O_z^{ac},
 \qquad p_z=p_x+p_y\pmod2.
 \label{eq:aseries_brane_ope}
\end{equation}
The notation $j,gj$ distinguishes even and odd BCOs. Their underlying Virasoro labels will be written
\[
 \nu(j)=j,\qquad \nu(gj)=Gj.
\]
Virasoro fusing matrices act on these underlying labels. In particular, equal Virasoro labels do not identify fields of different parity at a fixed point.

The crossing equations~\eqref{eq:open_4p_crossing} and~\eqref{eq:open_3p_crossing} retain their form after replacing the brane endpoints by $a,b,c,d$ and summing over the allowed BCO channels. In the four-point equation the channel parities obey $p_p=p_i+p_j$, $p_q=p_j+p_k$, and $p_l=p_i+p_j+p_k$ modulo $2$. The CP factors compose at the shared brane and do not require an extra sign in these ordered crossing equations.

A choice of normalization must be made for each BCO between each ordered pair of branes. Under a rescaling $x^{(ab)}\rightarrow\lambda^{(ab)}_x x^{(ab)}$, the structure constants transform as
\[
C^{(abc)z}_{xy} \rightarrow \frac{\lambda^{(ab)}_x \lambda^{(bc)}_y}{\lambda^{(ac)}_z} C^{(abc)z}_{xy}.
\]
These are OPE coefficients with one output index. To obtain three-point coefficients with all indices lowered, the output must be contracted with the boundary two-point pairing, including the disk vacuum normalization. This conversion will be used when assembling the triangle tensors below.

\subsubsection{Review of the bosonic boundary OPE solution}
\label{app:sec:bosonic_solution}

We review the derivation of the bosonic boundary OPE coefficients in Ref.~\cite{Runkel1998}, organizing the normalization choices and sewing relations into four steps. Here $a,b,c,d$ label Cardy boundaries of the bosonic parent. The $F$-symbols are the Virasoro fusing matrices in the conventions of Eqs.~(46)--(50) of that reference, as discussed in Section~\ref{app:subsec:Fsymbol_conventions}. The next subsection follows the same sequence for fermionic boundary fields.

\paragraph{Step 1: Normalization of Identity BCOs}
The first normalization choice fixes the unit field on each Cardy boundary $b$ by
\[
C^{(bbb)1}_{11} = 1.
\]
The crossing equation~\eqref{eq:open_3p_crossing} with $i=1$, $j=k$, $b=a$ yields $C^{(aac)j}_{1j} = C^{(aaa)1}_{11}$. The other placements of the identity give the same normalization for left and right multiplication. Consequently,
\[
C^{(abb)i}_{i1} = C^{(aab)i}_{1i} = 1.
\]

\paragraph{Step 2: Fixing $C^{(bcb)1}_{jj}$}
Applying~\eqref{eq:open_4p_crossing} with $a=1$, $d=b$, $i=l=b$, $j=k$, $q=1$, and using~\eqref{eq:open_3p_crossing} in the form $C^{(cb1)c}_{jb} C^{(1c1)1}_{cc} = C^{(1cb)b}_{c j}C^{(1b1)1}_{bb}$ leads to
\[
C^{(bcb)1}_{jj} {\left(\Fsymbol{b}{j}{j}{b}{c}{1}\right)}^{-1} =C^{(1bc)c}_{b j}C^{(1cb)b}_{c j}.
\]
Exchanging $b$ and $c$ and employing equation (49) from Runkel~\cite{Runkel1998} gives
\[
S_{1c}C^{(cbc)1}_{jj} = S_{1b}C^{(bcb)1}_{jj}.
\]
It is not possible to set both $C^{(cbc)1}_{jj}$ and $C^{(bcb)1}_{jj}$ to unity. The derivation instead chooses the normalization below by rescaling $j^{(bc)}$ and $j^{(cb)}$:
\begin{equation}
  C^{(cbc)1}_{jj} = {\left(\Fsymbol{c}{c}{b}{b}{1}{j}\right)}^{-1}\frac{S_{11}}{S_{1c}}.
  \label{eq:normalization_of_bco_bosonic}
\end{equation}
A residual gauge freedom remains:
\[
j^{(bc)} \rightarrow \lambda^{(bc)}_j j^{(bc)},\quad j^{(cb)} \rightarrow {(\lambda^{(bc)}_j)}^{-1} j^{(cb)} \quad (b \neq c),
\]
and
\[
j^{(bb)}\rightarrow -j^{(bb)} \quad (b = c,\ j\ne1).
\]
The unit field $1^{(bb)}$ is already fixed and is excluded from these remaining rescalings.

\paragraph{Step 3: Fixing $C^{(b1d)q}_{bd}$}
\label{app:subsubsec:bosgauge}
Setting $a=c=1$, $i=j=b$, $k=l=d$, $p=1$ in~\eqref{eq:open_4p_crossing} and using~\eqref{eq:open_3p_crossing} and~\eqref{eq:normalization_of_bco_bosonic} yields
\[
C^{(b1d)q}_{bd}C^{(d1b)q}_{db} = {\left(\Fsymbol{d}{d}{b}{b}{1}{q}\right)}^2.
\]
The next basis choice rescales $q^{(bd)}$ and $q^{(db)}$ to impose
\[
C^{(b1d)q}_{bd} = C^{(d1b)q}_{db} = \Fsymbol{d}{d}{b}{b}{1}{q} = \Fsymbol{b}{b}{d}{d}{1}{q}
\]

There remains a harmless reparametrization associated with the parent boundary objects,
\[
b^{(1b)} \to \lambda_b b^{(1b)},\quad
b^{(b1)} \to \lambda_b^{-1} b^{(b1)},\quad
q^{(bb')} \to \lambda_b^{-1} \lambda_{b'} q^{(bb')}.
\]
Here $\lambda_1=1$ keeps the unit fixed. The factors $\lambda_b^{-1}\lambda_{b'}$ cancel in every OPE coefficient. This is a redundancy of the field basis, not an additional family of structure constants.

\paragraph{Step 4: Determining general structure constants}
Using~\eqref{eq:open_4p_crossing} with $c=1$, $j=b$, $k=d$, $p=a$, and applying~\eqref{eq:open_3p_crossing} twice, one obtains
\[
C^{(abd)l}_{iq} =  \frac{C^{(a1d)l}_{pk}C^{(b1a)i}_{jp}C^{(aba)1}_{ii}}{C^{(a1a)1}_{pp}C^{(b1d)q}_{jk}}\Fsymbol{i}{j}{k}{l}{p}{q}.
\]
All OPE coefficients on the right-hand side have been fixed by the preceding steps. Substitution and simplification give
\[
C^{(abc)k}_{ij} = \Fsymbol{i}{a}{c}{j}{b}{k}.
\]
This is the bosonic solution in the field normalization just reviewed. The fermionic calculation below retains the distinction between even and odd BCOs when applying the same sewing argument.

\subsubsection{Fermionic OPE coefficients: all parity cases}
\label{app:sec:fermionic_solution}

We now apply the normalization and sewing steps to fermionic BCOs. Each brane endpoint $a,b,c,d$ ranges over the full bosonic label set $\mathcal I$, using the identification of Section~\ref{app:subsec:boundary_states_fermionic_minimal}. For a non-fixed representative $b_0\in B$, both $b_0$ and $Gb_0$ are available at each endpoint, encoding the two CP choices; fixed-point labels occur once. The endpoint labels are used with their actual values in all fusion rules and fusing matrices below. The brane labelled $1$ is $(1,0)$, with CP partner $G$, whereas $1^{(bb)}$ denotes the even unit field on brane $b$.

For each allowed fusion label $j$, the BCOs $j$ and $gj$ have total parities $0$ and $1$ and underlying Virasoro labels $j$ and $Gj$, respectively. In $F$-matrix arguments, $Gx$ denotes fusion of the primary label $x$ with $G$.

The first two steps fix the units and two-point OPE coefficients. Step~3 uses the eight parity assignments of a four-point function to determine the mixed coefficients on the distinguished brane. Step~4 then determines coefficients with this brane in the middle, and Step~5 reconstructs the general coefficients.

\paragraph{Step 1: Normalization of Identity BCOs}
As in the bosonic case, we rescale $1^{(bb)}$ so that $C^{(bbb)1}_{11}=1$. Consequently, for any bosonic BCO $i$ (and analogously for the fermionic BCO $gi$), the relations
\[
C^{(abb)i}_{i1} = C^{(aab)i}_{1i} = 1
\]
hold.
\paragraph{Step 2: Fixing $C^{(cbc)1}_{jj}$ and $C^{(cbc)1}_{gj,gj}$}
Generalizing the bosonic equations leads to
\begin{equation}
  \begin{aligned}
    S_{1c}C^{(cbc)1}_{jj} &= S_{1b}C^{(bcb)1}_{jj},\\
    S_{1c}C^{(cbc)1}_{gj,gj} &= S_{1b}C^{(bcb)1}_{gj,gj}
  \end{aligned}
\end{equation}
for $b,c\in\mathcal I$ and even BCOs $j\in b\times c$. The even BCOs are rescaled as in the purely bosonic case:
\[
C^{(cbc)1}_{jj} = {\left(\Fsymbol{c}{c}{b}{b}{1}{j}\right)}^{-1}\frac{S_{11}}{S_{1c}}.
\]
The fermionic BCOs are rescaled similarly:
\[
C^{(cbc)1}_{gj,gj} = {\left(\Fsymbol{c}{c}{b}{b}{1}{j}\right)}^{-1}\frac{S_{11}}{S_{1c}}.
\]
The remaining unfixed gauge transformations are
\begin{equation}\label{eq:remain_gauge_different_boundary_ferm}
  \begin{aligned}
    j^{(bc)} \rightarrow \lambda^{(bc)}_j j^{(bc)},&\quad
    j^{(cb)} \rightarrow {(\lambda^{(bc)}_j)}^{-1} j^{(cb)}\\
    gj^{(bc)} \rightarrow \lambda^{(bc)}_{gj} gj^{(bc)},&\quad
    gj^{(cb)} \rightarrow {(\lambda^{(bc)}_{gj})}^{-1} {gj}^{(cb)}
  \end{aligned}
\end{equation}
for $b \neq c$, and
\begin{equation}\label{eq:remain_gauge_same_boundary_ferm}
  \begin{aligned}
    j^{(bb)}&\rightarrow -j^{(bb)}\quad(j\ne1),\\
    gj^{(bb)}&\rightarrow -gj^{(bb)}
  \end{aligned}
\end{equation}
for $b = c$.
In the even line of~\eqref{eq:remain_gauge_same_boundary_ferm}, $j=1$ is excluded: the unit $1^{(bb)}$ was fixed in Step~1. The odd line has no such exclusion. In particular, an odd field with underlying Virasoro label $1$ is not the unit, whether it changes the CP label of an $m$-type brane or is the zero mode on a $q$-type brane.

\paragraph{Step 3: Fixing $C^{(1b1)g}_{b,gb}$ and $C^{(1b1)g}_{gb,b}$}
\label{app:subsubsec:step2_ferm}

Set $a=c=1$ in the four-point crossing equation~\eqref{eq:open_4p_crossing}. The fields $i,j$ lie in $\{b,gb\}$ and $k,l$ in $\{d,gd\}$. Their total parity must be even. The intermediate field $p$ on brane $1$ is $1$ or $g$, as fixed by $p_i+p_j$.

For the other channel, let $q\in b\times d$ denote an even BCO, with $gq$ its odd counterpart. The channel is $q$ when $p_j+p_k=0$ and $gq$ when $p_j+p_k=1$. The eight equations below are labelled by $(p_i,p_j,p_k,p_l)$:

\begin{align}
\text{$(0,0,0,0)$:}&\quad
C^{(b1d)q}_{bd} C^{(d1b)q}_{db} = {\left(\Fsymbol{d}{d}{b}{b}{1}{q}\right)}^2, \label{eq:step3_allbos}\\
\text{$(0,0,1,1)$:}&\quad
C^{(b1d)gq}_{b,gd} C^{(d1b)gq}_{gd,b} = \Fsymbol{b}{b}{Gd}{Gd}{1}{Gq}\Fsymbol{d}{d}{b}{b}{1}{q}, \label{eq:step3_klferm}\\
\text{$(1,1,0,0)$:}&\quad
C^{(b1d)gq}_{gb,d} C^{(d1b)gq}_{d,gb} = \Fsymbol{Gb}{Gb}{d}{d}{1}{Gq}\Fsymbol{d}{d}{b}{b}{1}{q}, \label{eq:step3_ijferm}\\
\text{$(1,1,1,1)$:}&\quad
C^{(b1d)q}_{gb,gd} C^{(d1b)q}_{gd,gb} = \Fsymbol{Gb}{Gb}{Gd}{Gd}{1}{q}\Fsymbol{d}{d}{b}{b}{1}{q}, \label{eq:step3_allferm}\\
\text{$(0,1,1,0)$:}&\quad
C^{(b1d)q}_{gb,gd} C^{(d1b)q}_{d,b} = \Fsymbol{b}{Gb}{Gd}{d}{G}{q}\Fsymbol{d}{d}{b}{b}{1}{q}
C^{(1b1)g}_{b,gb}C^{(1d1)g}_{gd,d}, \label{eq:step3_jkferm}\\
\text{$(1,0,0,1)$:}&\quad
C^{(b1d)q}_{b,d} C^{(d1b)q}_{gd,gb} = \Fsymbol{Gb}{b}{d}{Gd}{G}{q}\Fsymbol{d}{d}{b}{b}{1}{q}
C^{(1b1)g}_{gb,b}C^{(1d1)g}_{d,gd}, \label{eq:step3_ilferm}\\
\text{$(1,0,1,0)$:}&\quad
C^{(b1d)gq}_{b,gd} C^{(d1b)gq}_{d,gb} = \Fsymbol{Gb}{b}{Gd}{d}{G}{Gq}\Fsymbol{d}{d}{b}{b}{1}{q}
C^{(1b1)g}_{gb,b}C^{(1d1)g}_{gd,d}, \label{eq:step3_ikferm}\\
\text{$(0,1,0,1)$:}&\quad
C^{(b1d)gq}_{gb,d} C^{(d1b)gq}_{gd,b} = \Fsymbol{b}{Gb}{d}{Gd}{G}{Gq}\Fsymbol{d}{d}{b}{b}{1}{q}
C^{(1b1)g}_{b,gb}C^{(1d1)g}_{d,gd}. \label{eq:step3_jlferm}
\end{align}

The eight equations imply two compatibility constraints, which we derive below. The remaining equations will then determine the individual coefficients after fixing their basis normalization.
Multiply Eqs.~\eqref{eq:step3_allbos} and~\eqref{eq:step3_allferm}, and compare with the product of Eqs.~\eqref{eq:step3_jkferm} and~\eqref{eq:step3_ilferm}. Both products contain the same four OPE coefficients on the left. Equating the right-hand sides and cancelling the common factor ${\bigl(\Fsymbol{d}{d}{b}{b}{1}{q}\bigr)}^2$ gives
\begin{equation}\label{eq:C1b1_1_ferm}
  \Fsymbol{Gb}{Gb}{Gd}{Gd}{1}{q}\Fsymbol{b}{b}{d}{d}{1}{q} =
{\left(\Fsymbol{Gb}{b}{d}{Gd}{G}{q}\right)}^2
C^{(1b1)g}_{b,gb}C^{(1b1)g}_{gb,b}C^{(1d1)g}_{d,gd}C^{(1d1)g}_{gd,d},
\end{equation}
where we used the symmetry $\Fsymbol{b}{Gb}{Gd}{d}{G}{q} = \Fsymbol{Gb}{b}{d}{Gd}{G}{q}$ of the F-symbol.

Comparing the products of Eqs.~\eqref{eq:step3_klferm}, \eqref{eq:step3_ijferm} and of Eqs.~\eqref{eq:step3_ikferm}, \eqref{eq:step3_jlferm} similarly yields
\begin{equation}\label{eq:C1b1_2_ferm}
  \Fsymbol{b}{b}{Gd}{Gd}{1}{Gq}\Fsymbol{Gb}{Gb}{d}{d}{1}{Gq} =
{\left(\Fsymbol{Gb}{b}{Gd}{d}{G}{Gq}\right)}^2
C^{(1b1)g}_{b,gb}C^{(1b1)g}_{gb,b}C^{(1d1)g}_{d,gd}C^{(1d1)g}_{gd,d}.
\end{equation}

Setting $d=1$ in~\eqref{eq:C1b1_1_ferm} and~\eqref{eq:C1b1_2_ferm} gives
\[
\begin{aligned}
    \Fsymbol{Gb}{Gb}{G}{G}{1}{q}\Fsymbol{b}{b}{1}{1}{1}{q} &=
    {\left(\Fsymbol{Gb}{b}{1}{G}{G}{q}\right)}^2
    C^{(1b1)g}_{b,gb}C^{(1b1)g}_{gb,b},\\
    \Fsymbol{b}{b}{G}{G}{1}{Gq}\Fsymbol{Gb}{Gb}{1}{1}{1}{Gq} &=
    {\left(\Fsymbol{Gb}{b}{G}{1}{G}{Gq}\right)}^2
    C^{(1b1)g}_{b,gb}C^{(1b1)g}_{gb,b}.
\end{aligned}
\]
These equations allow the computation of $C^{(1b1)g}_{b,gb}C^{(1b1)g}_{gb,b}$ for all brane labels $b\in\mathcal I$. For $b=1$, both factors are $1$ by the unit normalization. For $b\ne1$, using the gauge transformations~\eqref{eq:remain_gauge_different_boundary_ferm} for the BCOs $b^{(1b)}$ and $b^{(b1)}$, we impose the symmetric condition $C^{(1b1)g}_{b,gb}=C^{(1b1)g}_{gb,b}$.
The sign ambiguity arising from taking square roots is not physically significant, as different sign choices are related by the gauge transformation
\[
b^{(1b)} \to -b^{(1b)},\qquad b^{(b1)} \to -b^{(b1)},
\]
for $b\ne1$, which changes the basis-dependent signs of the corresponding coefficients without changing the correlation functions of consistently transformed fields.
After taking the square root, the remaining continuous gauge transformation for $b\ne1$ is
\[
b^{(1b)} \to \lambda b^{(1b)},\quad b^{(b1)} \to \lambda^{-1} b^{(b1)},\quad
gb^{(1b)} \to \lambda gb^{(1b)},\quad gb^{(b1)} \to \lambda^{-1} gb^{(b1)},
\]
and
\[gb^{(1b)} \to - gb^{(1b)}, gb^{(b1)} \to - gb^{(b1)}, \forall b\neq 1,\quad  g^{(11)} \to -g^{(11)}.
\]

\paragraph{Step 4: Coefficients with the distinguished brane $1$ in the middle}
The bosonic operators $q^{(bd)}, q^{(db)}$ are rescaled as in the bosonic case, leading to
\[
C^{(b1d)q}_{bd} = C^{(d1b)q}_{db} = \Fsymbol{d}{d}{b}{b}{1}{q} = \Fsymbol{b}{b}{d}{d}{1}{q}.
\]
For $b=d$ and $q=1$, this is already fixed by the two-point normalization; the even identity is not rescaled again.
Substituting into Eqs.~\eqref{eq:step3_jkferm} and~\eqref{eq:step3_ilferm} yields
\[
C^{(b1d)q}_{gb,gd} = C^{(d1b)q}_{gd,gb} = \Fsymbol{b}{Gb}{Gd}{d}{G}{q}C^{(1b1)g}_{b,gb}C^{(1d1)g}_{d,gd}.
\]
Equations~\eqref{eq:step3_ijferm}, \eqref{eq:step3_klferm}, \eqref{eq:step3_jlferm}, and~\eqref{eq:step3_ikferm} determine the following four products, respectively:
\[
\begin{aligned}
    \text{(i)} &\equiv C^{(b1d)gq}_{gb,d} C^{(d1b)gq}_{d,gb},\\
    \text{(ii)} &\equiv C^{(b1d)gq}_{b,gd} C^{(d1b)gq}_{gd,b},\\
    \text{(iii)} &\equiv C^{(b1d)gq}_{gb,d} C^{(d1b)gq}_{gd,b},\\
    \text{(iv)} &\equiv C^{(b1d)gq}_{b,gd} C^{(d1b)gq}_{d,gb}.
\end{aligned}
\]
One finds that (i)$\times$(ii) = (iii)$\times$(iv). Since $C^{(1b1)g}_{b,gb} = C^{(1b1)g}_{gb,b}$ was chosen, the right-hand sides of Eqs.~\eqref{eq:step3_ikferm} and~\eqref{eq:step3_jlferm} are equal, implying (iii) = (iv).

For $b = d$, (i) = (ii) $= C^{(b1b)gq}_{gb,b} C^{(b1b)gq}_{b,gb}$, and (iii) = $(C^{(b1b)gq}_{gb,b})^2$, (iv) = $(C^{(b1b)gq}_{b,gb})^2$.
Taking square roots determines $C^{(b1b)gq}_{gb,b}$ and $C^{(b1b)gq}_{b,gb}$ up to an overall sign,
with the relative sign fixed by (i).
Flipping both signs corresponds to the reparametrization $gq^{(bb)} \to -gq^{(bb)}$.

For $b \neq d$, a rescaling of $gq^{(bd)}$ and $gq^{(db)}$ is performed such that $C^{(b1d)gq}_{gb,d} = C^{(d1b)gq}_{d,gb}$. Since (iii) = (iv),
it also follows that $C^{(b1d)gq}_{b,gd} = C^{(d1b)gq}_{gd,b}$.
The values are then determined by taking square roots of (i) and (ii).
The relative sign between $C^{(b1d)gq}_{gb,d}$ and $C^{(b1d)gq}_{b,gd}$ is fixed by (iii), equivalently (iv). Flipping both signs corresponds to $gq^{(bd)} \to -gq^{(bd)}, gq^{(db)} \to -gq^{(db)}$.

Thus, the coefficients $C^{(b1d)q}_{ij}$ and $C^{(b1d)gq}_{ij}$ are determined up to the stated basis transformations.

\paragraph{Step 5: General coefficients}
To reconstruct an arbitrary coefficient, write $x$ for the allowed BCO channel from brane $b$ to brane $d$, including its parity. Thus $x=q$ or $gq$ in the notation of the preceding steps. Using~\eqref{eq:open_4p_crossing} with $c=1$, $j\in\{b,gb\}$ and $k\in\{d,gd\}$ gives
\[
C^{(abd)l}_{ix} = \frac{C^{(a1d)l}_{pk}C^{(b1a)i}_{jp}C^{(aba)1}_{ii}}{C^{(a1a)1}_{pp}C^{(b1d)x}_{jk}}\Fsymbol{\nu(i)}{\nu(j)}{\nu(k)}{\nu(l)}{\nu(p)}{\nu(x)}.
\]
Here $\nu$ extracts the Virasoro label, as defined in Section~\ref{app:fermionic_bco_solution}. Choose intermediate fields for which all displayed channels exist and the denominator is nonzero; inadmissible coefficients vanish. For an admissible coefficient, the right-hand side is determined by the preceding steps. The square-root branches must satisfy the eight equations together. These formulas give the component solution of the boundary-module sewing problem in the chosen normalization; other cyclic orderings retain the corresponding parity and spin signs.

\subsubsection{Shrinkable brane weights and the fixed-point tensors}
\label{sec:shrinkable-weights}

The elimination of the central hole in a glued collection of triangles uses the NS vacuum Ishibashi state. In the parent theory,
\begin{equation}
 |0\rangle\rangle=\sum_{b\in\mathcal I} S_{11}^{1/2}S_{b1}|b\rangle^{\mathrm A},
 \qquad d_b:=\frac{S_{b1}}{S_{11}},
 \label{eq:app-bosonic-vacuum-resolution}
\end{equation}
where $\mathcal I$ is the full set of bosonic primary labels, hence of Cardy boundaries. The coefficients are proportional to the bosonic quantum dimensions $d_b$.

\paragraph{Weights in the full brane sum}
For each brane label $b\in\mathcal I$, Eq.~\eqref{eq:F_boundary_states} reads
\[
 |b\rangle^{\mathrm F}_{\mathrm{NS}}
 =\frac{|b\rangle^{\mathrm A}+|Gb\rangle^{\mathrm A}}{\sqrt2},
\]
including fixed points. Since $S_{Gb,1}=S_{b1}$ and $b\mapsto Gb$ permutes $\mathcal I$, the bosonic vacuum resolution becomes
\[
 |0\rangle\rangle
 =\sum_{b\in\mathcal I}\frac{S_{11}^{1/2}S_{b1}}{\sqrt2}
   |b\rangle^{\mathrm F}_{\mathrm{NS}}.
\]
Thus the symmetric weight of each individual brane is
\begin{equation}
 \boxed{\omega_b=\frac{S_{11}^{1/2}S_{b1}}{\sqrt2},
 \qquad b\in\mathcal I.}
 \label{eq:omega_b}
\end{equation}
Every non-fixed pair contributes two branes with these weights; a fixed-point brane is counted once.

\paragraph{Interpretation in terms of NS module dimensions}
To identify the corresponding module dimensions, use the induced graded modules $M_b=F\otimes b=b\oplus\Pi(Gb)$ with their canonical lifts $N_{M_b}=P_{M_b}$, where $P_{M_b}$ is parity. With the NS trace of Section~\ref{subsec:shrinkable-main},
\begin{equation}
 d_{M_b}^{\mathrm{NS}}
 =\operatorname{tr}_{\widehat{\mathcal C}}(N_{M_b})
 =d_b+d_{Gb}=2d_b.
 \label{eq:app-aseries-module-dimension}
\end{equation}
The last equality uses $d_G=1$ in this unitary family. The $N_{M_b}$ insertion cancels the minus sign of the odd summand in the categorical trace. In contrast, the ordinary signed dimension is $\operatorname{tr}_{\widehat{\mathcal C}}(\mathrm{id}_{M_b})=d_b-d_{Gb}=0$ and cannot supply the shrinking weights.

Writing $|M_b\rangle_{\mathrm{NS}}=|b\rangle^{\mathrm F}_{\mathrm{NS}}$, the weight is equivalently $\omega_b=S_{11}^{3/2}d_{M_b}^{\mathrm{NS}}/(2\sqrt2)$. Hence the complete brane sum takes the module form used in Section~\ref{subsec:shrinkable-main}:
\begin{equation}
 \boxed{
 |0\rangle\rangle
 =\frac{S_{11}^{3/2}}{2\sqrt2}
   \sum_{b\in\mathcal I}d_{M_b}^{\mathrm{NS}}|M_b\rangle_{\mathrm{NS}}.}
 \label{eq:app-full-module-shrinking}
\end{equation}

\paragraph{Combining parity-related NS states}
To express the same resolution using the representative set $B$, the two identical NS states of a non-fixed pair may be combined \emph{in the vacuum resolution}. Let $\Omega_b$ denote the combined weight for $b\in B$, so $\Omega_b=\omega_b+\omega_{Gb}$ for a non-fixed pair and $\Omega_b=\omega_b$ at a fixed point. Using $S_{Gb,1}=S_{b1}$ gives
\begin{equation}
 \sqrt2|0\rangle\rangle
 =2\sum_{\substack{b\in B\\Gb\ne b}}S_{11}^{1/2}S_{b1}|b\rangle^{\mathrm F}_{\mathrm{NS}}
 +\sum_{\substack{b\in B\\Gb=b}}S_{11}^{1/2}S_{b1}|b\rangle^{\mathrm F}_{\mathrm{NS}}.
 \label{eq:fermionic_shrinkable}
\end{equation}
Here $|b\rangle^{\mathrm F}_{\mathrm{NS}}$ in the reduced sum means the common NS state of the branes over $b$. Thus $|0\rangle\rangle=\sum_{b\in B}\Omega_b|b\rangle^{\mathrm F}_{\mathrm{NS}}$ with
\begin{equation}
 \Omega_b=\begin{cases}
 \sqrt2 S_{11}^{1/2}S_{b1},&Gb\ne b,\\[3pt]
 \dfrac{1}{\sqrt2}S_{11}^{1/2}S_{b1},&Gb=b.
 \end{cases}
 \label{eq:app-reduced-boundary-weights}
\end{equation}
Equivalently, these weights are
\begin{equation}
 \begin{gathered}
 \Omega_b=\frac{S_{11}^{3/2}}{\sqrt2}\,
             \frac{d_{M_b}^{\mathrm{NS}}}{e_b},
 \qquad
 e_b:=\dim_{\mathbb C}\operatorname{End}^{\mathrm{super}}_F(M_b),\\[3pt]
 e_b=\begin{cases}1,&Gb\ne b\quad(m\text{-type}),\\
                  2,&Gb=b\quad(q\text{-type}).\end{cases}
 \end{gathered}
 \label{eq:app-reduced-module-shrinking}
\end{equation}
Here $\operatorname{End}^{\mathrm{super}}$ includes both even and odd module endomorphisms. A parity orbit contains $2/e_b$ distinct graded simple modules: two in the $m$-type case, and one in the $q$-type case. Multiplying the weight of each brane in Eq.~\eqref{eq:omega_b} by this number gives Eq.~\eqref{eq:app-reduced-module-shrinking}. The relative factor of two in Eq.~\eqref{eq:app-reduced-boundary-weights} therefore comes from combining parity partners, not from a different definition of quantum dimension. This reduced NS resolution does not identify their brane endpoints in BCOs, OPE coefficients, or Ramond amplitudes.

\paragraph{Ising example}
For the three Ising branes $(0,1,f)$, corresponding to the modules in Eq.~\eqref{eq:main-ising-modules}, the NS dimensions are $(2,2,2\sqrt2)$ and Eq.~\eqref{eq:omega_b} gives the individual shrinking weights $(1/4,1/4,1/(2\sqrt2))$. Since the first two branes have identical NS states, the vacuum resolution fixes only their total weight $1/2$. The asymmetric assignment $(1/2,0,1/(2\sqrt2))$ used in Section~\ref{sec:FPtensor} is another distribution of that same total; all three branes remain available as BCO endpoints.

\paragraph{Assembly of the A-series tensors}
\label{sec:series-fptensor}
Equation~\eqref{eq:aseries_bco_spectrum} supplies the BCOs between branes, the preceding sewing calculation supplies their OPE coefficients, and Eq.~\eqref{eq:omega_b} supplies the weights for filling an NS hole. Each vertex carries a brane label in $\mathcal I$, including its CP choice, and each edge carries a BCO between its endpoint branes.

Contracting the OPE output with the boundary two-point pairing, including the disk vacuum amplitude, gives the three-point coefficients. Expressing them in the two-point-normalized field basis and combining them with the conformal blocks gives the triangle tensors of Section~\ref{sec:FPtensor}. Conversion from the blocks gauge to symmetrized $6j$ symbols uses the basis factors of Section~\ref{app:subsec:Fsymbol_conventions}.

The nonzero triangle components have even total BCO parity, including the CP actions. The CP parity at each vertex contributes twice around a triangle and hence cancels from the total parity. The graded contraction rules and the boundary-tensor construction of Section~\ref{sec:boundary_tensor_def} therefore apply with these A-series brane labels.

At the level of the untruncated CFT amplitudes, local changes of decomposition are governed by sewing and the shrinkable-boundary identity. Any finite descendant or bond-dimension truncation is a separate approximation. The numerical check presented in the main text concerns the free fermion; it should not be read as a numerical verification for every member of the A-series. The spin data used to assemble the triangles, including the boundary automorphisms, must be retained in applying the same construction to a prescribed closed spin surface.

\subsection{Combinatorial spin structures and sewing}\label{app:subsec:spin_structures}

This subsection fixes the geometric conventions used in the disk calculations below. We use the polygonal presentation of spin structures of Ref.~\cite[Section~2.2]{Runkel2022}, specialised to $\mathbb Z_2$. All edge-index equations in this subsection and the disk calculations are modulo $2$; negative representatives are retained when they make the orientation of an edge transparent. A free boundary is a physical conformal boundary. A gluing boundary is a parametrised circle or interval along which two world sheets are sewn. Removing a small disk around a bulk insertion produces a closed gluing boundary.

\paragraph{Admissibility and holonomy.}
Choose a marked edge of every polygon. Every edge outside the free boundary carries an orientation and an index $s_e\in\mathbb Z_2$. For a vertex $v$, let $H_v$ be its incident half-edges and $D_v\subset H_v$ the half-edges for which the face immediately anticlockwise has that edge marked. For an incident half-edge $h$ define
\begin{equation}
    \widehat s_h=
    \begin{cases}
        s_h, & h\text{ leaves }v,\\
        -1-s_h, & h\text{ enters }v.
    \end{cases}
    \label{eq:app-spin-half-edge}
\end{equation}
At an interior vertex, and at a non-distinguished vertex of a gluing boundary, impose
\begin{equation}
    \sum_{h\in H_v}\widehat s_h=|D_v|-|H_v|+1.
    \label{eq:app-spin-admissibility}
\end{equation}
There is no such condition at a free-boundary vertex. At the distinguished vertex $v$ of a closed gluing boundary, its prescribed type is instead imposed by
\begin{equation}
    \delta_v\left(|H_v|-|D_v|-1+\sum_{h\in H_v}\widehat s_h\right)=1-x,
    \qquad
    x=\begin{cases}0,&\mathrm{NS},\\1,&\mathrm{R},\end{cases}
    \label{eq:app-spin-holonomy}
\end{equation}
where $\delta_v=+1$ for ingoing and $-1$ for outgoing parametrisations. These signs agree modulo $2$, but their origin is useful when comparing orientations. The right-hand side is the spin holonomy in the convention of Ref.~\cite[Equations~(2.15)--(2.17)]{Runkel2022}. In particular, the bounding circle has holonomy $1$ and is NS; holonomy $0$ is R. A Ramond insertion is consequently a puncture across which the spin structure does not extend. The disk with three boundary insertions has the unique unpunctured-disk spin structure, whereas the punctured disk used for a bulk--boundary coefficient allows either choice of $x$.

\paragraph{Local changes of the marked graph.}
The following rules are the $r=2$ version of moves (M1)--(M5) in Ref.~\cite[Figures~6 and~7]{Runkel2022}. They preserve the represented spin structure. They are changes of presentation, as opposed to inserting a new spin seam around a non-contractible cycle.
\begin{enumerate}
\item[(M1)] \emph{Polygon gauge change.} For a polygon $p$ and $t\in\mathbb Z_2$, add $t$ to each indexed edge appearing once in $\partial p$. An edge appearing twice is unchanged. For general $r$, the two orientations give $+t$ and $-t$; they agree here.
\item[(M2)] \emph{Edge reversal.} Reverse an interior edge and set $s_e\mapsto-1-s_e=1+s_e$.
\item[(M3)] \emph{Move the marked corner.} Move the marked edge one step anticlockwise around a face, with the index update shown in Figure~6 of Ref.~\cite{Runkel2022}. For $r=2$ both orientation-dependent updates $s_e\mapsto s_e\pm1$ are the same toggle. Equivalently, the affected edge joins the two vertices at which the marked-corner count $|D_v|$ changes. If that edge lies on the free boundary, there is no index to change. Iterate for a longer displacement.
\item[(M4)] \emph{Merge or split polygons.} An interior edge separating distinct polygons may be removed after bringing its marking to the canonical form: it is marked for precisely one incident polygon, has the orientation shown in Figure~7 of Ref.~\cite{Runkel2022}, and has $s_e=0$. The merged polygon inherits the surviving marked edge. The inverse move inserts an edge with this marking and index. Use (M1)--(M3) to prepare the canonical form.
\item[(M5)] \emph{Remove or add a bivalent vertex.} For an interior bivalent vertex, put the edge orientations and markings in the form of Figure~7 of Ref.~\cite{Runkel2022}; the two edge indices are equal. Replace them by one edge with that common index. The analogous free-boundary move has no indices. This move does not alter the prescribed cell structure of a gluing boundary.
\end{enumerate}
The inverses of (M3)--(M5) are included. At fixed orientations and marked edges, quotienting admissible indices by (M1) gives precisely the isomorphism classes of spin structures, with the specified boundary parametrisations \cite[Theorem~2.2]{Runkel2022}.

These polygon moves also specify how to implement changes of triangulation. A $2$--$2$ move is an (M4) merge to a quadrilateral followed by an inverse (M4) split along the other diagonal. A $1$--$3$ subdivision can be implemented by polygon splits and the inverse of (M5), followed by (M1)--(M3) to restore the desired marking. Thus the edge signs must be transported together with the triangulation. Equivalently, in a patch with fixed exterior data the new labels can be obtained by solving the vertex equations. A convenient binary form is
\begin{equation}
  \sum_{h\in H_v}s_{e(h)}
    =1+|D_v|+n_{\mathrm{out}}(v)\pmod2,
  \label{eq:app-spin-binary-update}
\end{equation}
where $n_{\mathrm{out}}(v)$ counts outgoing half-edges, including multiplicity for loops. For example, if the new diagonal is the only unknown at an endpoint $v$, its index is
\begin{equation}
 s_{\mathrm{new}}
 =1+|D'_v|+n'_{\mathrm{out}}(v)
  +\sum_{h\in H'_v\setminus\{h_{\mathrm{new}}\}}s_{e(h)}.
 \label{eq:app-spin-diagonal-update}
\end{equation}
The equation at its other endpoint must agree. This is an efficient check on the choice of marked edges in a numerical implementation; simply copying the old diagonal index is not a spin-preserving rule. The equations at distinguished insertion vertices remain~\eqref{eq:app-spin-holonomy}.

For sewing, match the boundary parametrisations and edge orientations. The index on a glued edge is
\begin{equation}
 s_e=s_e^{\mathrm{in}}+s_e^{\mathrm{out}}\pmod2.
 \label{eq:app-spin-sewing-indices}
\end{equation}
Closed gluing circles must have the same type $x$. This is the sewing rule of Ref.~\cite[Equation~(2.20) and Appendix~A.1.3]{Runkel2022}. The annulus already provides an open--closed example of this formalism \cite[Equation~(2.19)]{Runkel2022}; our application below is to the boundary-mode contractions used to reconstruct the closed-channel spectrum.

\paragraph{The extra boundary datum and transport through an insertion.}
The edge indices determine powers of the Nakayama automorphism $N_F$ on the dual $F$-network. At a conformal boundary they determine powers of the chosen equivariant lift $N_M$ of the boundary module. The compatibility equation is
\begin{equation}
 N_M\rho_M=\rho_M(N_F\otimes N_M),\qquad N_M^2=\mathrm{id}_M.
 \label{eq:app-spin-lift-transport}
\end{equation}
Here $N_F$ is intrinsic to the Frobenius algebra, whereas $N_M$ is part of the boundary data; calling $N_M$ a Nakayama automorphism of a module obscures this distinction. Neither symbol denotes fermion parity in general. In the canonical parity lifts of the Ising calculation, $N_M=P_M$, with $P_M|_{\hat{m}}=(-1)^{|\hat{m}|}$.

A boundary insertion is an $F$-intertwiner $\psi_{\hat{\imath}}:M\otimes\hat{\imath}\to N$, with the specified chiral representation and parity on $\hat{\imath}$. Its spin-lift transform is
\begin{equation}
 \mathscr S_{MN}(\psi_{\hat{\imath}})
 :=N_N\circ\psi_{\hat{\imath}}\circ(N_M\otimes\mathrm{id}_{\hat{\imath}}),
 \qquad \mathscr S_{MN}^{2}=\mathrm{id}.
 \label{eq:app-spin-coupon-action}
\end{equation}
Equation~\eqref{eq:app-spin-lift-transport} ensures that the transformed map is again $F$-linear. A homogeneous basis may therefore be chosen with $\mathscr S_{MN}(\psi)=\eta_\psi\psi$, $\eta_\psi=\pm1$. In that basis,
\begin{equation}
 N_N\circ\psi_{\hat{\imath}}
 =\eta_\psi\,\psi_{\hat{\imath}}\circ(N_M\otimes\mathrm{id}_{\hat{\imath}}).
 \label{eq:app-spin-coupon-sliding}
\end{equation}
For the canonical parity lifts, evenness of the morphism in $\widehat{\mathcal C}$ gives $\eta_\psi=(-1)^{|\hat{\imath}|}$. This is the sign needed when moving a spin seam through a boundary field. Imposing $\eta_\psi=+1$ on all BCOs would incorrectly discard the odd fields. It is the complete decorated contraction---the cyclic boundary fields, the module lifts, the edge labels, and the sewing---that determines the closed spin sector. Boundary field labels by themselves do not do so.

\subsection{Disk correlation functions}\label{app:subsec:disk_correlators}

In this subsection, we use the TFT construction reviewed above to compute the structure constants entering disk correlation functions in the fermionic CFT. We first compute the three-boundary and bulk--boundary structure constants. We then consider one bulk field with an arbitrary number of boundary insertions and show how the spin sector of the bulk field is encoded into the boundary.

\begingroup
\definecolor{RunkelGray}{rgb}{0.50195,0.50195,0.50195}
\definecolor{RunkelBlack}{rgb}{0.11374,0.11374,0.11374}
\definecolor{RunkelGreen}{rgb}{0.09021,0.61569,0.14511}
\definecolor{RunkelBlue}{rgb}{0.09021,0.14511,0.65490}
\definecolor{RunkelRed}{rgb}{1,0,0}

\tikzset{
  Rline/.style={line width=.996pt},
  boundary/.style={Rline,draw=RunkelGray},
  greenedge/.style={Rline,draw=RunkelGreen},
  blueedge/.style={Rline,draw=RunkelBlue},
  blackedge/.style={Rline,draw=RunkelBlack},
  greenarrow/.style={
    greenedge,
    postaction={decorate},
    decoration={markings,mark=at position .52 with {\arrow{Stealth[length=2.9pt,width=4.6pt]}}}
  },
  blackarrow/.style={
    blackedge,
    postaction={decorate},
    decoration={markings,mark=at position .55 with {\arrow{Stealth[length=2.9pt,width=4.6pt]}}}
  },
  bluecircle/.style={
    blueedge,
    postaction={decorate},
    decoration={markings,
      mark=at position .10 with {\arrow{Stealth[length=2.9pt,width=4.6pt]}},
      mark=at position .27 with {\arrow{Stealth[length=2.9pt,width=4.6pt]}},
      mark=at position .44 with {\arrow{Stealth[length=2.9pt,width=4.6pt]}},
      mark=at position .60 with {\arrow{Stealth[length=2.9pt,width=4.6pt]}},
      mark=at position .77 with {\arrow{Stealth[length=2.9pt,width=4.6pt]}},
      mark=at position .94 with {\arrow{Stealth[length=2.9pt,width=4.6pt]}}
    }
  },
  blackcircle/.style={
    blackedge,
    postaction={decorate},
    decoration={markings,
      mark=at position .07 with {\arrow{Stealth[length=2.9pt,width=4.6pt]}},
      mark=at position .23 with {\arrow{Stealth[length=2.9pt,width=4.6pt]}},
      mark=at position .40 with {\arrow{Stealth[length=2.9pt,width=4.6pt]}},
      mark=at position .57 with {\arrow{Stealth[length=2.9pt,width=4.6pt]}},
      mark=at position .73 with {\arrow{Stealth[length=2.9pt,width=4.6pt]}},
      mark=at position .90 with {\arrow{Stealth[length=2.9pt,width=4.6pt]}}
    }
  },
  redtangent/.style={Rline,draw=RunkelRed,-{Stealth[length=3.3pt,width=5.0pt]}},
  gdot/.style={circle,fill=RunkelGreen,draw=RunkelGreen,inner sep=0pt,minimum size=3pt},
  bdot/.style={circle,fill=RunkelBlack,draw=RunkelBlack,inner sep=0pt,minimum size=3pt},
  bluedot/.style={circle,fill=RunkelBlue,draw=RunkelBlue,inner sep=0pt,minimum size=3pt},
  odot/.style={circle,fill=white,draw=RunkelBlack,line width=.996pt,inner sep=0pt,minimum size=4pt}
}

\subsubsection{Three boundary fields}\label{app:subsubsec:three_boundary}

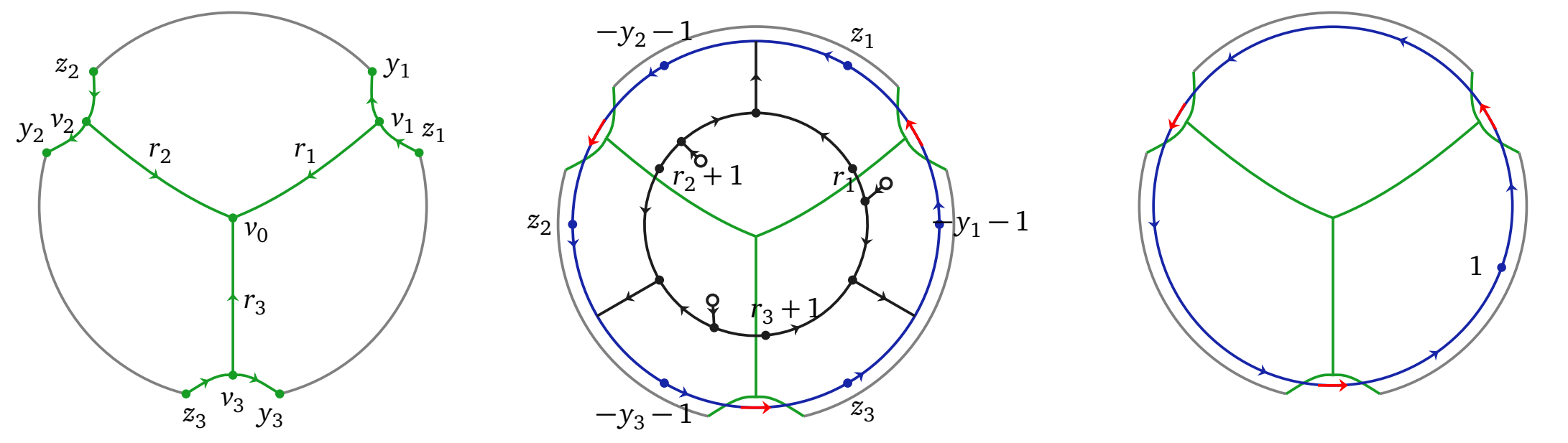
\begin{figure}[!htbp]
\centering
\begin{subfigure}[t]{0.325\textwidth}
\centering
\resizebox{\linewidth}{!}{%
\begin{tikzpicture}[scale=1.05]
  \def\R{2.45}
  \def\Rin{2.14}

  \coordinate (A1) at (16:\R);
  \coordinate (B1) at (44:\R);
  \coordinate (V1) at (30:\Rin);
  \coordinate (A2) at (136:\R);
  \coordinate (B2) at (164:\R);
  \coordinate (V2) at (150:\Rin);
  \coordinate (A3) at (256:\R);
  \coordinate (B3) at (284:\R);
  \coordinate (V3) at (270:\Rin);
  \coordinate (V0) at (0,-0.15);

  \draw[boundary] (0,0) circle (\R);
  \draw[white,line width=3.4pt] (A1) arc[start angle=16,end angle=44,radius=\R];
  \draw[white,line width=3.4pt] (A2) arc[start angle=136,end angle=164,radius=\R];
  \draw[white,line width=3.4pt] (A3) arc[start angle=256,end angle=284,radius=\R];

  \draw[greenarrow] (A1) .. controls (21:2.25) and (25:2.12) .. (V1);
  \draw[greenarrow] (V1) .. controls (35:2.12) and (39:2.25) .. (B1);
  \draw[greenarrow] (A2) .. controls (141:2.25) and (145:2.12) .. (V2);
  \draw[greenarrow] (V2) .. controls (155:2.12) and (159:2.25) .. (B2);
  \draw[greenarrow] (A3) .. controls (261:2.25) and (265:2.12) .. (V3);
  \draw[greenarrow] (V3) .. controls (275:2.12) and (279:2.25) .. (B3);

  \draw[greenarrow] (V1) .. controls (1.25,0.55) and (0.65,0.10) .. (V0);
  \draw[greenarrow] (V2) .. controls (-1.25,0.55) and (-0.65,0.10) .. (V0);
  \draw[greenarrow] (V3) -- (V0);

  \foreach \P in {A1,B1,A2,B2,A3,B3,V1,V2,V3,V0} \node[gdot] at (\P) {};

  \node at (0.92,0.67) {$r_1$};
  \node at (-0.92,0.67) {$r_2$};
  \node at (0.28,-1.25) {$r_3$};

  \node at (20:2.72) {$z_1$};
  \node at (40:2.72) {$y_1$};
  \node at (140:2.72) {$z_2$};
  \node at (160:2.72) {$y_2$};
  \node at (260:2.72) {$z_3$};
  \node at (280:2.72) {$y_3$};

  \node at ($(V0)+(0.30,-0.18)$) {$v_0$};
  \node at ($(V1)+(0.30,-0.02)$) {$v_1$};
  \node at ($(V2)+(-0.30,-0.02)$) {$v_2$};
  \node at ($(V3)+(0,-0.34)$) {$v_3$};
\pgfresetboundingbox
  \path[use as bounding box] (-3.35,-2.95) rectangle (3.35,2.95);
\end{tikzpicture}%
}
\caption{World-sheet decomposition}
\label{fig:app-spin-three-decomposition}
\end{subfigure}\hfill
\begin{subfigure}[t]{0.325\textwidth}
\centering
\resizebox{\linewidth}{!}{%
\begin{tikzpicture}[scale=1.05]

  \def\R{2.45}
  \def\Rin{2.14}
  \def\Rb{2.27}
  \def\Rk{1.38}

  \coordinate (A1) at (16:\R);
  \coordinate (B1) at (44:\R);
  \coordinate (V1) at (30:\Rin);

  \coordinate (A2) at (136:\R);
  \coordinate (B2) at (164:\R);
  \coordinate (V2) at (150:\Rin);

  \coordinate (A3) at (256:\R);
  \coordinate (B3) at (284:\R);
  \coordinate (V3) at (270:\Rin);

  \coordinate (V0) at (0,-0.15);

  \draw[boundary] (0,0) circle (\R);

  \draw[white,line width=3.4pt] (A1) arc[start angle=16,end angle=44,radius=\R];
  \draw[white,line width=3.4pt] (A2) arc[start angle=136,end angle=164,radius=\R];
  \draw[white,line width=3.4pt] (A3) arc[start angle=256,end angle=284,radius=\R];

  \draw[greenedge] (A1) .. controls (21:2.25) and (25:2.12) .. (V1);
  \draw[greenedge] (V1) .. controls (35:2.12) and (39:2.25) .. (B1);

  \draw[greenedge] (A2) .. controls (141:2.25) and (145:2.12) .. (V2);
  \draw[greenedge] (V2) .. controls (155:2.12) and (159:2.25) .. (B2);

  \draw[greenedge] (A3) .. controls (261:2.25) and (265:2.12) .. (V3);
  \draw[greenedge] (V3) .. controls (275:2.12) and (279:2.25) .. (B3);

  \draw[greenedge] (V1) .. controls (1.25,0.55) and (0.65,0.10) .. (V0);
  \draw[greenedge] (V2) .. controls (-1.25,0.55) and (-0.65,0.10) .. (V0);
  \draw[greenedge] (V3) -- (V0);

  \draw[bluecircle] (90:\Rb) arc[start angle=90,end angle=450,radius=\Rb];

  \coordinate (T1) at (30:\Rb);
  \coordinate (T2) at (150:\Rb);
  \coordinate (T3) at (270:\Rb);

  \draw[redtangent] ($(T1)+(300:0.19)$) -- ($(T1)+(120:0.19)$);
  \draw[redtangent] ($(T2)+(60:0.19)$) -- ($(T2)+(240:0.19)$);
  \draw[redtangent] ($(T3)+(180:0.19)$) -- ($(T3)+(0:0.19)$);

  \coordinate (X1) at (60:\Rb);
  \coordinate (Y1) at (0:\Rb);
  \coordinate (Y2) at (120:\Rb);
  \coordinate (X2) at (180:\Rb);
  \coordinate (Y3) at (240:\Rb);
  \coordinate (X3) at (300:\Rb);

  \foreach \P in {X1,Y1,Y2,X2,Y3,X3} \node[bluedot] at (\P) {};

  \node at ($(X1)+(60:0.40)$) {$z_1$};
  \node at ($(Y1)+(0:0.52)$) {$-y_1-1$};
  \node at ($(Y2)+(120:0.46)$) {$-y_2-1$};
  \node at ($(X2)+(180:0.40)$) {$z_2$};
  \node at ($(Y3)+(240:0.48)$) {$-y_3-1$};
  \node at ($(X3)+(300:0.40)$) {$z_3$};

  \coordinate (K1) at (90:\Rk);
  \coordinate (BK1) at (90:\Rb);

  \coordinate (K2) at (210:\Rk);
  \coordinate (BK2) at (210:\Rb);

  \coordinate (K3) at (330:\Rk);
  \coordinate (BK3) at (330:\Rb);

  \draw[blackarrow] (K1) -- (BK1);
  \draw[blackarrow] (K2) -- (BK2);
  \draw[blackarrow] (K3) -- (BK3);

  \coordinate (P1) at (12:\Rk);
  \coordinate (Q1) at ($(P1)+(0.26,0.22)$);

  \coordinate (P2) at (132:\Rk);
  \coordinate (Q2) at ($(P2)+(0.24,-0.24)$);

  \coordinate (P3) at (248:\Rk);
  \coordinate (Q3) at ($(P3)+(-0.02,0.34)$);

  \draw[blackarrow] (Q1) -- (P1);
  \draw[blackarrow] (Q2) -- (P2);
  \draw[blackarrow] (Q3) -- (P3);

  \draw[blackarrow]
      (12:\Rk) arc[start angle=12,end angle=90,radius=\Rk];

  \draw[blackarrow]
      (132:\Rk) arc[start angle=132,end angle=90,radius=\Rk];

  \draw[blackarrow]
      (132:\Rk) arc[start angle=132,end angle=210,radius=\Rk];

  \draw[blackarrow]
      (248:\Rk) arc[start angle=248,end angle=210,radius=\Rk];

  \draw[blackarrow]
      (248:\Rk) arc[start angle=248,end angle=330,radius=\Rk];

  \draw[blackarrow]
      (12:\Rk) arc[start angle=12,end angle=-30,radius=\Rk];

  \foreach \P in {K1,K2,K3} \node[bdot] at (\P) {};

  \foreach \P in {P1,P2,P3} \node[bdot] at (\P) {};

  \foreach \P in {Q1,Q2,Q3} \node[odot] at (\P) {};

  \coordinate (R1p) at (30:\Rk);
  \coordinate (R2p) at (150:\Rk);
  \coordinate (R3p) at (275:\Rk);

  \foreach \P in {R1p,R2p,R3p} \node[bdot] at (\P) {};

  \node at ($(R1p)+(-0.10,-0.16)$) {$r_1$};
  \node at ($(R2p)+(0.60,-0.12)$) {$r_2+1$};
  \node at ($(R3p)+(0.25,0.30)$) {$r_3+1$};

\end{tikzpicture}
}
\caption{Full $F$-defect network}
\label{fig:app-spin-three-full}
\end{subfigure}\hfill
\begin{subfigure}[t]{0.325\textwidth}
\centering
\resizebox{\linewidth}{!}{%
\begin{tikzpicture}[scale=1.05]
  \def\R{2.45}
  \def\Rin{2.14}
  \def\Rb{2.27}

  \coordinate (A1) at (16:\R);
  \coordinate (B1) at (44:\R);
  \coordinate (V1) at (30:\Rin);
  \coordinate (A2) at (136:\R);
  \coordinate (B2) at (164:\R);
  \coordinate (V2) at (150:\Rin);
  \coordinate (A3) at (256:\R);
  \coordinate (B3) at (284:\R);
  \coordinate (V3) at (270:\Rin);
  \coordinate (V0) at (0,-0.15);

  \draw[boundary] (0,0) circle (\R);
  \draw[white,line width=3.4pt] (A1) arc[start angle=16,end angle=44,radius=\R];
  \draw[white,line width=3.4pt] (A2) arc[start angle=136,end angle=164,radius=\R];
  \draw[white,line width=3.4pt] (A3) arc[start angle=256,end angle=284,radius=\R];

  \draw[greenedge] (A1) .. controls (21:2.25) and (25:2.12) .. (V1);
  \draw[greenedge] (V1) .. controls (35:2.12) and (39:2.25) .. (B1);
  \draw[greenedge] (A2) .. controls (141:2.25) and (145:2.12) .. (V2);
  \draw[greenedge] (V2) .. controls (155:2.12) and (159:2.25) .. (B2);
  \draw[greenedge] (A3) .. controls (261:2.25) and (265:2.12) .. (V3);
  \draw[greenedge] (V3) .. controls (275:2.12) and (279:2.25) .. (B3);

  \draw[greenedge] (V1) .. controls (1.25,0.55) and (0.65,0.10) .. (V0);
  \draw[greenedge] (V2) .. controls (-1.25,0.55) and (-0.65,0.10) .. (V0);
  \draw[greenedge] (V3) -- (V0);

  \draw[bluecircle] (90:\Rb) arc[start angle=90,end angle=450,radius=\Rb];

  \coordinate (T1) at (30:\Rb);
  \coordinate (T2) at (150:\Rb);
  \coordinate (T3) at (270:\Rb);
  \draw[redtangent] ($(T1)+(300:0.19)$) -- ($(T1)+(120:0.19)$);
  \draw[redtangent] ($(T2)+(60:0.19)$) -- ($(T2)+(240:0.19)$);
  \draw[redtangent] ($(T3)+(180:0.19)$) -- ($(T3)+(0:0.19)$);

  \coordinate (M) at (340:\Rb);
  \node[bluedot] at (M) {};
  \node at ($(M)+(-0.32,0.02)$) {$1$};
\pgfresetboundingbox
  \path[use as bounding box] (-3.35,-2.95) rectangle (3.35,2.95);
\end{tikzpicture}%
}
\caption{Reduced defect network}
\label{fig:app-spin-three-reduced}
\end{subfigure}
\caption{Three boundary insertions on a disk: the decomposition, the full
$F$-defect network, and the reduced network. }
\label{fig:app-spin-three-boundary}
\end{figure}

\paragraph{Worldsheet and defect network.}
Consider a disk with three boundary insertions. We decompose the disk as Figure~\ref{fig:app-spin-three-boundary}\subref{fig:app-spin-three-decomposition}. For the orientation convention of the working decomposition, the vertex conditions are
\begin{align}
 v_{0}:& \qquad r_1+r_2+r_3 = 0-3+1 = -2 ,\\
 v_{i}:& \qquad (-1-r_i)+(-1-y_i)+z_i = 0-3+1 \\
 &  \qquad \qquad \implies z_{i} = y_i + r_i.
 \label{eq:app-spin-three-local}
\end{align}
Choose the labeling as 
\begin{equation}
 (r_1,r_2,r_3)=(0,-1,-1),\qquad
 (y_1,y_2,y_3)=(0,0,0),\qquad
 (z_1,z_2,z_3)=(0,-1,-1).
 \label{eq:app-spin-three-gauge}
\end{equation}

The corresponding defect network becomes Figure~\ref{fig:app-spin-three-boundary}\subref{fig:app-spin-three-full}. Applying the local reduction rules of~\cite{Runkel2022}, it then gives Figure~\ref{fig:app-spin-three-boundary}\subref{fig:app-spin-three-reduced}.

In the reduced-loop convention, name the segment carrying this residual insertion $M$. The remaining boundary coupons are read in the order
\begin{equation}
 M\xrightarrow{\ \psi_{\hat{\imath}}\ }N
  \xrightarrow{\ \psi_{\hat{\jmath}}\ }L
  \xrightarrow{\ \psi_{\hat{k}}\ }M.
 \label{eq:app-spin-three-cycle}
\end{equation}
The arrows abbreviate the maps $\psi_{\hat{\imath}}:M\otimes\hat{\imath}\to N$, $\psi_{\hat{\jmath}}:N\otimes\hat{\jmath}\to L$, and $\psi_{\hat{k}}:L\otimes\hat{k}\to M$. The closed $M$-ribbon carries the equivariant lift $N_M$ once, which is the dot labelled $1$ in Figure~\ref{fig:app-spin-three-boundary}\subref{fig:app-spin-three-reduced}.

\paragraph{Structure constant.}

The structure constant is obtained by inserting the chosen trivalent basis of $\widehat{\mathcal C}$; the three boundary components are 
\begin{equation}
 \psi_{\hat{\imath}}^{\hat{m}\hat{n}}\,\psi_{\hat{\jmath}}^{\hat{n}\hat{l}}\,\psi_{\hat{k}}^{\hat{l}\hat{m}},\qquad
 \hat{m}\in M,\quad \hat{n}\in N,\quad \hat{l}\in L.
 \label{eq:app-spin-three-components}
\end{equation}
Evaluating the corresponding ribbon graph gives the structure constant
\begin{align}
 C^{(MNL)}_{\hat{\imath}\hat{\jmath}\hat{k}}
 &=
\raisebox{-0.5\height}{%
\begin{tikzpicture}[
baseline=(current bounding box.center),
scale=0.72,
transform shape,
blueedge/.style={
draw=RunkelBlue,
line width=.996pt
},
greenedge/.style={
draw=RunkelGreen,
line width=.996pt
},
bluecircle/.style={
blueedge,
postaction={decorate},
decoration={
markings,
mark=at position .18 with {\arrow{Stealth[length=2.9pt,width=4.6pt]}},
mark=at position .52 with {\arrow{Stealth[length=2.9pt,width=4.6pt]}},
mark=at position .84 with {\arrow{Stealth[length=2.9pt,width=4.6pt]}}
}
},
greenarrow/.style={
greenedge,
postaction={decorate},
decoration={
markings,
mark=at position .55 with {\arrow{Stealth[length=2.9pt,width=4.6pt]}}
}
},
bluedot/.style={
circle,
fill=RunkelBlue,
draw=RunkelBlue,
inner sep=0pt,
minimum size=3pt
},
gvtx/.style={
rectangle,
draw=RunkelGreen,
fill=yellow,
line width=.8pt,
inner sep=0pt,
minimum width=1.25pt,
minimum height=1.25pt
}
]
\def\Rb{2.25}
\draw[bluecircle] (90:\Rb) arc[start angle=90,end angle=450,radius=\Rb];
\coordinate (L) at (220:\Rb);
\coordinate (T) at (270:\Rb);
\coordinate (R) at (300:\Rb);
\coordinate (Td) at (335:\Rb);
\node[bluedot] at (Td) {};
\node[blue] at ($(Td)+(0.26,-0.06)$) {$1$};
\node[blue] at (1.72,1.08) {$\hat{m}$};
\node[blue] at (-0.65,-1.72) {$\hat{n}$};
\node[blue] at (0.35,-1.92) {$\hat{l}$};
\coordinate (N) at (-1.30,-3.52);
\coordinate (B) at (-1.46,-5.00);
\draw[greenarrow] (N) -- (L);
\draw[greenarrow] (N) -- (T);
\draw[greenarrow] (B) -- (N);
\draw[greenarrow] (B) -- (R);
\node[gvtx] at (L) {};
\node[gvtx] at (T) {};
\node[gvtx] at (R) {};
\node[gvtx] at (N) {};
\node[gvtx] at (B) {};
\node[green!50!black] at (-1.92,-2.55) {$\hat{\imath}$};
\node[green!50!black] at (-0.62,-2.55) {$\hat{\jmath}$};
\node[green!50!black] at (-1.80,-4.60) {$\overline{\hat{k}}$};
\node[green!50!black] at (-0.80,-4.60) {$\hat{k}$};
\end{tikzpicture}%
}
\notag\\[1.5ex]
&=\sum_{\substack{\hat{m}\in M,\,\hat{n}\in N,\,\hat{l}\in L}}
\psi_{\hat{\imath}}^{\hat{m}\hat{n}}
\psi_{\hat{\jmath}}^{\hat{n}\hat{l}}
\psi_{\hat{k}}^{\hat{l}\hat{m}}
\bigl[(F^{ m i j}_{ l})^{-1}\bigr]_{ \overline{ k}\, n}\,
\bigl[(F^{ m\,\overline{ k}\, k}_{ m})^{-1}\bigr]_{1l}\,
d_{\hat{m}} (-1)^{|\hat{m}|}.
\label{eq:app-spin-three-result}
\end{align}

\subsubsection{One bulk and one boundary field}
\label{app:subsubsec:bulk_boundary}

\begin{figure}[!htbp]
\centering
\begin{subfigure}[t]{0.325\textwidth}
\centering
\resizebox{\linewidth}{!}{%
\begin{tikzpicture}[
  scale=1.05,
  boundary/.style={draw=RunkelGray, line width=.996pt},
  greenarrow/.style={
    draw=RunkelGreen,
    line width=.996pt,
    postaction={decorate},
    decoration={
      markings,
      mark=at position .55 with {\arrow{Stealth[length=2.9pt,width=4.6pt]}}
    }
  }
]

  \def\R{2.45}
  \def\Rin{2.14}
  \def\Rs{0.52}

  \draw[boundary] (0,0) circle (\R);

  \coordinate (A) at (256:\R);
  \coordinate (B) at (284:\R);
  \coordinate (V) at (270:\Rin);

  \draw[white,line width=3.4pt]
      (A) arc[start angle=256,end angle=284,radius=\R];

  \draw[greenarrow]
      (V) .. controls (261:2.25) and (265:2.12) .. (A);
  \draw[greenarrow]
      (B) .. controls (275:2.12) and (279:2.25) .. (V);

  \coordinate (C) at (0,0.72);
  \coordinate (S) at ($(C)+(270:\Rs)$);

  \draw[greenarrow]
      ($(C)+(20:\Rs)$)
      arc[start angle=20,end angle=380,radius=\Rs];

  \draw[greenarrow] (S) -- (V);

  \node at ($(C)+(0.72,0.12)$) {$u$};
  \node at ($(S)+(0.28,-0.78)$) {$r$};
  \node at ($(V)+(-0.45,0.08)$) {$z$};
  \node at ($(V)+(0.45,0.08)$) {$y$};

\pgfresetboundingbox
  \path[use as bounding box] (-3.35,-2.95) rectangle (3.35,2.95);
\end{tikzpicture}%
}
\caption{World-sheet decomposition}
\label{fig:app-spin-bb-decomposition}
\end{subfigure}\hfill
\begin{subfigure}[t]{0.325\textwidth}
\centering
\resizebox{\linewidth}{!}{%
\begin{tikzpicture}[
  scale=1.05,
  boundary/.style={
    draw=RunkelGray,
    line width=.996pt
  },
  greenedge/.style={
    draw=RunkelGreen,
    line width=.996pt
  },
  blueedge/.style={
    draw=RunkelBlue,
    line width=.996pt
  },
  blackedge/.style={
    draw=RunkelBlack,
    line width=.996pt
  },
  blackarrow/.style={
    blackedge,
    postaction={decorate},
    decoration={
      markings,
      mark=at position .55 with
      {\arrow{Stealth[length=2.9pt,width=4.6pt]}}
    }
  },
  bluecircle/.style={
    blueedge,
    postaction={decorate},
    decoration={
      markings,
      mark=at position .18 with
      {\arrow{Stealth[length=2.9pt,width=4.6pt]}},
      mark=at position .52 with
      {\arrow{Stealth[length=2.9pt,width=4.6pt]}},
      mark=at position .84 with
      {\arrow{Stealth[length=2.9pt,width=4.6pt]}}
    }
  },
  blackcircle/.style={
  blackedge,
  postaction={decorate},
  decoration={
    markings,
    mark=at position .55 with
    {\arrow{Stealth[length=2.9pt,width=4.6pt]}}
  }
},
  odot/.style={
    circle,
    draw=RunkelBlack,
    fill=white,
    line width=.8pt,
    inner sep=0pt,
    minimum size=3.5pt
  },
  bdot/.style={
    circle,
    draw=RunkelBlack,
    fill=RunkelBlack,
    inner sep=0pt,
    minimum size=3pt
  },
  bluedot/.style={
    circle,
    draw=RunkelBlue,
    fill=RunkelBlue,
    inner sep=0pt,
    minimum size=3pt
  },
  redarrow/.style={
    draw=red,
    line width=1.1pt,
    -{Stealth[length=3.2pt,width=4.8pt]}
  }
]

  \def\R{2.45}
  \def\Rin{2.14}
  \def\Rb{2.27}
  \def\Rs{0.53}
  \def\Rk{1.22}


  \draw[boundary] (0,0) circle (\R);

  \coordinate (A) at (256:\R);
  \coordinate (B) at (284:\R);
  \coordinate (V) at (270:\Rin);

  \draw[white,line width=3.4pt]
      (A) arc[start angle=256,end angle=284,radius=\R];

  \draw[greenedge]
      (A) .. controls (261:2.25) and (265:2.12) .. (V);

  \draw[greenedge]
      (V) .. controls (275:2.12) and (279:2.25) .. (B);


  \draw[bluecircle]
      (90:\Rb)
      arc[start angle=90,end angle=450,radius=\Rb];

  \coordinate (Z1) at (238:\Rb);
  \coordinate (Y1) at (302:\Rb);

  \node[bluedot] at (Z1) {};
  \node[bluedot] at (Y1) {};

  \node at ($(Z1)+(-0.18,-0.30)$) {$-z-1$};
  \node at ($(Y1)+(0.18,-0.30)$) {$y$};


  \coordinate (C) at (0,0.68);
  \draw[greenedge] (C) circle (\Rs);

  \coordinate (S) at ($(C)+(270:\Rs)$);
  \draw[greenedge] (S) -- (V);


  \coordinate (RT) at ($(C)+(-0.28,0)$);
  \coordinate (RH) at ($(C)+(0.18,0)$);

  \draw[redarrow] (RT) -- (RH);

  \coordinate (T3) at (270:\Rb);
  \draw[redtangent] ($(T3)+(180:0.19)$) -- ($(T3)+(0:0.19)$);


  \coordinate (KC) at (0,0.52);

  \coordinate (Kright) at ($(KC)+(15:\Rk)$);
  \coordinate (Kbot)   at ($(KC)+(288:\Rk)$);
  \coordinate (Ktop)   at ($(KC)+(90:\Rk)$);

  \draw[blackcircle]
      ($(KC)+(90:\Rk)$)
      arc[start angle=90,end angle=15,radius=\Rk];

  \draw[blackcircle]
      ($(KC)+(288:\Rk)$)
      arc[start angle=288,end angle=90,radius=\Rk];

  \draw[blackcircle]
      ($(KC)+(288:\Rk)$)
      arc[start angle=-72,end angle=15,radius=\Rk];

  \node[bdot] at (Kright) {};
  \node[bdot] at (Kbot) {};
  \node[bdot] at (Ktop) {};

  \coordinate (Rp) at ($(KC)+(225:\Rk)$);
  \node[bdot] at (Rp) {};
  \node at ($(Rp)+(0.26,0.12)$) {$r$};


  \coordinate (BR) at (15:\Rb);
  \draw[blackarrow] (Kright) -- (BR);


  \coordinate (Qb) at ($(Kbot)+(0.20,-0.34)$);
  \draw[blackarrow] (Qb) -- (Kbot);
  \node[odot] at (Qb) {};


  \coordinate (J) at ($(RT)+(0.18,0.18)$);

  \draw[blackedge]
      (RT)
      .. controls ($(RT)+(0.13,0)$)
               and ($(J)+(0,-0.13)$) ..
      (J);

  \coordinate (Nin)  at ($(J)!0.30!(Ktop)$);
  \coordinate (Nout) at ($(J)!0.73!(Ktop)$);

  \draw[blackarrow] (J) -- (Nin);

  \draw[blackarrow] (Nout) -- (Nin);

  \draw[blackarrow] (Nout) -- (Ktop);

  \node[bdot] at (Nin) {};
  \node[bdot] at (Nout) {};


  \coordinate (Ol) at ($(Nout)+(-0.28,0.16)$);
  \draw[blackarrow] (Ol) -- (Nout);
  \node[odot] at (Ol) {};

  \coordinate (Or) at ($(Nin)+(0.27,-0.02)$);
  \draw[blackarrow] (Nin) -- (Or);
  \node[odot] at (Or) {};

\pgfresetboundingbox
  \path[use as bounding box] (-3.35,-2.95) rectangle (3.35,2.95);
\end{tikzpicture}%
}
\caption{Full $F$-defect network}
\label{fig:app-spin-bb-full}
\end{subfigure}\hfill
\begin{subfigure}[t]{0.325\textwidth}
\centering
\resizebox{\linewidth}{!}{%
\begin{tikzpicture}[
  scale=1.05,
  boundary/.style={
    draw=RunkelGray,
    line width=.996pt
  },
  greenedge/.style={
    draw=RunkelGreen,
    line width=.996pt
  },
  blueedge/.style={
    draw=RunkelBlue,
    line width=.996pt
  },
  blackedge/.style={
    draw=RunkelBlack,
    line width=.996pt
  },
  blackarrow/.style={
    blackedge,
    postaction={decorate},
    decoration={
      markings,
      mark=at position .62 with
      {\arrow{Stealth[length=2.9pt,width=4.6pt]}}
    }
  },
  bluecircle/.style={
    blueedge,
    postaction={decorate},
    decoration={
      markings,
      mark=at position .22 with
      {\arrow{Stealth[length=2.9pt,width=4.6pt]}},
      mark=at position .78 with
      {\arrow{Stealth[length=2.9pt,width=4.6pt]}}
    }
  },
  bluedot/.style={
    circle,
    draw=RunkelBlue,
    fill=RunkelBlue,
    inner sep=0pt,
    minimum size=3pt
  },
  redarrow/.style={
    draw=red,
    line width=1.1pt,
    -{Stealth[length=3.2pt,width=4.8pt]}
  }
]

  \def\R{2.45}
  \def\Rin{2.14}
  \def\Rb{2.27}
  \def\Rs{0.53}


  \draw[boundary] (0,0) circle (\R);

  \coordinate (A) at (256:\R);
  \coordinate (B) at (284:\R);
  \coordinate (V) at (270:\Rin);

  \draw[white,line width=3.4pt]
      (A) arc[start angle=256,end angle=284,radius=\R];

  \draw[greenedge]
      (A) .. controls (261:2.25) and (265:2.12) .. (V);

  \draw[greenedge]
      (V) .. controls (275:2.12) and (279:2.25) .. (B);


  \draw[bluecircle]
      (90:\Rb)
      arc[start angle=90,end angle=450,radius=\Rb];

  \coordinate (Z1) at (238:\Rb);
  \coordinate (Y1) at (302:\Rb);

  \node[bluedot] at (Z1) {};
  \node[bluedot] at (Y1) {};

  \node[blue] at ($(Z1)+(-0.18,-0.30)$) {$-z-1$};
  \node[blue] at ($(Y1)+(0.18,-0.30)$) {$y$};


  \coordinate (C) at (0,0.68);
  \draw[greenedge] (C) circle (\Rs);

  \coordinate (S) at ($(C)+(270:\Rs)$);
  \draw[greenedge] (S) -- (V);


  \coordinate (RT) at ($(C)+(-0.28,0)$);
  \coordinate (RH) at ($(C)+(0.18,0)$);
  \draw[redarrow] (RT) -- (RH);
  \coordinate (T3) at (270:\Rb);
  \draw[redtangent] ($(T3)+(180:0.19)$) -- ($(T3)+(0:0.19)$);


  \coordinate (P1) at ($(RT)+(0.38,0.00)$);
  \coordinate (P2) at ($(RT)+(1.00,0.40)$);
  \coordinate (BR) at (28:\Rb);

  \draw[blackarrow]
      (RT)
      .. controls (P1) and (P2) ..
      (BR);

\pgfresetboundingbox
  \path[use as bounding box] (-3.35,-2.95) rectangle (3.35,2.95);
\end{tikzpicture}%
}
\caption{Reduced defect network}
\label{fig:app-spin-bb-reduced}
\end{subfigure}
\caption{One bulk and one boundary insertion on a disk: the decomposition,
 the full $F$-defect network, and the reduced network. }
\label{fig:app-spin-bulk-boundary}
\end{figure}
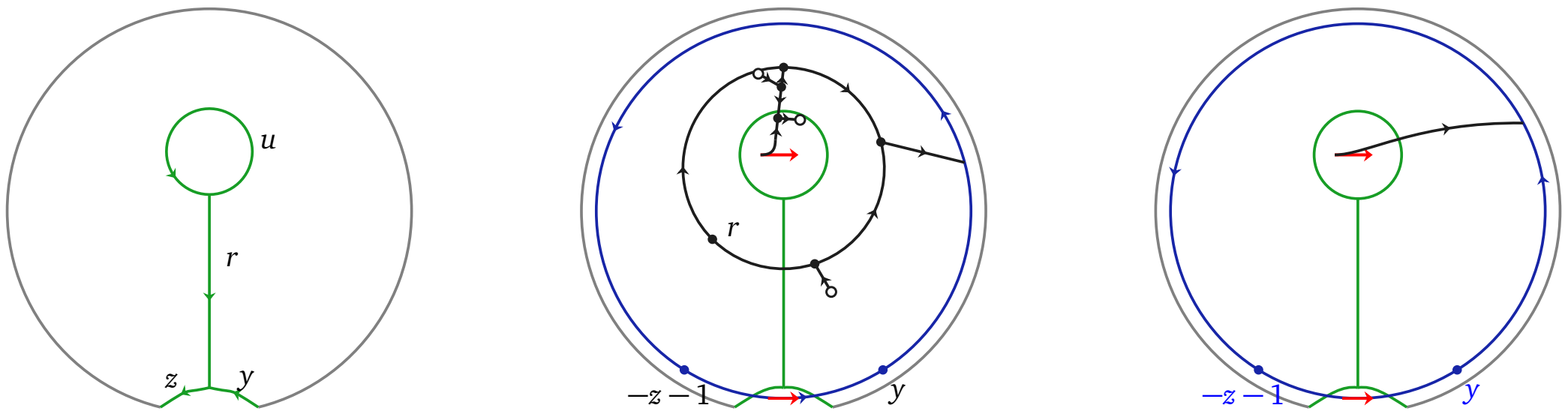

\paragraph{Worldsheet and defect network.}

Consider a disk with one bulk insertion $\phi_{\hat{\imath}\hat{\jmath}}^{x}$ and one boundary insertion $\psi_{\hat{k}}\in\mathrm{Hom}_F(M\otimes\hat{k},M)$. We decompose the disk as in Figure~\ref{fig:app-spin-bulk-boundary}\subref{fig:app-spin-bb-decomposition}. Let $u$ be the index on the bulk insertion circle, $r$ the index on the edge joining it to the boundary, and $y,z$ the two adjacent boundary-edge indices. The vertex $v_0$ lies at the junction with the bulk circle, and $v_1$ is the boundary vertex. The bulk holonomy and the boundary admissibility condition give
\begin{align}
 v_0:\quad &u+(-1-u)+r+3-0-1=1-x,
       &\Longleftrightarrow\quad&r=-x,
       \label{eq:app-spin-bb-bulk-vertex}\\
 v_1:\quad &(-1-r)+(-1-y)+z=0-3+1,
       &\Longleftrightarrow\quad&z=y-x.
       \label{eq:app-spin-bb-boundary-vertex}
\end{align}
Choose the labeling as
\begin{equation}
 u=0,\qquad r=-x,\qquad y=0,\qquad z=-x.
 \label{eq:app-spin-bb-labeling}
\end{equation}

The corresponding defect network is Figure~\ref{fig:app-spin-bulk-boundary}\subref{fig:app-spin-bb-full}. Applying the local reduction rules of~\cite{Runkel2022} gives Figure~\ref{fig:app-spin-bulk-boundary}\subref{fig:app-spin-bb-reduced}.

The two boundary labels in the reduced network leave a power of the module lift $N_M$ with total exponent
\begin{equation}
  y-z-1=x-1=1-x\pmod2.
 \label{eq:app-spin-bb-residual}
\end{equation}
For an NS bulk field, $x=0$, the choice $(y,z)=(0,0)$ leaves one $N_M$ insertion on the boundary. For an R bulk field, $x=1$, the choice $(y,z)=(0,-1)$ leaves no such insertion.

\paragraph{Structure constant.}

Cut the reduced $M$-ribbon immediately next to the residual lift and denote the remaining endomorphism-valued ribbon graph by $\mathcal A^{x}_{\phi,\psi}$. This includes the field coupons, the $F$-action, braidings, and duality maps, but not that residual lift. The diagram fixes the coefficient as
\begin{equation}
 C^{M}_{\phi^{x},\psi}
 =\operatorname{tr}^{\widehat{\mathcal C}}_M
  \!\left(N_M^{1-x}\mathcal A^{x}_{\phi,\psi}\right).
 \label{eq:app-spin-bb-invariant}
\end{equation}
The trace is categorical, so the dimension of an odd simple summand is already negative. For a canonical parity lift, cutting in a simple-component basis gives $\sum_{\hat{m}}d_{\hat{m}}(-1)^{(1-x)|\hat{m}|}(\mathcal A^{x}_{\phi,\psi})_{\hat{m}}$, with an additional matrix trace if a summand has multiplicity.

In the chosen trivalent basis, with the $\hat{\imath},\hat{\jmath}$ legs fused to $\overline{\hat{k}}$ and paired with $\hat{k}$, the corresponding expansion is obtained from the ribbon evaluation of~\cite[Equations~(4.23)--(4.25)]{FRS4}. The closure factor is $(-1)^{(1-x)|\hat{m}|}d_{\hat{m}}=(-1)^{x|\hat{m}|}d_m$, where $d_m$ is the dimension in $\mathcal C$:
\begin{align}
 C^{M}_{\phi_{\hat{\imath}\hat{\jmath}}^{x},\psi_{\hat{k}}}
 &=
 \sum_{\substack{\hat{m},\hat{n}\in M,\,\hat{f}\in F}}
 \psi_{\hat{k}}^{\hat{m}\hat{n}}\,
 \phi_{\hat{\imath}\hat{\jmath}}^{x;\hat{f}}\,
 \rho_M^{(\hat{f}\hat{n})\hat{m}}\,
 \notag\\[-0.2ex]
 &\qquad\times \bigl[(F^{\hat{m}\hat{k}\,\overline{\hat{k}}}_{\hat{m}})^{-1}\bigr]_{1\hat{n}}
 \,R^{\hat{n}\hat{f}}_{\hat{m}}\,
 (-1)^{(1-x)|\hat{m}|}d_{\hat{m}} \sum_{\hat{p}}
 \frac{\theta_{\hat{n}}\theta_{\hat{\imath}}}{\theta_{\hat{p}}}\,
 [F^{\hat{n}\hat{\imath}\hat{\jmath}}_{\hat{m}}]_{\hat{f}\hat{p}}\,
 \bigl[(F^{\hat{n}\hat{\imath}\hat{\jmath}}_{\hat{m}})^{-1}\bigr]_{\hat{p}\,\overline{\hat{k}}}.
 \label{eq:app-spin-bb-result}
\end{align}
The full bulk--boundary correlator is obtained by multiplying this coefficient by the corresponding conformal block. As a normalization check, taking the bulk and boundary insertions to be identities in the NS sector closes the canonical lift on $M_1=\mathbf1\oplus\Pi\epsilon$, giving $\operatorname{tr}(N_{M_1})=1+1=2$ before the common field normalization. Omitting that lift would instead give the incorrect cancellation $1-1=0$. The component expression uses the supplied trivalent and ribbon-bending conventions. In particular, $d_{\hat{m}}$ is the categorical superdimension, and the parity factors in a component formula must be read together with its braidings and duality maps; the convention-independent sector test is the geometric exponent~\eqref{eq:app-spin-bb-residual}.

\subsubsection{One bulk field and multiple boundary fields}
\label{app:subsubsec:multiple}

The world-sheet decomposition for one bulk insertion and $n$ boundary
insertions is shown in Figure~\ref{fig:app-spin-multiple-decomposition}. 

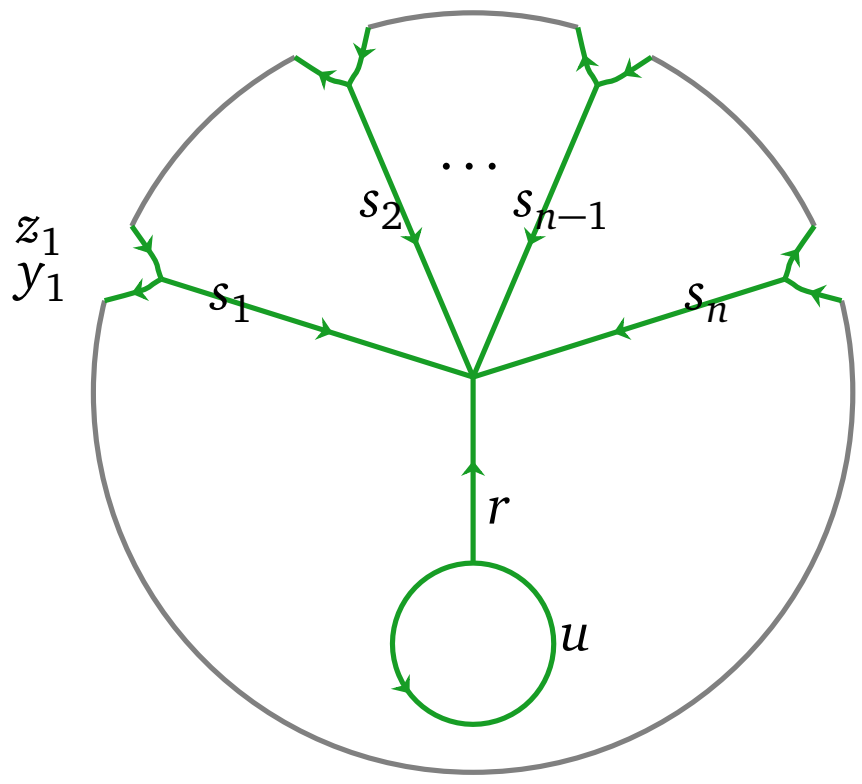
\begin{figure}[!htbp]
\centering
\resizebox{0.38\textwidth}{!}{%
\begin{tikzpicture}[
  scale=1.05,
  boundary/.style={draw=RunkelGray, line width=.996pt},
  greenarrow/.style={
    draw=RunkelGreen,
    line width=.996pt,
    postaction={decorate},
    decoration={
      markings,
      mark=at position .55 with {\arrow{Stealth[length=2.9pt,width=4.6pt]}}
    }
  }
]

  \def\R{2.45}
  \def\Rin{2.14}
  \def\Rs{0.52}


  \draw[boundary] (0,0) circle (\R);


  \coordinate (O) at (0,0.10);


  \coordinate (A1) at (166:\R);
  \coordinate (B1) at (154:\R);
  \coordinate (V1) at (160:\Rin);

  \coordinate (A2) at (118:\R);
  \coordinate (B2) at (106:\R);
  \coordinate (V2) at (112:\Rin);

  \coordinate (A3) at (74:\R);
  \coordinate (B3) at (62:\R);
  \coordinate (V3) at (68:\Rin);

  \coordinate (A4) at (26:\R);
  \coordinate (B4) at (14:\R);
  \coordinate (V4) at (20:\Rin);

  \draw[white,line width=3.4pt]
      (A1) arc[start angle=166,end angle=154,radius=\R];
  \draw[white,line width=3.4pt]
      (A2) arc[start angle=118,end angle=106,radius=\R];
  \draw[white,line width=3.4pt]
      (A3) arc[start angle=74,end angle=62,radius=\R];
  \draw[white,line width=3.4pt]
      (A4) arc[start angle=26,end angle=14,radius=\R];


  \draw[greenarrow]
      (V1) .. controls (163:2.25) and (161:2.12) .. (A1);
  \draw[greenarrow]
      (B1) .. controls (159:2.12) and (157:2.25) .. (V1);

  \draw[greenarrow]
      (V2) .. controls (115:2.25) and (113:2.12) .. (A2);
  \draw[greenarrow]
      (B2) .. controls (111:2.12) and (109:2.25) .. (V2);

  \draw[greenarrow]
      (V3) .. controls (71:2.25) and (69:2.12) .. (A3);
  \draw[greenarrow]
      (B3) .. controls (67:2.12) and (65:2.25) .. (V3);

  \draw[greenarrow]
      (V4) .. controls (23:2.25) and (21:2.12) .. (A4);
  \draw[greenarrow]
      (B4) .. controls (19:2.12) and (17:2.25) .. (V4);


  \draw[greenarrow] (V1) -- (O);
  \draw[greenarrow] (V2) -- (O);
  \draw[greenarrow] (V3) -- (O);
  \draw[greenarrow] (V4) -- (O);


  \coordinate (C) at (0,-1.62);
  \coordinate (Ct) at ($(C)+(90:\Rs)$);

  \draw[greenarrow]
      ($(C)+(20:\Rs)$)
      arc[start angle=20,end angle=380,radius=\Rs];

  \draw[greenarrow] (Ct) -- (O);


  \node at (-1.55,0.58) {$s_1$};
  \node at (-0.58,1.18) {$s_2$};
  \node at (0.58,1.18) {$s_{n-1}$};
  \node at (1.52,0.58) {$s_n$};

  \node at (0,1.43) {$\cdots$};

  \node at (0.17,-0.76) {$r$};
  \node at (0.66,-1.58) {$u$};

  \node at ($(V1)+(-0.78,0.28)$) {$z_1$};
  \node at ($(V1)+(-0.78,-0.02)$) {$y_1$};

\end{tikzpicture}%
}
\caption{Decomposition for one bulk field and $n$ boundary fields. }
\label{fig:app-spin-multiple-decomposition}
\end{figure}

The residual boundary-lift exponent is again fixed by the bulk NS/R type.
The admissibility conditions give 
\begin{align}
    (-1-u) + u + r + 3 - 0 - 1 = 1 - x &\Leftrightarrow& r = -x,\\
    \sum_{i = 1}^{n} (-1 - s_{i}) -r - 1 = 0- (n+1) +1 &\Leftrightarrow& \sum_{i = 1}^{n} s_{i} = -1 + x,\\
    (-1-y_{i}) + z_{i} + s_{i} = 0 - 3 + 1 &\Leftrightarrow& z_{i} = y_{i} - s_{i} - 1.
\end{align}
We can choose a representative with $s_{1} = -1+x$ and $s_{i} = 0 , \ i \in \{2,\dots, n\}$. We set $y_{i} = 0$. 
We have
\begin{align}
    y_{1} = 0, z_{1} = x &\implies y_{1} = 0, -z_{1}-1 = 1-x,\\
    y_{i} = 0, z_{i} = -1 &\implies y_{i} = 0, -z_{i}-1 = 0, \quad i \in \{2,\dots,n\}.  
\end{align}

Thus, after reduction in this representative, an NS bulk field leaves one boundary-lift insertion and an R bulk field leaves none, independently of $n$. 

Apply the boundary OPE to adjacent fields, following the cyclic sequence of boundary modules and keeping the fermionic insertions in order. The sequence closes with $M_n=M_0$. The conformal-block coefficients are obtained from the boundary OPE data of~\eqref{eq:app-spin-three-result}, in the convention of Section~\ref{sec:bco-structure-constants}, and the bulk--boundary ribbon coefficient~\eqref{eq:app-spin-bb-invariant}, summing over intermediate channels. Combining these coefficients with the conformal blocks gives the full correlator.

\endgroup

\subsection{Three-dimensional interpretation of boundary tensors}\label{app:subsec:3d_boundary}

We consider here the bosonic Cardy boundary construction used in Section~\ref{sec:boundary-tensors}. The strip partition function admits a strange-correlator description $\langle\Omega|\Psi\rangle$, where $|\Psi\rangle$ is a ground state of the Levin--Wen model with boundaries.

\begin{figure}[htbp]
    \centering
    \begin{subfigure}{0.49\textwidth}
        \includegraphics[width=0.9\textwidth]{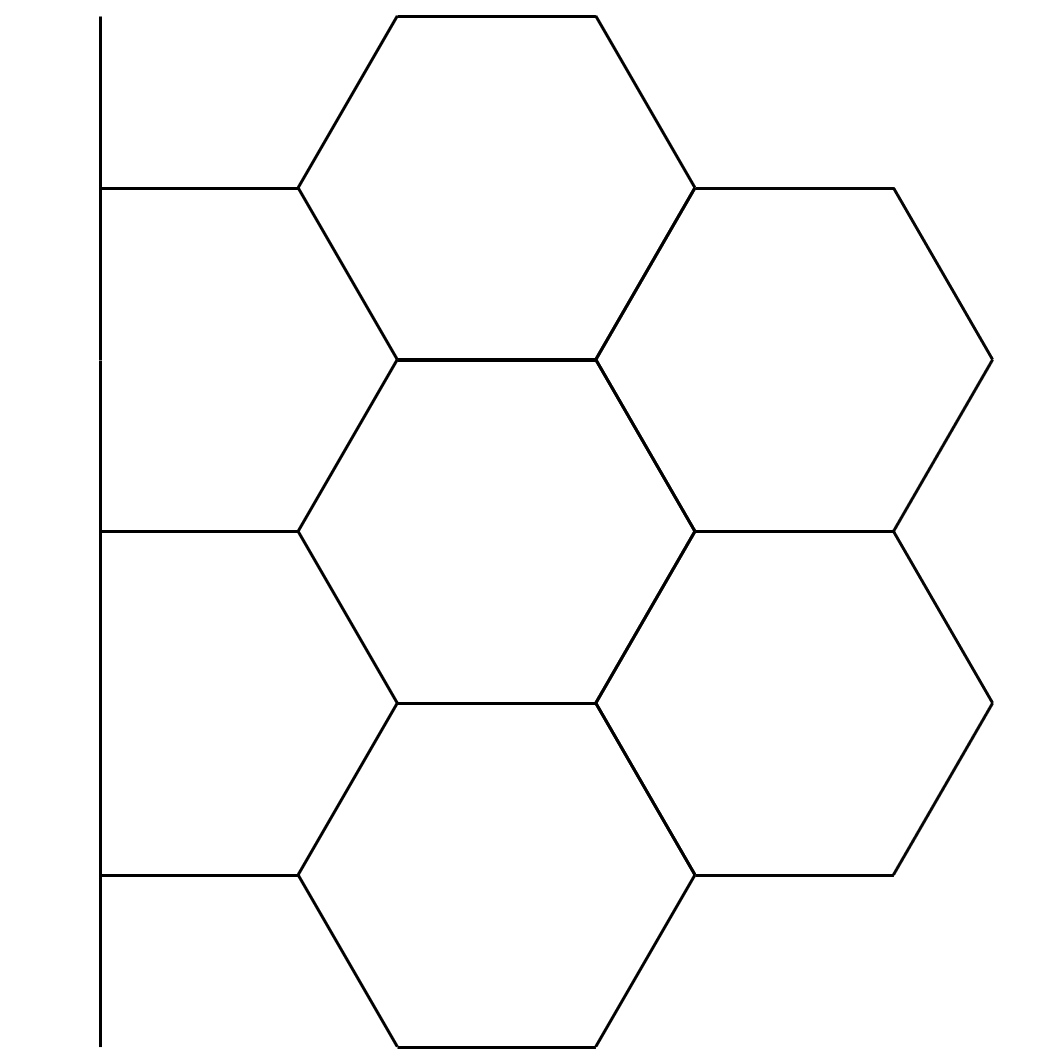}
        \caption{}\label{fig:boundary_LevinWen}
    \end{subfigure}
    \begin{subfigure}{0.49\textwidth}
        \includegraphics[width=0.9\textwidth]{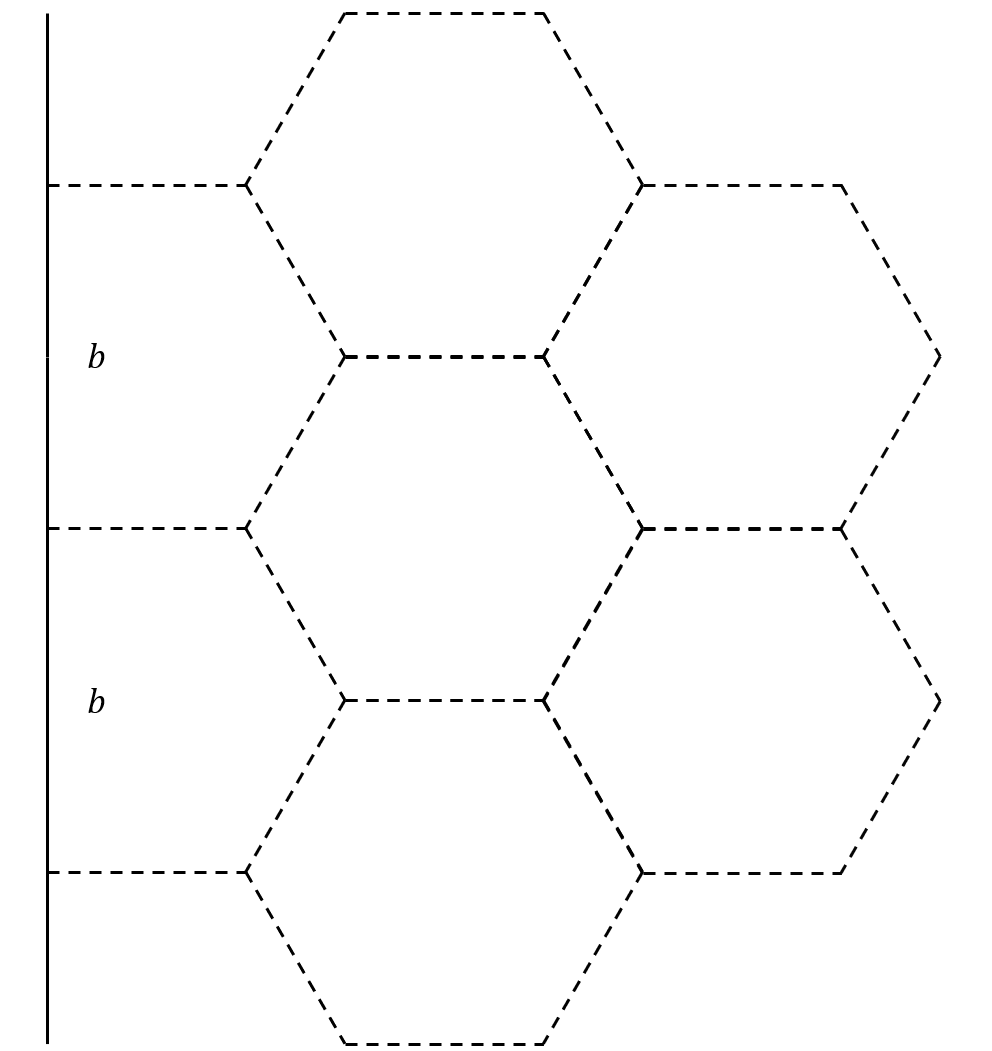}
        \caption{}\label{fig:eigenstate_of_A}
    \end{subfigure}
    \caption{(a) A lattice with boundary. (b) An eigenstate of all $A_v$ operators.}
\end{figure}

Consider the Levin--Wen model on a graph with boundary (Fig.~\ref{fig:boundary_LevinWen}), with prescribed boundary labels and plaquette operators that preserve them. The Hamiltonian $H=-\sum_v A_v-\sum_p B_p$ involves vertex projectors $A_v$ that enforce the fusion rules at each vertex, and plaquette projectors $B_p=\frac{1}{\sum_i d_i^2}\sum_s d_s B_p^s$, where $B_p^s$ adds an $s$-type loop around the plaquette $p$. The $A_v$ and $B_p$ are mutually commuting projectors. For any state satisfying $A_v|\phi\rangle=|\phi\rangle$ at every vertex, $\prod_p B_p|\phi\rangle$ is a ground state of $H$, provided the projected state is nonzero.

Locally, assigning the vacuum label $1$ to the bulk edges and a simple-object label $b$ to the adjacent boundary edges satisfies the vertex constraints, as illustrated in Fig.~\ref{fig:eigenstate_of_A}. Here $b$ labels the boundary shown in the local diagram; different physical boundaries may carry different labels. Let $|\Phi\rangle$ be an initial state satisfying all vertex constraints and compatible with the prescribed boundary labels. A ground state is then obtained as $|\Psi\rangle\propto\prod_p B_p|\Phi\rangle$, provided the projection is nonzero.

These ground states admit a tensor-network representation~\cite{Buerschaper2008}.

\begin{figure}[htbp]
    \centering
    \begin{subfigure}{0.3\textwidth}
        \includegraphics[width=\textwidth]{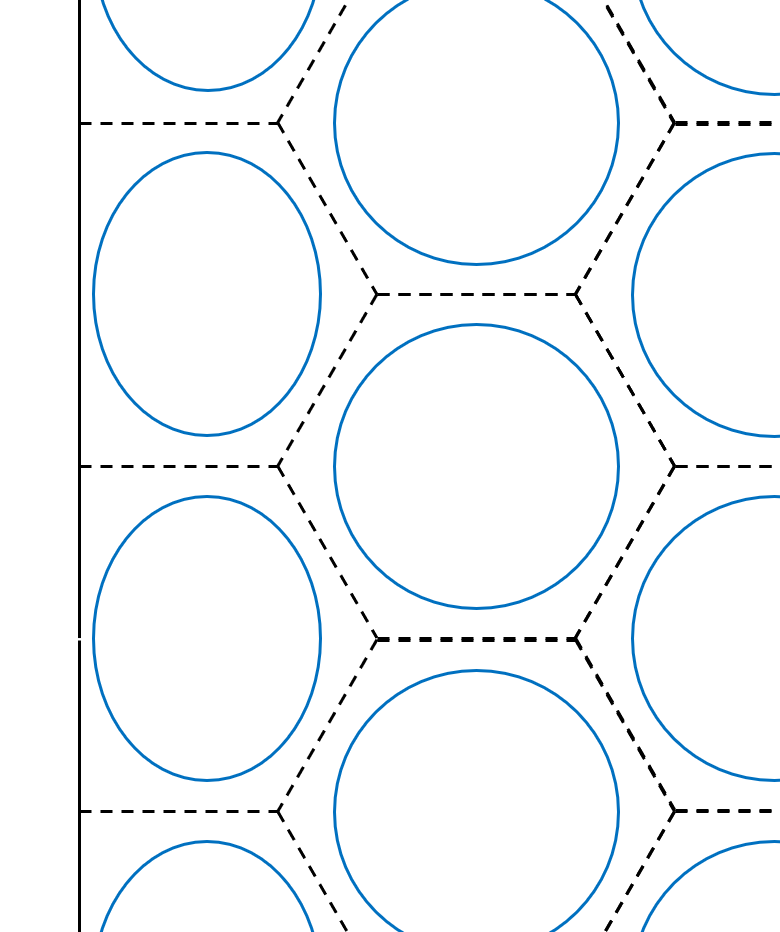}
        \caption{}\label{fig:boundary_a}
    \end{subfigure}
    \begin{subfigure}{0.3\textwidth}
        \includegraphics[width=\textwidth]{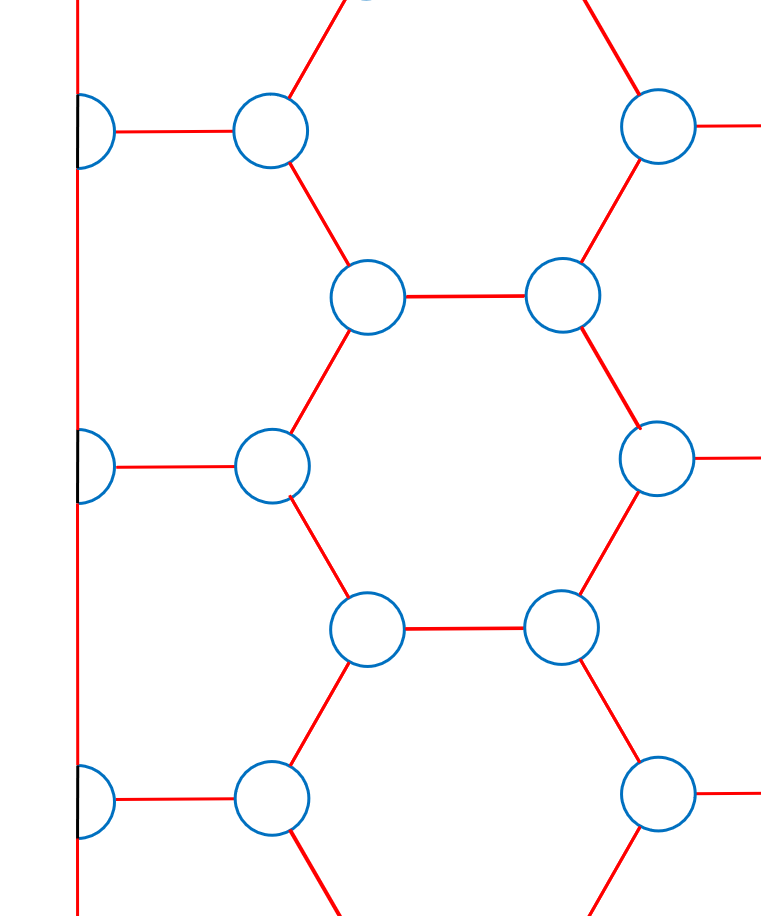}
        \caption{}\label{fig:boundary_b}
    \end{subfigure}
    \begin{subfigure}{0.3\textwidth}
        \includegraphics[width=\textwidth]{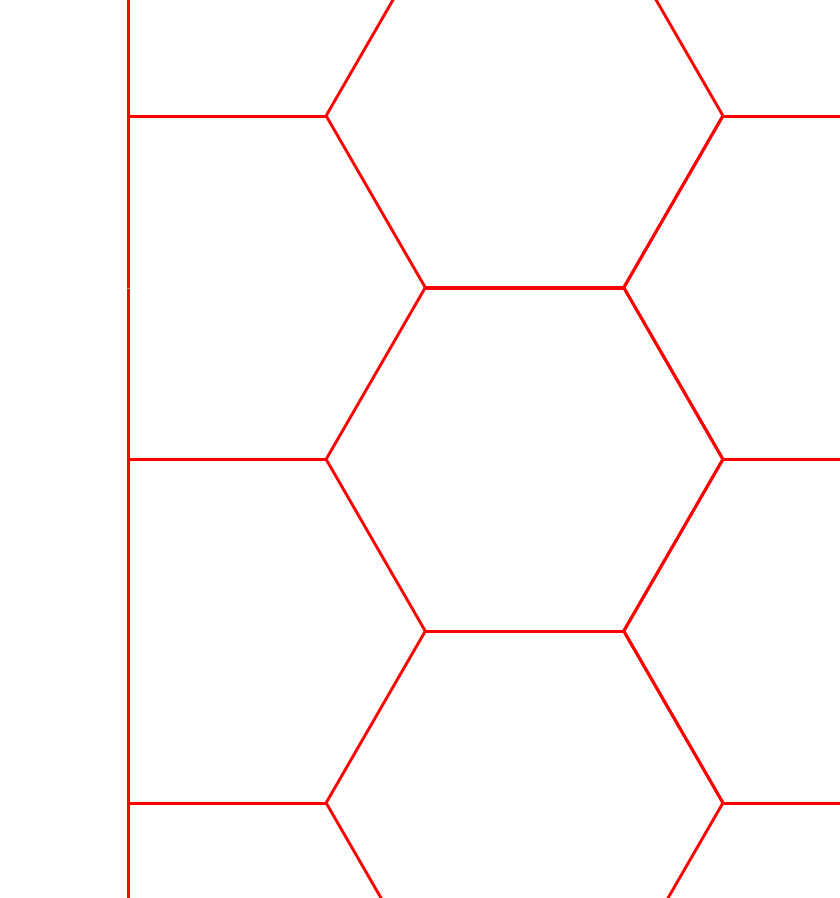}
        \caption{}\label{fig:boundary_c}
    \end{subfigure}
    \caption{F-move transformations on the $B_p$-decorated lattice.}
\end{figure}

As shown in Fig.~\ref{fig:boundary_a}--\ref{fig:boundary_c}, applying F-moves to the $B_p$-decorated lattice successively transforms the configuration into a purely combinatorial tensor network, introducing factors of the $6j$ symbol at each triangle. The resulting ground state admits the following tensor-network representation, with the prescribed boundary labels held fixed:
\begin{equation}
|\Psi\rangle = \sum_{\{i_e\}}\left[\sum_{\{a_v\}}\prod_e d_{i_e}^{1/2}\prod_v\omega_{a_v}\prod_{\triangle}\sixj{i}{j}{k}{c}{a}{b}\right]|\{i_e\}\rangle .
\end{equation}
Here $|\{i_e\}\rangle$ denotes an edge-label basis state compatible with the prescribed boundary labels. In each triangle, $i,j,k$ denote its edge labels and $a,b,c$ its vertex labels. The sums run over the unfixed internal labels, while the labels on each physical boundary are held fixed and need not coincide between different boundaries. Since the boundary tensors of Section~\ref{sec:boundary_tensor_def} factorize into three-point coefficients $C^{abc}_{ijk}$, expressible in terms of the chiral $6j$ symbols in the present Cardy setting, and conformal blocks, the strip partition function is recovered as $\langle\Omega|\Psi\rangle$.
\nolinenumbers
\end{document}